\documentclass[twocolumn]{aastex631}

\usepackage{}
\usepackage{gensymb}  % \degree
\usepackage{multirow}  % https://tex.stackexchange.com/questions/72945/how-to-merge-cells-vertically

\shorttitle{EMPG impostors}
\shortauthors{Hsiao et al.}

\begin{document}

% \title{A \HeIIw\ Emitter with Extremely Dense ISM or Carbon-Enhanced Metal-Poor at $z\sim5.79$?}

\title{An Optical Illusion:\\
High Electron Densities Create Extremely Metal-Poor Galaxy Impostors}

\correspondingauthor{Tiger Hsiao}
\email{tiger.hsiao@utexas.edu}

%%%%%%%%%%%%%%
% Affiliations
\newcommand{\CfA}{\affiliation{Center for Astrophysics \text{\textbar} Harvard \& Smithsonian, 60 Garden Street, Cambridge, MA 02138, USA}}

\newcommand{\STScI}{\affiliation{Space Telescope Science Institute (STScI), 3700 San Martin Drive, Baltimore, MD 21218, USA}}

\newcommand{\JHU}{\affiliation{Center for Astrophysical Sciences, Department of Physics and Astronomy, The Johns Hopkins University, 3400 N Charles St. Baltimore, MD 21218, USA}}

\newcommand{\ESAAURA}{\affiliation{Association of Universities for Research in Astronomy (AURA), Inc.~for the European Space Agency (ESA)}}

\newcommand{\Austin}{\affiliation{Department of Astronomy, The University of Texas at Austin, 2515 Speedway, Austin, Texas 78712, USA}}

\newcommand{\BGU}{\affiliation{Physics Department, Ben-Gurion University of the Negev, P.O. Box 653, Be'er-Sheva 84105, Israel}}

\newcommand{\kapteyn}{\affiliation{Kapteyn Astronomical Institute, University of Groningen, 9700 AV Groningen, The Netherlands}}

\newcommand{\CAB}{\affiliation{Centro de Astrobiología (CAB), CSIC-INTA, Ctra. de Ajalvir km 4, Torrejón de Ardoz, E-28850, Madrid, Spain}}

\newcommand{\Cambridge}{\affiliation{Kavli Institute for Cosmology, University of Cambridge, Madingley Road, Cambridge CB3 0HA, UK}}

\newcommand{\Cavendish}{\affiliation{Cavendish Laboratory, University of Cambridge, 19 JJ Thomson Avenue, Cambridge CB3 0HE, UK}}

\newcommand{\UCL}{\affiliation{Department of Physics and Astronomy, University College London, Gower Street, London WC1E 6BT, UK}}

\newcommand{\Arizona}{\affiliation{Department of Astronomy / Steward Observatory, University of Arizona, 933 N Cherry Ave, Tucson, AZ 85721}}

\newcommand{\TsingHua}{\affiliation{Department of Astronomy, Tsinghua University, Beijing 100084, China}}

\newcommand{\Chiba}{\affiliation{Center for Frontier Science, Chiba University, 1-33 Yayoi-cho, Inage-ku, Chiba 263-8522, Japan}}

\newcommand{\ICRR}{\affiliation{Institute for Cosmic Ray Research, The University of Tokyo, 5-1-5 Kashiwanoha, Kashiwa, Chiba 277-8582, Japan}}

\newcommand{\NAOJ}{\affiliation{National Astronomical Observatory of Japan, 2-21-1 Osawa, Mitaka, Tokyo 181-8588, Japan}}

\newcommand{\Sokendai}{\affiliation{Department of Astronomical Science, SOKENDAI (The Graduate University for Advanced Studies), Osawa 2-21-1, Mitaka, Tokyo, 181-8588, Japan}}

\newcommand{\IPMU}{\affiliation{Kavli Institute for the Physics and Mathematics of the Universe (WPI), University of Tokyo, Kashiwa, Chiba 277-8583, Japan}}

\newcommand{\Dawn}{\affiliation{Cosmic Dawn Center (DAWN), Denmark}}

\newcommand{\Cop}{\affiliation{Niels Bohr Institute, University of Copenhagen, Jagtvej 128, DK-2200 Copenhagen N, Denmark}}

\newcommand{\Tokyo}{\affiliation{Department of Astronomy, Graduate School of Science, the University of Tokyo, 7-3-1 Hongo, Bunkyo, Tokyo 113-0033, Japan}}

\newcommand{\Oxford}{\affiliation{Department of Physics, University of Oxford, Denys Wilkinson Building, Keble Road, Oxford OX1 3RH, UK}}

\newcommand{\MIT}{\affiliation{MIT Kavli Institute for Astrophysics and Space Research, 70 Vassar Street, Cambridge, MA 02139, USA}}

\newcommand{\Dunlap}{\affiliation{David A. Dunlap Department of Astronomy and Astrophysics, University of Toronto, 50 St. George Street, Toronto, Ontario, M5S 3H4, Canada}}

\newcommand{\IAP}{\affiliation{Institut d'Astrophysique de Paris, CNRS, Sorbonne Université, 98bis Boulevard Arago, 75014, Paris, France}}

\newcommand{\CFC}{\affiliation{Cosmic Frontier Center, The University of Texas at Austin, Austin, TX 78712}}

\newcommand{\Rutgers}{\affiliation{Rutgers University, Department of Physics and Astronomy, 136 Frelinghuysen Road, Piscataway, NJ 08854, USA}}

\newcommand{\YaleAstro}{\affiliation{Department of Astronomy, Yale University, New Haven, CT 06511, USA}}

\newcommand{\YalePhys}{\affiliation{Department of Physics, Yale University, New Haven, CT 06511, USA}}
\newcommand{\BHI}{\affiliation{Black Hole Initiative, Harvard University, 20 Garden Street, Cambridge, MA 02138, USA}}

\newcommand{\unige}{\affiliation{Department of Astronomy, University of Geneva, Chemin Pegasi 51, 1290 Versoix, Switzerland}}

%%%%%%%%%%%%%%%%%

\author[0000-0003-4512-8705]{Tiger Yu-Yang Hsiao} \Austin \CFC
\author[0000-0002-4153-053X]{Danielle A. Berg} \Austin \CFC
\author[0000-0001-8519-1130]{Steven L. Finkelstein} \Austin \CFC
\author[0000-0003-4242-8606]{Ansh R. Gupta} \altaffiliation{NSF Graduate Research Fellow} \Austin \CFC
\author[0009-0000-2997-7630]{Zorayda Martinez} \Austin \CFC
\author[0000-0003-1282-7454]{Anthony J. Taylor} \Austin \CFC
\author[0000-0003-3596-8794]{Hollis B. Akins} \Austin \CFC
\author[0000-0003-2332-5505]{\'{O}scar A. Ch\'{a}vez Ortiz} \Austin \CFC
\author[0000-0002-0302-2577]{John Chisholm} \Austin \CFC
\author[0000-0001-6278-032X]{Lukas J. Furtak} \Austin \CFC
\author[0000-0002-5588-9156]{Vasily Kokorev}
\Austin \CFC

%\collaboration{20}{(AAS Journals Data Editors)}

%% Note that the \and command from previous versions of AASTeX is now
%% depreciated in this version as it is no longer necessary. AASTeX 
%% automatically takes care of all commas and "and"s between authors names.

%% AASTeX 6.31 has the new \collaboration and \nocollaboration commands to
%% provide the collaboration status of a group of authors. These commands 
%% can be used either before or after the list of corresponding authors. The
%% argument for \collaboration is the collaboration identifier. Authors are
%% encouraged to surround collaboration identifiers with ()s. The 
%% \nocollaboration command takes no argument and exists to indicate that
%% the nearby authors are not part of surrounding collaborations.

%% Mark off the abstract in the ``abstract'' environment. 

\newcommand{\LCDM}{$\Lambda$CDM}

\newcommand{\red}[1]{{\color{red} #1}}
\newcommand{\redss}[1]{{\color{red} ** #1}}
\newcommand{\redbf}[1]{{\color{red}\bf #1 \color{black}}}

\newcommand{\ny}{$\tilde {\rm n}$}
\newcommand{\about}{$\sim$}
\newcommand{\appr}{$\approx$}
\newcommand{\gt}{$>$}
\newcommand{\um}{$\mu$m}
\newcommand{\uJy}{$\mu$Jy}
\newcommand{\sig}{$\sigma$}
\newcommand{\Lya}{Lyman-$\alpha$}
\renewcommand{\th}{$^{\rm th}$}
\newcommand{\lam}{$\lambda$}

\newcommand{\tentothe}[1]{$10^{#1}$}
\newcommand{\tentotheminus}[1]{$10^{-#1}$}
\newcommand{\e}[1]{$\times 10^{#1}$}
\newcommand{\en}[1]{$\times 10^{-#1}$}
\newcommand{\cgsfluxunits}{erg$\,$s$^{-1}\,$cm$^{-2}$}
\newcommand{\linefluxunits}{\tentotheminus{20} \cgsfluxunits}

\newcommand{\logU}{$\log(U)$}
\newcommand{\logOH}{12+log(O/H)}

\newcommand{\sinv}{s$^{-1}$}
\newcommand{\kms}{km\,s$^{-1}$}

\newcommand{\footnoteurl}[1]{\footnote{\url{#1}}}

\newcommand{\tnm}[1]{\tablenotemark{#1}}
\newcommand{\super}[1]{$^{\rm #1}$}
\newcommand{\supa}{$^{\rm a}$}
\newcommand{\supb}{$^{\rm b}$}
\newcommand{\supc}{$^{\rm c}$}
\newcommand{\supd}{$^{\rm d}$}
\newcommand{\supe}{$^{\rm e}$}
\newcommand{\supf}{$^{\rm f}$}
\newcommand{\supg}{$^{\rm g}$}
\newcommand{\suph}{$^{\rm h}$}
\newcommand{\supi}{$^{\rm i}$}
\newcommand{\supj}{$^{\rm j}$}
\newcommand{\supk}{$^{\rm k}$}
\newcommand{\supl}{$^{\rm l}$}
\newcommand{\supm}{$^{\rm m}$}
\newcommand{\supn}{$^{\rm n}$}
\newcommand{\supo}{$^{\rm o}$}

\newcommand{\squared}{$^2$}
\newcommand{\cubed}{$^3$}

\newcommand{\sqarcmin}{arcmin\squared}

\newcommand{\supcomma}{$^{\rm ,}$}

\newcommand{\rhalf}{$r_{1/2}$}

\newcommand{\chisq}{$\chi^2$}

\newcommand{\Zgas}{$Z_{\rm gas}$}  % gas-phase metallicity
\newcommand{\Zstar}{$Z_*$}  % stellar metallicity

\newcommand{\per}{$^{-1}$}
\newcommand{\inv}{\per}
\newcommand{\Mstar}{$M^*$}
\newcommand{\Lstar}{$L^*$}
\newcommand{\phistar}{$\phi^*$}

\newcommand{\logM}{log($M_*$/\Msun)}

\newcommand{\LUV}{$L_{UV}$}
\newcommand{\MUV}{$M_{UV}$}

\newcommand{\Msun}{$M_\odot$}
\newcommand{\Lsun}{$L_\odot$}
\newcommand{\Zsun}{$Z_\odot$}

\newcommand{\Mvir}{$M_{vir}$}
\newcommand{\Mt}{$M_{200}$}
\newcommand{\Mf}{$M_{500}$}

\newcommand{\Ndotion}{$\dot{N}_{\rm ion}$}
\newcommand{\xiion}{$\xi_{\rm ion}$}
\newcommand{\logxiion}{log(\xiion)}
\newcommand{\fesc}{$f_{\rm esc}$}

\newcommand{\XHI}{$X_{\rm HI}$}
\newcommand{\XHII}{$X_{\rm HII}$}
\newcommand{\RHII}{$R_{\rm HII}$}

\newcommand{\Halpha}{H$\alpha$}
\newcommand{\Hbeta}{H$\beta$}
\newcommand{\Hgamma}{H$\gamma$}
\newcommand{\Hdelta}{H$\delta$}
\newcommand{\Halphaw}{\Halpha\,$\lambda$6563}
\newcommand{\Hbetaw}{\Hbeta\,$\lambda$4861}
\newcommand{\Hgammaw}{H$\gamma$\,$\lambda$4340}
\newcommand{\Hdeltaw}{H$\delta$\,$\lambda$4101}
\newcommand{\Ha}{\Halpha}
\newcommand{\Hb}{\Hbeta}

\newcommand{\I}{\,{\sc i}}
\newcommand{\II}{\,{\sc ii}}
\newcommand{\III}{\,{\sc iii}}
\newcommand{\IV}{\,{\sc iv}}
\newcommand{\V}{\,{\sc v}}
\newcommand{\VI}{\,{\sc vi}}
\newcommand{\VII}{\,{\sc vii}}
\newcommand{\VIII}{\,{\sc viii}}

\newcommand{\HI}{H\I}
\newcommand{\HII}{H\II}
\newcommand{\HeI}{He\I}
\newcommand{\HeII}{He\II}

\newcommand{\CII}{[C\II]}
\newcommand{\CIIw}{\CII\,$\lambda$2325 (blend)}
\newcommand{\CIII}{[C\III]}
\newcommand{\CIIIw}{\CIII\,$\lambda$1908}
\newcommand{\CIIIwa}{\CIII\,$\lambda$1907}
\newcommand{\CIIIwb}{\CIIId\,$\lambda$1909}
\newcommand{\CIIId}{C\III]}
\newcommand{\CIIIdw}{C\III]\,$\lambda\lambda$1907,1909}
\newcommand{\CIV}{C\IV}
\newcommand{\CIVwa}{\CIV\,$\lambda$1548}
\newcommand{\CIVwb}{\CIV\,$\lambda$1550}
\newcommand{\OII}{[O\II]}
\newcommand{\OI}{[O\I]}
\newcommand{\OIw}{\OI\,$\lambda$6300}
\newcommand{\OIIw}{\OII\,$\lambda$3727}
\newcommand{\OIIdw}{\OII\,$\lambda\lambda$3727,3729}
\newcommand{\OIIdwa}{\OII\,$\lambda\lambda$7322,7332}
\newcommand{\SiIII}{Si\III]}
\newcommand{\SiIIIdw}{\SiIII\,$\lambda\lambda$1883,1892}
\newcommand{\OIII}{[O\III]}
\newcommand{\OIIIs}{O\III]}
\newcommand{\OIIIw}{\OIII\,$\lambda$5008}
\newcommand{\OIIIww}{\OIII\,$\lambda$4960,$\lambda$5008}
\newcommand{\OIIIdw}{\OIIIww}
\newcommand{\OIIIwa}{\OIII\,$\lambda$4364}
\newcommand{\OIIIwb}{O\III]\,$\lambda$1666}
\newcommand{\OIIIwbb}{O\III]\,$\lambda$1661}
\newcommand{\OIIIwc}{\OIII\,$\lambda$4960}
\newcommand{\NeIII}{[Ne\III]}
\newcommand{\NeV}{[Ne\V]}
\newcommand{\NeVw}{[Ne\V]$\lambda$3426}
\newcommand{\NeIIIw}{\NeIII\,$\lambda$3870}
\newcommand{\NeIIIwb}{\NeIII\,$\lambda$3969}
\newcommand{\NII}{[N\II]}
\newcommand{\NIIw}{\NII\,$\lambda$6585}
\newcommand{\NIIww}{\NII\,$\lambda$6550,$\lambda$6585}
\newcommand{\SII}{[S\II]}
\newcommand{\SIIdw}{\SII\,$\lambda\lambda$6718,6733}
\newcommand{\HeIw}{\HeI\,$\lambda$3889}
\newcommand{\HeIwa}{\HeI\,$\lambda$4473}
\newcommand{\HeIIw}{\HeII\,$\lambda$1640}
\newcommand{\HeIIwb}{\HeII\,$\lambda$4687}
\newcommand{\NIII}{N\III]}
\newcommand{\NIV}{N\IV]}
\newcommand{\NV}{N\V}
\newcommand{\NVdw}{\NV\,$\lambda$1238,1242}
\newcommand{\NIIIw}{\NIII\,$\lambda$1748}
\newcommand{\NIVw}{\NIV\,$\lambda$1486}
\newcommand{\NIVdw}{\NIV\,$\lambda\lambda$1483,1487}
\newcommand{\MgII}{Mg\II}
\newcommand{\MgIIw}{\MgII\,$\lambda$2800}

\newcommand{\Lyaw}{Ly$\alpha$\,$\lambda$1216}

%\NIVw, \CIVw, \HeIIw, \NIIIw, \MgIIw.

% Stark17:
%\newcommand{\ciiid}{\hbox{[C\,{\sc iii}]$\lambda1907$+C\,{\sc iii}]$\lambda1909$}}
%\newcommand{\ciiit}{\hbox{C\,{\sc iii}]$\lambda1908$}}
%\newcommand{\Hii}{\mbox{H\,{\sc ii}}}

\newcommand{\Om}{\Omega_{\rm M}}
\newcommand{\OL}{\Omega_\Lambda}

\newcommand{\etal}{et al.}

\newcommand{\citeps}{\citep}

\newcommand{\HST}{{\em HST}}
\newcommand{\SST}{{\em SST}}
\newcommand{\Hubble}{{\em Hubble}}
\newcommand{\Spitzer}{{\em Spitzer}}
\newcommand{\Chandra}{{\em Chandra}}
\newcommand{\JWST}{{\em JWST}}
\newcommand{\Planck}{{\em Planck}}

\newcommand{\Bradac}{{Brada\v{c}}}

\newcommand{\citepeg}[1]{\citep[e.g.,][]{#1}}

\newcommand{\range}[2]{\! \left[ _{#1} ^{#2} \right] \!}  % range of values in brackets

\newcommand{\grizli}{\textsc{grizli}}
\newcommand{\eazypy}{\textsc{eazypy}}
\newcommand{\msaexp}{\textsc{msaexp}}
\newcommand{\trilogy}{\textsc{trilogy}}
\newcommand{\bagpipes}{\textsc{bagpipes}}
\newcommand{\beagle}{\textsc{beagle}}
\newcommand{\photutils}{\textsc{photutils}}
\newcommand{\SEP}{\textsc{sep}}
\newcommand{\piXedfit}{\textsc{piXedfit}}
\newcommand{\pyneb}{\textsc{pyneb}}
\newcommand{\HIIC}{\textsc{hii-chi-mistry}}
\newcommand{\astropy}{\textsc{astropy}}
\newcommand{\astrodrizzle}{\textsc{astrodrizzle}}
\newcommand{\multinest}{\textsc{multinest}}
\newcommand{\cloudy}{\textsc{Cloudy}}
\newcommand{\jdaviz}{\textsc{Jdaviz}}

\renewcommand{\tt}[1]{\texttt{#1}}

\newcommand{\SE}{\tt{SourceExtractor}}

\newcommand{\PD}[1]{\textcolor{blue}{[PD: #1\;]}}

%%%%%%%%%%%%%%%%%%%%%%%%%
% ABSTRACT

\begin{abstract}
JWST has enabled the discovery of dozens of extremely metal-poor galaxies (EMPGs) with metallicities below $5\%\,Z_{\odot}$, representing a significant leap toward detecting the first galaxies without metals. 
However, accurate metallicity measurements require careful determination of physical conditions in the ionized gas.
In this paper, we study four galaxies that appear to be EMPGs when analyzed using the {direct-$T_e$} method with the common low-density assumption ($n_e=10^3\,{\rm cm}^{-3}$).
To test whether their metal-poor status holds when accurately measuring the density, we apply the direct method with a self-consistent determination of electron temperature ($T_e$) and density ($n_e$) in the high-ionization zone using the [O\,{\sc iii}]\,$\lambda$5008, [O\,{\sc iii}]\,$\lambda$4364, and [O\,{\sc iii}]\,$\lambda$1666 lines, using JWST/NIRSpec data from the SPURS program in the Abell 2744 lensed field.
We find that three out of four galaxies in our sample have extremely high electron densities ($n_e \sim 10^{5}-10^{6}\,{\rm cm^{-3}}$), which suppress the [O\,{\sc iii}]\,$\lambda$5008 line and lead to underestimates of metallicity by up to $\sim1.1\,$dex when this density is not accounted for, and have true metallicities of 12+log(O/H) $\sim7.3-8.2$.
Failure to account for high densities can therefore lead to systematic misclassification of metal-poor galaxies and biased conclusions about early chemical enrichment.

\end{abstract}
%% Keywords should appear after the \end{abstract} command. 
%% The AAS Journals now uses Unified Astronomy Thesaurus concepts:
%% https://astrothesaurus.org
%% You will be asked to selected these concepts during the submission process
%% but this old "keyword" functionality is maintained in case authors want
%% to include these concepts in their preprints.
\keywords{
%Population III stars (1285),
Metallicity (1031),
Early universe (435),
Chemical abundances (224),
Galaxies (573),
High-redshift galaxies (734), 
Galaxy spectroscopy (2171)
%Galaxy clusters (584), 
}

%% From the front matter, we move on to the body of the paper.
%% Sections are demarcated by \section and \subsection, respectively.
%% Observe the use of the LaTeX \label
%% command after the \subsection to give a symbolic KEY to the
%% subsection for cross-referencing in a \ref command.
%% You can use LaTeX's \ref and \label commands to keep track of
%% cross-references to sections, equations, tables, and figures.
%% That way, if you change the order of any elements, LaTeX will
%% automatically renumber them.
%%
%% We recommend that authors also use the natbib \citep
%% and \citet commands to identify citations.  The citations are
%% tied to the reference list via symbolic KEYs. The KEY corresponds
%% to the KEY in the \bibitem in the reference list below. 

\section{Introduction} \label{sec:intro}

The first generation of stars (Population III, or Pop III, stars) initiated the chemical enrichment of the Universe and marked the beginning of stellar and galaxy evolution.
Formed almost entirely from hydrogen and helium, Pop III stars synthesized the first elements heavier than helium (i.e., metals).
To understand how the first stars and galaxies formed, JWST \citep{Gardner2006,Rigby2023,Gardner2023} has made significant progress in identifying not only the earliest galaxies at $z\gtrsim10$ \citep[e.g.,][]{Bunker2023,CurtisLake2023,Wang2023,Hsiao2024,Carniani2024,Finkelstein2024,Zavala2025,Kokorev2025,Naidu2026}, but also extremely metal-poor galaxies \citep[EMPGs; $Z<5\%\,Z_{\odot}$; e.g.,][]{Vanzella2023,Chemerynska2024,Fujimoto2025,Fujimoto2025b,Willott2025,Hsiao2025,Maiolino2025,Nakajima2025,Morishita2025,Vanzella2025,Cai2025,Asada2026} in the early Universe. 
{However, many EMPG candidates have been identified using ``strong-line'' diagnostics, which calibrate metallicity against bright nebular emission-line ratios through empirical  \citep[e.g.,][]{Pettini2004,Nagao2006,Marino2013,Curti2017,Curti2020,Nakajima2022} or theoretical/photoionization models \citep[e.g.,][]{Kewley2002,Kewley2008}.
Although empirical calibrations were historically based largely on local samples, recent studies have expanded their coverage to high-redshift and metal-poor galaxies \citep[e.g.,][]{Cataldi2025,Chakraborty2025,Sanders2026,Isobe2026}.
Whether these diagnostics remain valid in the high-density regime is a key unanswered question.}

One way to directly measure gas-phase metallicity, {commonly expressed in terms of the oxygen abundance,} in star-forming galaxies is the {``direct-$T_e$"} method, which requires measuring both the electron temperature ($T_e$) and density ($n_e$) \citep[e.g.,][]{Peimbert1967,Peimbert1969,Aller1984,Osterbrock1989,Izotov2006}.
{In this approach, the interstellar medium (ISM) of a galaxy is commonly modeled using three to four ionization zones, with ions assigned to each zone according to their ionization potentials \citep{Garnett1992,Berg2021}.
In a given zone,}
the electron temperature is determined from the ratio of two emission lines arising from the same ion but from transitions with substantially different excitation energies.
For example, the high-ionization zone temperature is typically measured from the ratio of \OIIIwa\ to \OIIIw, both emitted by doubly ionized oxygen (O$^{++}$), whose upper levels have excitation energies of $5.35\,{\rm eV}$ and $2.51\,{\rm eV}$, respectively.
Because the excitation rates of these transitions depend exponentially on $T_e$, their line ratio provides a sensitive probe of the electron temperature.

The electron density, on the other hand, is determined from the ratio of two emission lines arising from the same ion with similar excitation energies but different critical densities, such as \OIIdw\ or \SIIdw\ for the low-ionization zone and \CIIIdw\ for the intermediate ionization zone.
Because these transitions have comparable excitation energies, their ratio is largely insensitive to $T_e$; instead, as $n_e$ approaches the critical density of the lower-critical-density transition, collisional de-excitation begins to suppress that line relative to the other, making the ratio a sensitive probe of $n_e$ from $\sim0.1\times$ the lower critical density to $\sim10\times$ the higher critical density.
Ionic oxygen abundances, O$^+$/H$^+$ and O$^{++}$/H$^+$, for example, are then determined using the corresponding observed emission line flux and emissivity, which is calculated with the appropriate ionization zone $T_e$ and $n_e$.
The total oxygen abundance is then obtained by summing over the observed ionic species, as O$^0$ and O$^{3+}$ contribute negligibly to the total oxygen abundance \citep[e.g.,][]{Berg2021}. {The contribution from O$^{3+}$ is expected to remain small even in extremely metal-poor or highly ionized systems, such as EXCELS-63107 \citep{Cullen2025} and I Zw 18 \citep{Rickards2025}.}
Owing to the intrinsic faintness of auroral lines such as \OIIIwa, direct-$T_e$ method metallicity measurements remain extremely challenging at high redshift.
To date, only one EMPG has been claimed with a direct metallicity measurement {at high redshift \citep[$z>5$; ][]{Cullen2025}}.

Accurate measurements of $n_e$ are essential for the direct-$T_e$ method, but are challenging in their own right: some density-sensitive doublets require high spectral resolution to resolve them (e.g., \OIIdw), while others are typically faint (e.g., \SiIIIdw).
Consequently, when such doublets are unavailable, a low electron density ($n_e=10^{2}\,{\rm cm^{-3}}$ typically for local and $n_e=10^{3}\,{\rm cm^{-3}}$ for high-z galaxies) has traditionally been assumed, as this approximation has generally been considered adequate since local galaxies have low electron density based on rest-frame optical diagnostics.
In the low-density regime typical of local star-forming galaxies, the derived $T_e$ is only weakly dependent on $n_e$, because the critical density of \OIIIw\ is high ($\sim7\times10^5\,{\rm cm^{-3}}$).
However, recent observations suggest that high-redshift galaxies systematically exhibit higher electron densities than their local counterparts \citep[e.g.,][]{Isobe2023,Abdurrouf2024,Topping2025,Martinez2025,Li2025,Berg2026,Mendez2026,Papovich2026}.

At high redshift, when $n_e$ is measured, it is typically inferred from low/intermediate-ionization doublets such as \OIIdw, \SIIdw, or \CIIIdw, owing to the wavelength coverage of JWST/NIRSpec and the prevalence of these emission lines.
However, the assumption of a homogeneous ISM is likely an oversimplification.
The \OIIIw\ emission originates in highly ionized regions, whereas \OIIdw\ and \SIIdw\ arise from lower-ionized gas, and \CIIIdw\ is emitted from the intermediate-ionization zone \citep[e.g.,][]{Berg2021,Mingozzi2022,Harikane2025,Mendez2026}.
The density is found to be higher in the highly ionized regions compared to the low/intermediate ones {across redshifts}.
Consequently, adopting electron densities measured from low-ionization zones for high-ionization abundance diagnostics, or simply assuming the low-density limit, can systematically bias direct metallicity measurements \citep[e.g.,][]{Hayes2025,Martinez2025,Arellano2026,Hsiao2026}. 

% In particular, the abundance derived from \OIIIw\ becomes increasingly underestimated as $n_e$ approaches the critical density of \OIIIw\ ($n_e \gtrsim 10^5\,{\rm cm^{-3}}$), resulting in artificially low oxygen abundances due to the inaccurate assumption of $n_e$ \citep[e.g.,][]{Hayes2025,Martinez2025,Arellano2026,Hsiao2026}.
% Accurate metallicity measurements via the direct-$T_e$ method require simultaneous determination of both $T_e$ and $n_e$.
In particular, since \OIIIw\ becomes collisionally de-excited at high densities ($n_e\gtrsim10^{5}\,{\rm cm^{-3}}$), in this regime the \OIII$\lambda5008/\lambda4364$ ratio depends on both $T_e$ and $n_e$ rather than $T_e$ alone.
A self-consistent method to determine both $T_e$ and $n_e$ within the same high-ionization zone traced by \OIII\ emissions, using \OIIIwb, \OIIIwa, and \OIIIw, was first proposed by \citet{Berg2025_GISM} and \citet{Arellano2026}.
\OIIIwa\ and \OIIIwb\ have higher critical densities {of $n_e\sim2\times10^{7}$ and $\sim4\times10^{10}\,{\rm cm^{-3}}$, respectively}.
Therefore, since $T_e$ and $n_e$ are thus two unknowns entangled in a single line ratio, instead, combining three measurements: \OIIIw, \OIIIwa, and \OIIIwb, one can solve for both simultaneously.
% While $T_e$ is commonly derived from the \OIII$\lambda5008/\lambda4364$ line ratio, $n_e$ is more challenging to measure and is often assumed or measured from different ionization zones than the metallicity tracer itself.

In this paper, we analyze four extremely metal-poor galaxy candidates that exhibit low R3 (=\OIIIw/H$\beta$) ratios, with detected \OIIIwb, \OIIIwa\ and \OIIIw, enabling self-consistent direct metallicity measurements.
This paper is organized as follows.
In \S\ref{Sec:odm}, we describe the observations and data reduction, and in \S\ref{sec:fitting&sample}, we present the emission-line measurements and sample selection.
In \S\ref{sec:metallicity_big}, we describe our measurements of the physical conditions and metallicities.
We discuss the discovery and implications of high-density EMPG impostors in \S\ref{sec:discussion} and summarize our conclusions in \S\ref{sec:conclusion}.

Throughout this article, we adopt the solar oxygen abundance \logOH\ = 8.69 from \citet{Asplund2021}.
Where needed, we adopt the {\em Planck} 2018 flat \LCDM\ cosmology \citep{Planck18_cosmo}
with $H_0 = 67.7$ km s\inv\ Mpc\inv, $\Om = 0.31$, and $\OL = 0.69$.

\section{Observations and Data}
\label{Sec:odm}

\subsection{NIRSpec Spectroscopy}
\label{sec:spec}

We use JWST/NIRSpec observations obtained as part of the Cycle 4: SPectroscopic Ultra-deep Reionization-era Survey (SPURS; GO 9214; PIs: Mason $\&$ Stark; \citealt{Chen2026,Tang2026}).
SPURS obtained ultra-deep, medium-resolution ($R\sim1000$) spectroscopy with the three NIRSpec gratings G140M/F100LP ($\sim29$ hr), G235M/F170LP ($\sim8$ hr), and G395M/F290LP ($\sim3$ hr).
The broad wavelength coverage provided by these gratings is well suited to this study, as our self-consistent abundance analysis requires simultaneous coverage of both the rest-frame UV \OIIIwb\ and the rest-frame optical \OIIIwa\ and \OIIIw\ emission lines.
Consequently, SPURS provides complete coverage of these three \OIII\ transitions over the redshift range $z\sim4.8-9.5$. 
SPURS targets four legacy fields: Abell 2744, the Extended Groth Strip (EGS), GOODS-South, and GOODS-North.
In this work, we focus on the Abell 2744 field to maximize the likelihood of identifying intrinsically faint EMPGs through gravitational lensing.
The observations were carried out in the NIRSpec Multi-Object Spectroscopy (MOS) mode using the standard three-shutter slitlet nodding pattern. 
The data were obtained on 2025 November 6 and 9.
Further details of the SPURS survey are presented by \citet{Chen2026} and \citet{Tang2026}.

\subsection{Data Reduction}
\label{sec:reduction}

We reduce the data using a custom wrapper around the standard JWST Data Calibration Pipeline v1.20.2 with the default parameters for Calibration Reference Data System (CRDS) mapping \texttt{pmap-1481}.
In between the \texttt{detector1} and \texttt{spec2} stages of the pipeline, we apply a custom routine to more aggressively remove the 1/f noise and the ``picture frame'' effect from individual exposures (see H. Akins in prep. for details).
In \texttt{spec2}, we employ custom flat-field reference files for G140M/F100LP and G235M/F170LP to extend their spectral coverage from the nominal 0.97--1.89~$\mu$m and 1.66–3.17~$\mu$m to 0.97--3.2~$\mu$m and 1.66–5.2~$\mu$m, respectively.
This is possible due to the use of long pass filters in NIRSpec's design, which permit light redder than the nominal red limit for each grating to still reach the gratings and diffract onto the detector.
These extended wavelength ranges are excluded by the standard reference files provided with v1.20.2 of the JWST Calibration Pipeline to prevent second-order spectra from being included in the final spectral products.
While second-order spectra are of concern for some analyses, they are inconsequential for the emission line-based analyses used in this work.

We calibrate the extended wavelength ranges using the SPURS dataset.
SPURS is ideal for these calibrations due to its tiered depths (29 hrs in G140M, 8 hrs in G235M, 3 hrs in G395M).
We first extend the flat-field data (and associated photometry and pipeline configuration files) to 3.2~$\mu$m and 5.2~$\mu$m for G140M and G235M, respectively, by simply repeating the last non-zero value in each flat-field calibration file out to the desired maximum wavelength.
We reduce SPURS data using these extended files to produce spectra that are well-calibrated in their canonical wavelength ranges, and effectively uncalibrated in the extended ranges.
We measure integrated emission line fluxes for strong lines (e.g., H$\alpha$ at $z=3$ and observed frame 2.62~$\mu$m) present in both the extended range of a bluer grating (e.g., G140M) and the canonical range of a redder grating (e.g., G235M).
By comparing the line fluxes measured from both the uncalibrated bluer grating and the calibrated redder grating, we derive a flux calibration factor for the bluer grating at the observed wavelength of the line.
Crucially, this process is unaffected by second-order spectra, unless the wavelength of a line is exactly twice the wavelength of another strong bluer line.
Fortunately, no such lines of significant strength exist in this dataset.
We repeat this process for strong emission lines across all of the SPURS objects, and build up significant samples of correction factors across the extended wavelength ranges {for 202 galaxies}.
We then fit simple quadratic functions to the resulting correction factors as a function of wavelength and use these quadratics to calibrate the extended flat-field files.
We then re-reduce SPURS using the now-calibrated extended flat-fields. Finally, we repeat our measurement and comparison of emission lines present in multiple gratings and find only a $\sim$10\% 1$\sigma$ scatter in the residual line flux ratios with no strong trends as a function of redshift, {while we do not propagate the $10\%$ error into the analysis}.
We therefore conclude that our systematic flux calibrations in the extended wavelength ranges are accurate to $\sim10\%$.
If a calibration offset is present, we identify which grating is discrepant by comparing against a third, independent grating, and adopt the flux only from gratings confirmed to be consistent.

Our final spectroscopic dataset is produced using these extended, calibrated flat-fields and extracted using a S/N weighted optimal extraction \citep{horne86} for the final 1D spectra.
We fit preliminary redshifts using a non-negative-least-squares fitter to combine templates derived from galaxy continuum models, emission lines, and blackbodies, returning the redshift that minimizes the $\chi^2$ between the fitted combination of templates and the spectral data. 
We then inspected all spectra by eye, manually assigning any incorrect redshifts and flagging objects in the sample with poorly fit redshifts.
% If there is a calibration issue among gratings, we utilize the unaffected gratings throughout the analysis (e.g., if the line fluxes of the same line in G235M and G395M are not consistent, we check G140M and G235M lines, if they're consistent we use G235M) 
% For the same line appearing in both gratings (e.g., H$\gamma$), we observe a systematic offset in the measured line fluxes, whereas no such offset is seen between G140M and G235M, nor in the other two candidates (415 and 437).
% Since the extended wavelength coverage of G235M for these two candidates already extends redward to H$\alpha$ (see \S\,\ref{sec:reduction}), we simply exclude G395M spectrum.
% For any emission line covered by multiple gratings, we fit the line in each grating and adopt the highest-SNR measurement as the fiducial value for this work.

% All the reduced data, the full data reduction pipeline with custom steps, and the extended wavelength flat-field files, are hosted on the COSMOS Archive of MultiPle-Field Internal Reductions \& Extractions (CAMPFIRE): a platform for accessing and inspecting reduced JWST imaging \& spectroscopic data, primarily in the COSMOS field but inclusive of other fields as well (H. Akins in prep.).

\section{Selection of a High-Redshift EMPG Sample}
\label{sec:fitting&sample}

\subsection{Emission Line Fitting}
\label{sec:fitting}
The goal of this work is to measure robust direct-method metallicities of candidate EMPGs that account for density using significant detections of multiple O$^{++}$ emission lines.
In order to select a sample with the requisite line detections, we need to construct a catalog of observed emission line fluxes. 
We fit all emission lines using Markov Chain Monte Carlo (MCMC) methods implemented with \texttt{emcee} \citep{Foreman-Mackey2013}, which is also used for the measurements of physical properties later (\S\ref{sec:metallicity_big}).
Prior to line fitting, we subtract the continuum using a single linear fit around the lines of interest using a continuum mask that excludes regions within $\pm0.007\,\mu$m of each emission line center, which is visually inspected to ensure the mask extends far enough beyond the line.
% The B-spline is fit to the masked regions with a smoothing parameter proportional to the number of continuum points, producing a smooth continuum estimate that is subtracted from the observed spectrum.
For isolated emission lines, we model each line as a Gaussian profile.
We use 32 walkers with 5000 MCMC steps and discard the first 3000 steps as burn-in.
We first utilize the preliminary redshift measured in \S\ref{sec:reduction}, and search for \OIIIw.
We then treat \OIIIw\ as the anchor line, and fix the redshift from the fits of \OIIIw, and adopt the updated redshift to fit the remaining emission lines.
We also visually inspect the fitted emission lines to verify the redshift.
For doublets such as \CIIIdw, we fit both components simultaneously using a double Gaussian model with the rest-frame wavelength separation fixed (redshifted to the observed frame), and both components share the same line width. 

For any emission line covered by multiple gratings, we fit the line in each grating and adopt the highest-SNR measurement as the fiducial value, provided the gratings agree.
{Most emission lines covered by multiple gratings have measured fluxes consistent within $1–2\,\sigma$. However, SPURS-A2744-422 and SPURS-A2744-544 show systematic discrepancies between G235M and G395M: the G395M fluxes are approximately $60\%$ and $20\%$ lower, respectively. 
In contrast, their G140M and G235M measurements are generally consistent within $1\,\sigma$. 
As described in \S\ref{sec:reduction}, we therefore exclude the G395M measurements for these two EMPG candidates and adopt the G140M or G235M measurements instead.
}
We report line fluxes and uncertainties from the MCMC posterior distributions in Table \ref{tab:lines}.
For each parameter, we quote the median value and 16th-84th percentile ranges as $1\sigma$ uncertainties.
The MCMC samples provide full flux distributions for each line, which we propagate through subsequent calculations of dust correction, electron density, temperature, and metallicity (see \S\ref{sec:metallicity_big}).

\subsection{Sample Selection}
\label{sec:selection}
\begin{figure*}
\centering
\includegraphics[width=\textwidth]{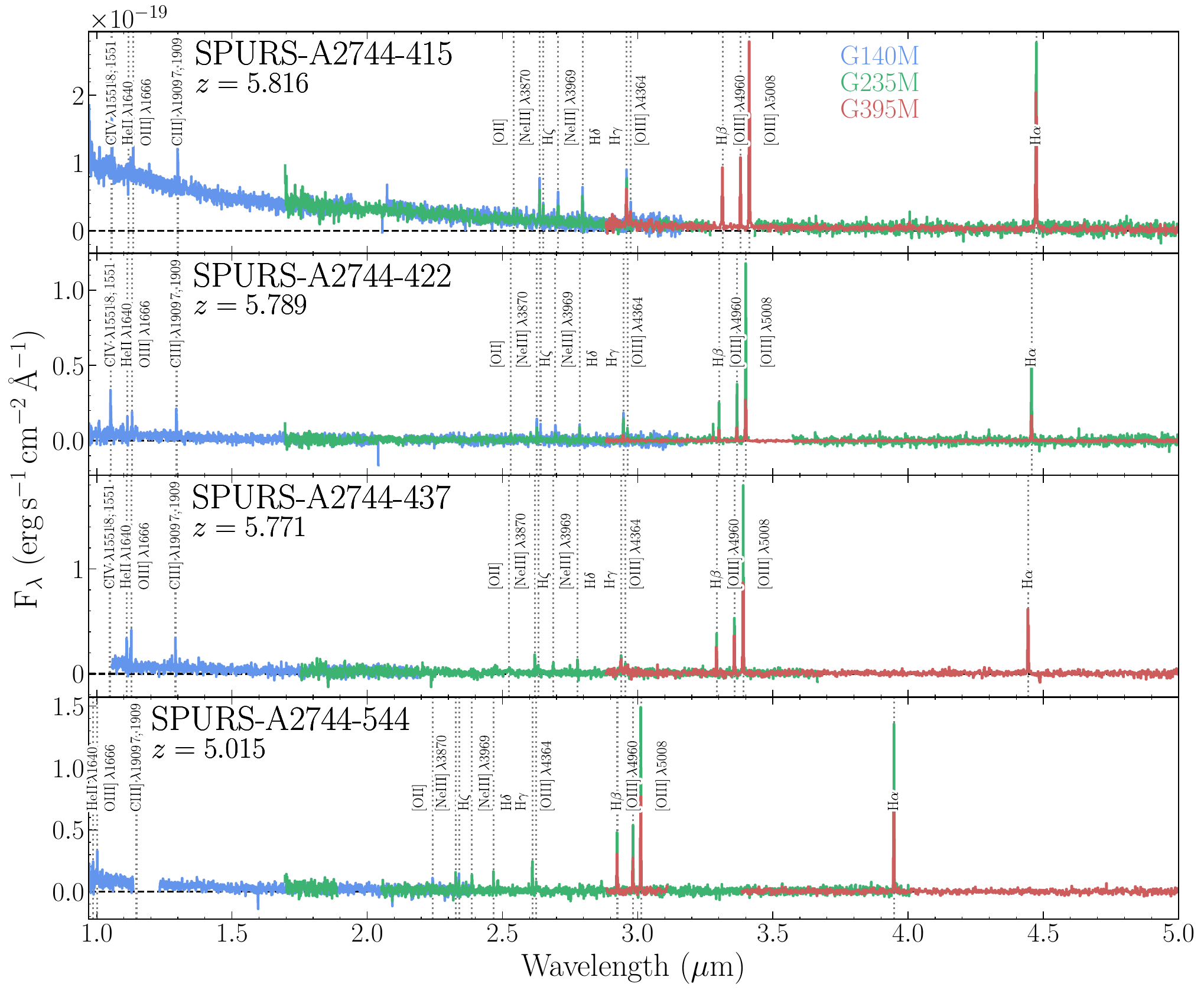}
\caption{NIRSpec G140M (blue), G235M (green), and G395M (red) spectroscopy of four EMPG candidates from the JWST SPURS survey in the Abell 2744 lensed field.
The deep spectroscopic coverage from $1-5\,{\rm \mu m}$ enables a detailed study of rest-frame UV and optical lines of high-z galaxies, allowing the self-consistent $T_e$ and $n_e$ measurements from \OIIIwb, \OIIIwa, and \OIIIw.
}
\label{fig:spec}
\end{figure*}

\begin{figure*}
\centering
\includegraphics[width=\textwidth]{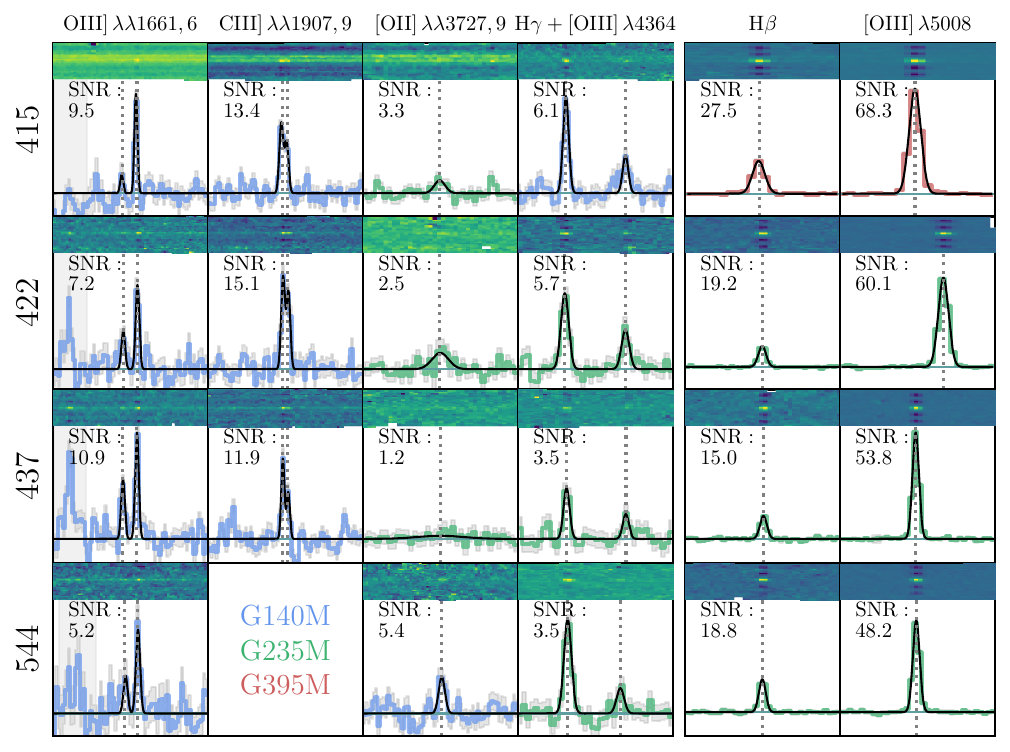}
\caption{Key emission lines and the best fits of NIRSpec G140M (blue), G235M (green), and G395M (red) spectroscopy of four EMPG candidates.
The best fits are shown in black curves.
The SNR of each fit is shown in the upper-right corner of each panel.
{The 2D spectra are shown in the upper parts of each panel.}
For H$\gamma$ and \OIIIwa\ panels, the SNRs are shown for \OIIIwa, {and for \OIII$\,\lambda\lambda1661,6$, the SNRs are shown for \OIIIwb}.
The detections of \OIIIwb, \OIIIwa, and \OIIIw\ allow us to simultaneously constrain $T_e$ and $n_e$.
{For emission lines covered by multiple gratings (e.g., \OIIdw\ and \OIIIwa), we show and adopt the highest-SNR measurements.
}
}
\label{fig:lines}
\end{figure*}

% We access all spectra in Abell2744 from SPURS on CAMPFIRE.
We selected our candidate EMPGs in three steps.
First, using preliminary redshift estimates (\S\ref{sec:reduction}), we identified 47 out of 75 spectra at $z\sim4.8-9.5$, where all three \OIII\ emission lines are covered by the NIRSpec wavelength range, which allows us to conduct the self-consistent direct metallicity.
Second, we identified EMPG candidates based on their R3 ratio ($\mathrm{R3}\equiv$\OIIIw/H$\beta$), where a low R3 ratio, corresponding to relatively strong H$\beta$ and weak \OIIIw\ emission, is assumed to be indicative of low metallicity.
We selected galaxies with $\mathrm{R3}<5$, corresponding to a metallicity of $Z\lesssim5\%\,Z_{\odot}$ according to the empirical calibrations of \citet{Hsiao2026} and \citet{Isobe2026}, resulting in a sample of $16$ EMPG candidates.
% Finally, we visually inspected these spectra and selected galaxies with detections of all three \OIII\ lines, yielding a sample of 12 galaxies.
Finally, we selected galaxies with detections ($>3\sigma$) of all three \OIII\ lines (\OIIIw, \OIIIwa, and \OIIIwb), yielding a final sample of 4 candidate EMPGs.
The SPURS NIRSpec spectra of our 4 candidate EMPGs are shown in  Figure \ref{fig:spec}.
The fits of essential lines are presented in Figure \ref{fig:lines}, where three different O$^{++}$ emission lines are measured in each of four candidate EMPGs at $z>5$.

{We do not explicitly exclude active galactic nuclei (AGN) during the sample selection.
We select galaxies with low R3 ratios that would be identified as EMPG candidates based on their emission-line properties, regardless of potential AGN contributions, and characterize the physical conditions of their line-emitting gas using integrated spectra.
These nebular diagnostics remain applicable whether the ionizing radiation originates from stars, an AGN, or a combination of both.
The uncertain ionizing source, therefore, does not invalidate the inferred gas conditions, although it may affect their physical interpretation and the assumptions underlying the abundance analysis.
We therefore retain these sources in our initial sample and assess possible AGN contributions separately in \S\ref{sec:impostors}.
}
% We again visually inspect individual fits to ensure they are valid, and the fits of essential lines are presented in Figure \ref{fig:lines}.

\subsection{Dust Reddening Correction} \label{sec:dust}

Before measuring the physical conditions and the metallicity, we first correct the emission lines for dust attenuation.
All the dust correction is conducted using \pyneb\ \citep{Luridiana2015}.
We adopt the \citet{Calzetti2000} dust attenuation law, and measure the color excess $E(B-V)$ with $R_V = 4.05$.
We derive $E(B-V)$ from the observed Balmer decrement (H$\alpha$/H$\beta$) for each object.
Rather than adopting a fixed intrinsic H$\alpha$/H$\beta$ ratio (e.g., the canonical value of 2.75 at 
$T_e=10^4\,{\rm K}$ and $n_e=10^2\,{\rm cm^{-3}}$), we iteratively update the assumed Case B ratio using our own measured $T_e$ and $n_e$
\citep[][]{Osterbrock1989,Storey1995}.
At each step, we recompute the dust correction using the current Case B ratio, re-derive $T_e$ and $n_e$ from the corrected fluxes, and update the Case B ratio accordingly.
We repeat this until the derived $T_e$, $n_e$, and the Case B ratio converge (typically within five iterations, given the weak dependence of the Case B ratio on $T_e$ and $n_e$).
For galaxies with the observed Balmer decrement consistent with the Case B scenario within 1$\sigma$, we set $E(B-V) = 0$ and apply no dust correction.
We then use the measured $E(B-V)$ and the Calzetti dust law to deredden all emission line fluxes.
The measured $E(B-V)$ is summarized in Table \ref{tab:prop}.

% Filters
\begin{deluxetable*}{lcccc}
\tablecaption{\label{tab:lines}
The redshift, RA, DEC, and the measured emission line fluxes of four candidate EMPGs.}
\tablewidth{\columnwidth}
\tablehead{
\colhead{Line} &
\colhead{SPURS-A2744-415} &
\colhead{SPURS-A2744-422} &
\colhead{SPURS-A2744-437} &
\colhead{SPURS-A2744-544} 
}
\startdata
$z$ & $5.816$ & $5.789$ & $5.771$ & $5.015$ \\
RA (deg)$^{a}$ & 3.56797 & 3.56117 & 3.62025 & 3.60819 \\
DEC (deg)$^{a}$ & -30.39193 & -30.41380 & -30.38927 & -30.42201 \\
\hline
\CIVwa & $ 70\pm 35$ & $ 50\pm  5$ & \ldots & \ldots \\
\CIVwb & $133\pm 37$ & $ 31\pm  4$ & \ldots & \ldots \\
\HeIIw & \ldots & $ 15\pm  3$ & $ 45\pm  5$ & \ldots \\
\OIIIwbb & $ 16\pm  8$ & $  9\pm  3$ & $ 23\pm  4$ & $ 14\pm  5$ \\
\OIIIwb & $ 82\pm  9$ & $ 19\pm  3$ & $ 41\pm  4$ & $ 30\pm  6$ \\
\CIIIwa & $ 76\pm  8$ & $ 22\pm  2$ & $ 28\pm  3$ & $ 35\pm 37$ \\
\CIIIwb & $ 56\pm  8$ & $ 18\pm  2$ & $ 16\pm  3$ & $ 36\pm 39$ \\
\OIIdw & $ 39\pm 12$ & \ldots & \ldots & $ 19\pm  4$ \\
\NeIIIw & $ 97\pm  9$ & $ 25\pm  3$ & $ 36\pm  5$ & $ 21\pm  4$ \\
$\mathrm{H\zeta}$ & $ 52\pm  9$ & $ 20\pm  4$ & \ldots & $ 18\pm  3$ \\
\NeIIIwb & $ 34\pm 13$ & $ 10\pm  5$ & $ 15\pm  8$ & $ 14\pm  9$ \\
$\mathrm{H\epsilon}$ & $ 81\pm 17$ & $ 24\pm  5$ & $ 23\pm 10$ & $ 23\pm 10$ \\
$\mathrm{H\delta}$ & $ 86\pm 11$ & $ 21\pm  3$ & $ 26\pm  6$ & $ 21\pm  5$ \\
$\mathrm{H\gamma}$ & $174\pm 12$ & $ 37\pm  4$ & $ 33\pm  5$ & $ 61\pm  5$ \\
\OIIIwa & $ 61\pm 10$ & $ 19\pm  3$ & $ 16\pm  5$ & $ 16\pm  5$ \\
$\mathrm{H\beta}$ & $346\pm 13$ & $ 70\pm  4$ & $ 87\pm  6$ & $122\pm  6$ \\
\OIIIwc & $369\pm 12$ & $ 94\pm  4$ & $129\pm  6$ & $106\pm  6$ \\
\OIIIw & $1045\pm 15$ & $293\pm  5$ & $375\pm  7$ & $342\pm  7$ \\
$\mathrm{H\alpha}$ & $848\pm 14$ & $175\pm  8$ & $250\pm  7$ & $315\pm  8$ \\
\hline
\enddata
\tablenotetext{}{Line fluxes are in units of $10^{-20}\,{\rm erg\,s^{-1}\,cm^{-2}}$. Fluxes presented here are not dust-corrected.}
\tablenotetext{a}{J2000}
%\textbf{Note.} 
%Different codes assume different IMFs (Table \ref{tab:sedfitting}). 
%For a fair comparison, we suggest readers to multiply 0.61 for stellar mass and 0.63 for SFR from \citet{Salpeter1955} to \citet{Chabrier2003} \citep{Madau2014}. 
%For a fair comparison, we \textcolor{red}{multiply} 0.94 for stellar mass from \citet{1993MNRAS.262..545K} to \citet{Chabrier2003} \citep{Madau2014}. 
\end{deluxetable*}
%\tablecomments{}
% https://docs.google.com/spreadsheets/d/1k5Xgf7vtZjnpvPa5rycbFL_WlGYhXsqBX5VMnrCTFos

\section{Physical Conditions and Metallicities}
\label{sec:metallicity_big}
The goal of this paper is to measure the most reliable metallicity of EMPG candidates selected based on the strong-line diagnostic (low-R3).
We start by measuring the metallicity with the traditional direct method assumption of a low $n_e$ in \S \ref{sec:te_old}.
Then we adopt the self-consistent method \citep{Berg2025_GISM,Arellano2026} to simultaneously account for $T_e$ and $n_e$ in \S\ref{sec:tene}, and study how biased it is if $n_e$ is not correctly accounted for.
{Throughout the analysis, we adopt transition probabilities for \OII\ \citep{Zeippen1982,Wiese1996}, \OIII\ \citep{Storey2000,Froese2004}, and \CIII\ \citep{Glass1983,Nussbaumer1978,Wiese1996}, and collision strengths for \OII\ \citep{Kisielius2009}, \OIII\ \citep{Aggarwal1999}, and \CIII\ \citep{Berrington1985}.}

\subsection{$n_e$ in Low- and Intermediate-Ionization Zones} \label{sec:den_ciii}
We start by measuring the densities in lower ionization zones.
The classical density-sensitive doublets in the low-ionization zone include \OIIdw\ (13.6\,eV) and \SIIdw\ (10.4\,eV).
However, none of the EMPG candidates show \SIIdw\ detections.
SPURS-A2744-415 and SPURS-A2744-544 have detected ($>3\,\sigma$) \OIIdw\ (see Figure \ref{fig:lines}), but the medium grating spectra lack sufficient resolution to resolve the doublet and measure a reliable $n_e$(\OII).

On the other hand, \CIIIdw\ provides access to the density of intermediate-ionization gas (24.4\,eV).
In three EMPGs (SPURS-A2744-415, 422, and 437), although blended, \CIIIdw\ shows clear double peak features.
Thus, we fit the \CIIIdw\ (see \S\ref{sec:fitting})
%to access the $n_e$(\CIIId) of intermediate ionization gases.
and estimate the $n_e$(\CIIId) of the intermediate-ionization gas using the \CIIIwa\ / \CIIIwb\ line ratio.
Two of the EMPG candidates (SPURS-A2744-415 and 422) have $n_e$(\CIIId) $\sim10^{4}{\, \rm cm^{-3}}$ while SPURS-A2744-437 has unconstrained $n_e$(\CIIId) $\lesssim10^{4}{\, \rm cm^{-3}}$ since the ratio of \CIIIwa\ / \CIIIwb\ is close to 1.5 where it becomes insensitive to lower densities.
The \CIIId\ densities of the EMPG candidates are presented in Table \ref{tab:prop}.
% Figure \ref{fig:nn} also demonstrates the comparison between $n_e$ in high- (\OIII) and intermediate- (\CIIId) ionization zones.

\subsection{The Traditional Direct Method: Uniform, Low Densities} \label{sec:te_old}

We begin by inferring $T_e$ (and the corresponding $T_e$-based metallicity) under two assumptions: (1) an assumed low $n_e$, and (2) a measured $n_e$ from \CIIIdw\ with an assumption of homogeneous $n_e$ across the galaxies.
% In addition to the high-ionization $T_e$ and $n_e$ measurements, we also test the $T_e$ derived using the typical low-density assumption; and therefore we can test whether the traditional method without the involvement of the additional \OIIIwb\ is biased.
$T_e$(\OIII) is measured using \OIII$\lambda5008/\lambda4364$, with an input $n_e$.
We adopt two values of $n_e$.
First, we assume a uniform, low density of $n_e= 10^3\,{\rm cm^{-3}}$, adopting the average low-ionization value of high-z galaxies from \citet{Abdurrouf2024}.
Second, we utilize $n_e$(\CIIId) derived from \CIIIdw.
In both scenarios, we assume the ISM is homogeneous. 
As a result, we determine the $T_e$ using the measured \OIII$\lambda5008/\lambda4364$ ratio with the adopted $n_e$.
All of the candidate EMPGs have high electron temperatures of $T_e\sim24000-30000\,{\rm K}$ if low $n_e$ is assumed, as presented in Table \ref{tab:prop}.
Such high temperatures are possible as metallicities approach zero, but even at low metallicity, the decline in metal-line cooling is compensated for by increasingly efficient hydrogen cooling via collisional excitation of residual H$^0$, such that the equilibrium electron temperature tends to be constrained to $\lesssim2.5\times10^4$ K \citep[e.g.,][]{Stasinska2004,Osterbrock2006}.

\subsection{The Self-Consistent Direct Method: Simultaneous \OIII\ $T_e$ and $n_e$}
\label{sec:tene}
\begin{figure*}
\centering
\includegraphics[width=\textwidth]{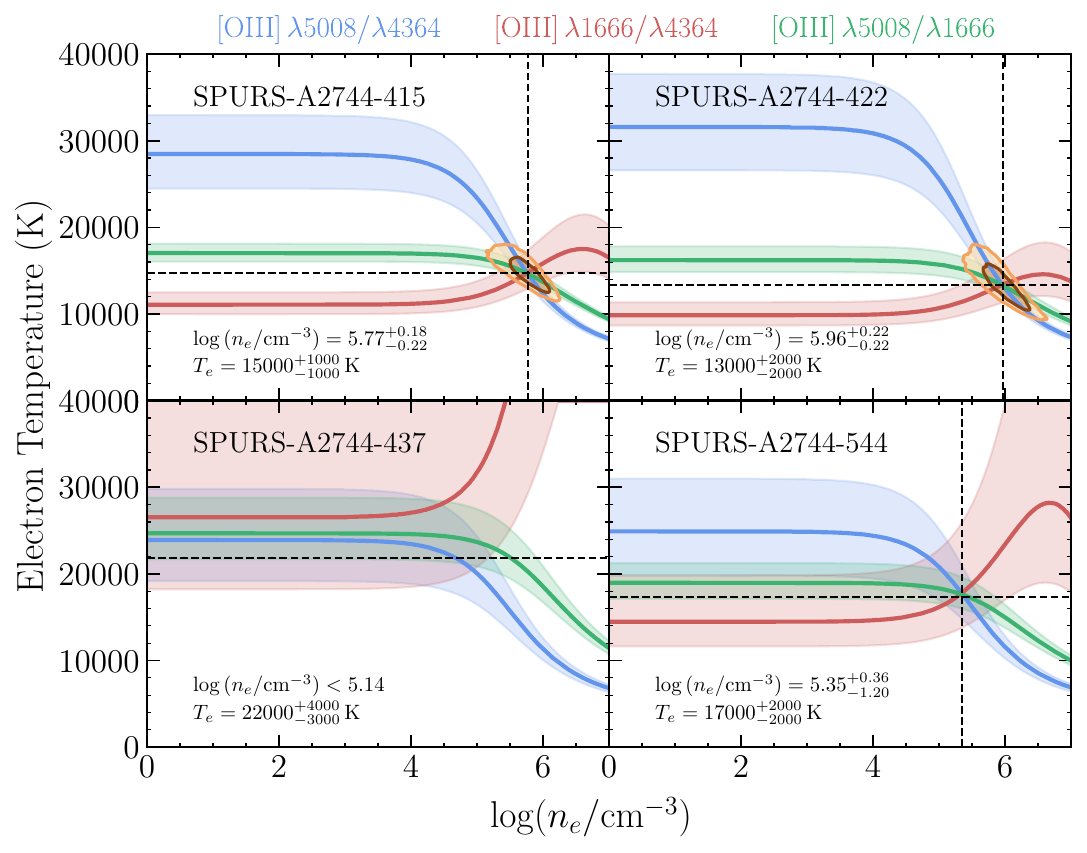}
\caption{
Simultaneous constraints of $T_e$ and $n_e$ with the self-consistent method for the EMPG candidates.
The blue, red, and green curves represent the $T_e-n_e$ relation for the given observational ratio of \OIIIw/$\lambda4364$, \OIIIwb/$\lambda4364$, and \OIIIw/$\lambda1666$, respectively, and the shaded bands indicate the $1\sigma$ uncertainty associated with each ratio.
The black dotted lines mark the best $T_e$ and $n_e$, and orange contours highlight the best $1\sigma$ (dark orange) and $2\sigma$ (light orange) in SPURS-A2744-415 and 422 when $n_e$ is constrained in the high-density regime.
For 437, the three $T_e-n_e$ relations do not converge, favoring a low-$n_e$ solution.
For 544, although $T_e-n_e$ relations do converge, the large uncertainties in the line ratios still allow a low $n_e$ solution.
This figure demonstrates that combining the three lines allows $T_e$ and $n_e$ to be constrained simultaneously, breaking the degeneracy present when only a single ratio is used.
}
\label{fig:den}
\end{figure*}

% As introduced in \S\ref{sec:te_old}, the traditional method for measuring  $T_e$ uses the \OIII$\lambda5008/\lambda4364$ line ratio, combined with either a density measured from other lines such as \OIIdw, or an assumed (low) density.
% However, \OIIIw\ becomes collisionally de-excited at high densities ($n_e\gtrsim10^{5}\,{\rm cm^{-3}}$), so in this regime the \OIII$\lambda5008/\lambda4364$ ratio depends on both $T_e$ and $n_e$ rather than $T_e$ alone (e.g., the blue curves in Figure \ref{fig:den}).
% \OIIIwa\ and \OIIIwb, in contrast, have higher critical densities and remain primarily sensitive to $n_e$.
% Since $T_e$ and $n_e$ are thus two unknowns entangled in a single line ratio, we instead combine three measurements: \OIIIw, \OIIIwa, and \OIIIwb, to solve for both simultaneously, as illustrated in Figure \ref{fig:den}.
{Here}, we adopt the self-consistent direct method \citep[][]{Berg2025_GISM,Arellano2026}, to simultaneously measure $T_e$ and $n_e$ in the high-ionization zone.
Figure \ref{fig:den} illustrates how the temperature vs. density sensitivities of the \OIII$\lambda5008/\lambda4364$, \OIIIwb/\OIIIwa, and \OIIIw/\OIIIwb\ line ratios can be used together to find a self-consistent solution.
The blue curves in Figure \ref{fig:den} demonstrate how an \OIII$\lambda5008/\lambda4364$ line ratio alone (\S\ref{sec:te_old}) cannot uniquely constrain both $T_e$ and $n_e$, especially when $n_e$ is high.
The inclusion of the rest-frame UV line \OIIIwb\ breaks the degeneracy since all three transitions (\OIIIwb, \OIIIwa, and \OIIIw) have different excitation energies and critical densities.
The density is determined from the intersection of the $T_e(n_e)$ curves derived from \OIII$\lambda5008/\lambda4364$ and \OIIIwb/\OIIIwa, where both ratios can be produced from the same $T_e$ and $n_e$.

We construct temperature-density curves $T_e(n_e)$ for each ratio (i.e., \OIII$\lambda5008/\lambda4364$, \OIIIwb/\OIIIwa, and \OIIIw/\OIIIwb) over a density grid spanning $\log(n_e/{\rm cm^{-3}})=0-7$ (with 1,000 grid steps).
The best $n_e$ and $T_e$ are determined from the intersection of the curves, with uncertainties propagated through the MCMC posterior samples.
When no converged result is found, we determine the best $n_e$ and $T_e$ as the closest values.
The $n_e$ and $T_e$ measured in the four EMPG candidates are presented in Table \ref{tab:prop}.

\begin{figure}
\centering
\includegraphics[width=\columnwidth]{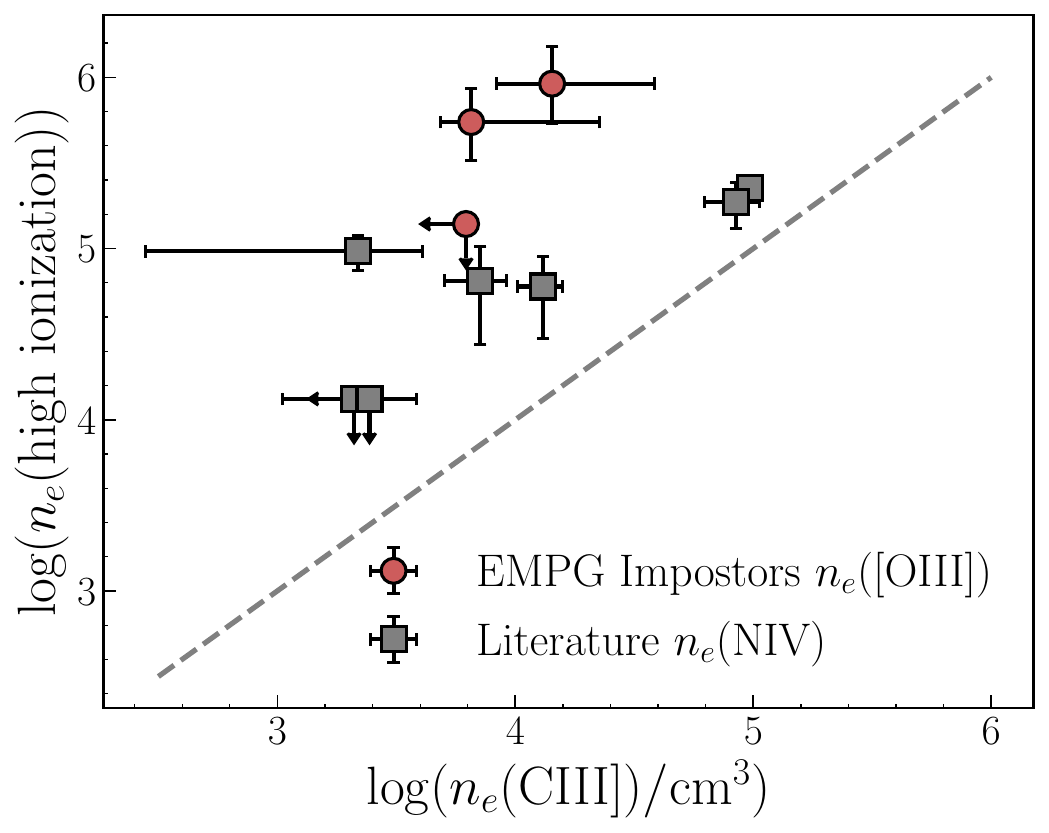}
\caption{Density stratification of the intermediate ionization zone probed by \CIIIdw\ and the high-ionization zone probed by O$^{++}$ lines \OIIIwb, \OIIIwa, and \OIIIw.
The red data show the EMPG Impostors, and the gray dashed-line depicts ${\rm log}(n_e({\rm [OIII]})) = {\rm log}(n_e({\rm CIII]}))$.
The gray squares are measurements of literature galaxies {from \citet{Martinez2025}}, for which $n_e({\rm [OIII]})$ is not available; we therefore compare to $n_e({\rm NIV})$ as a representative tracer of the high-ionization zone electron density.
These EMPG impostors further suggest the inhomogeneity in the ISM.
}
\label{fig:nn}
\end{figure}

With the self-consistent direct method, these candidate EMPG have lower $T_e$ of $\sim13000-22000{\, \rm K}$, and three of them show high $n_e$(\OIII) of $\sim10^{5-6}{\, \rm cm^{-3}}$ while one of them (437) has a low $n_e$(\OIII) of $\lesssim10^{5.14}{\, \rm cm^{-3}}$.
This method is particularly powerful because all three lines trace the same ionization zone (O$^{++}$), which is also used to derive the oxygen abundance (see \S\ref{sec:metallicity}), minimizing systematic biases from different elements/ions using, for example, \OIIdw, \SIIdw, or \CIIIdw\ doublets, which trace species from different regions.
We show the comparison between $n_e$(\OIII) and $n_e$(\CIIId) of EMPG candidates in Figure \ref{fig:nn} (see more discussion in \S\ref{sec:ISM}).

\subsection{direct-$T_{e}$ based metallicity} \label{sec:metallicity}

To determine the direct ($T_{e}$-based) oxygen abundances,
we use \pyneb\ \citep{Luridiana2015} with the collision strengths from \citet{Aggarwal1999} to calculate the corresponding oxygen abundances and physical conditions ($T_e$).
The total oxygen abundance can be approximated by the sum of $\rm{O^{++}}/\rm{H^{+}}$ and $\rm{O^{+}}/\rm{H^{+}}$, assuming that higher ionization states and neutral oxygen make negligible contributions \citep[e.g.,][]{Berg2021}:
\begin{equation}
    \rm{\frac{O}{H}}\simeq\frac{\rm{O^{+}}}{\rm{H^{+}}}+\frac{\rm{O^{++}}}{\rm{H^{+}}},
\end{equation}
where the ionic abundances can be obtained from:
\begin{equation}
    \frac{\rm O^{++}}{\rm H^{+}}=\frac{I_{\rm [OIII]}}{I_{\rm H\beta}}\frac{j_{\rm H\beta}}{j_{\rm [OIII]}},
\end{equation}
where $I$ is the emission line flux, and $j$ is the $T_e$- and $n_e$-dependent emissivity.
For the singly ionized O$^{+}$, we estimate the temperature of the low-ionization zone (O$^+$), by applying the relation $T_{\rm e}$(\OII) = 0.7 $\times$ $T_{\rm e}$(\OIII) + 3000\,K \citep{Campbell1986} and assume a density of $10^{3}\,{\rm cm^{-3}}$.
When \OIIdw\ is not detected in the sample {(${\rm SNR<3}$)}, we simply approximate $\rm{O/H}\sim\rm{O^{++}}/\rm{H^{+}}$ as the total oxygen abundance.
{We also do not detect \OIIdwa\ in any of our EMPG candidates.}

To finally measure the metallicity, we first adopt the $T_e$ and $n_e$ from low-ionization gases with the assumption of homogeneous ISM described in \S\ref{sec:te_old}.
We cap the maximum $T_e$ to be $30,000\,{\rm K}$ to avoid unphysical temperature when deriving the oxygen abundance.
We find these EMPG candidates, with the direct method with low-$n_e$ assumption (regardless of assumed $n_e=10^{3}\,{\rm cm^{-3}}$ or measured $n_e$(\CIII)), have extremely metal-poor signatures of 12+log(O/H) $\sim7.0-7.2$.
Second, we use the $T_e$ and $n_e$ from the self-consistent method with multi-zone models described in \S\ref{sec:tene}.
In contrast, the self-consistent method yields substantially higher metallicities, 12+log(O/H) $\sim7.3-8.2$.
These measurements are presented in Table \ref{tab:prop}.

\begin{figure}
\centering
\includegraphics[width=\columnwidth]{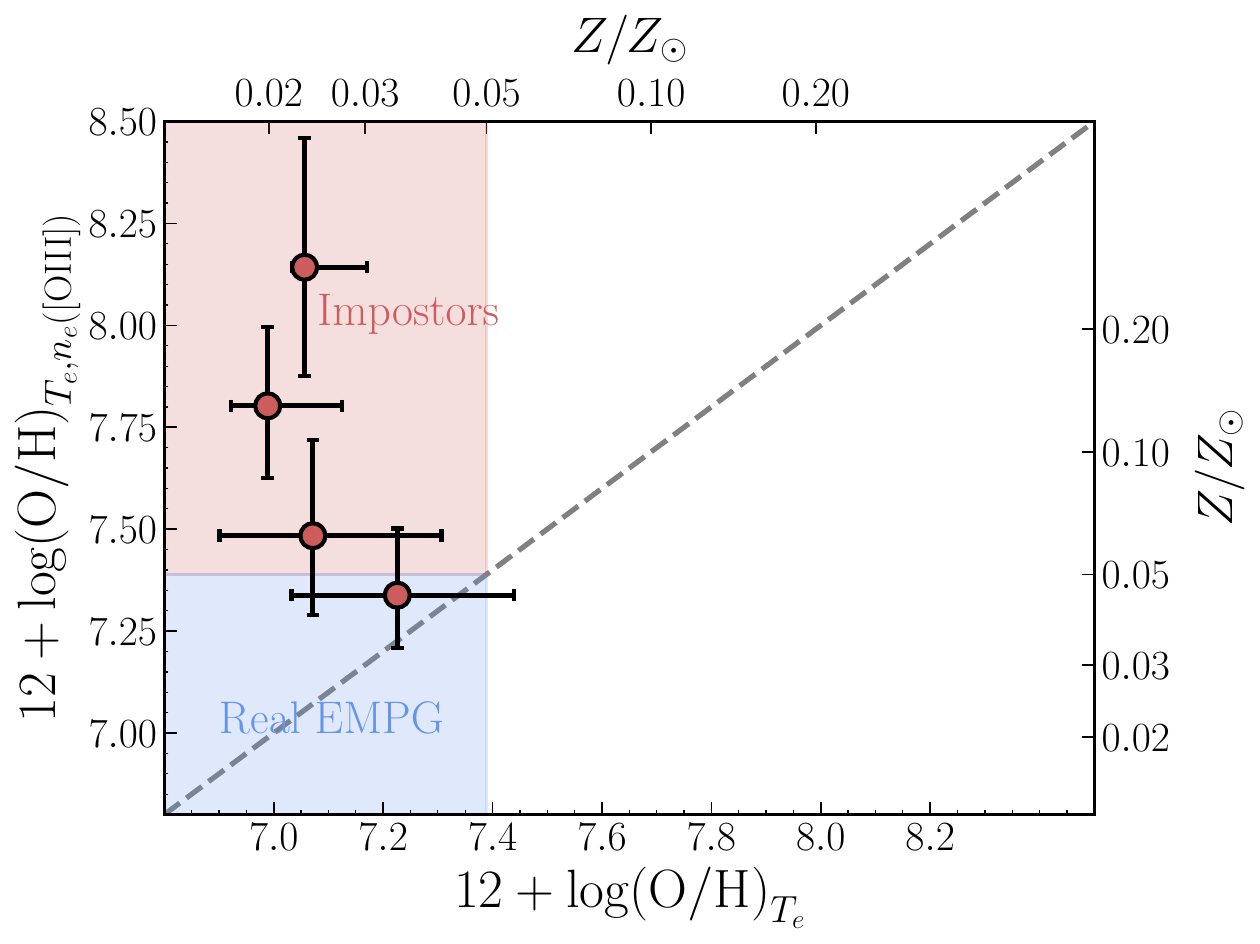}
\caption{Direct metallicities from the traditional direct-$T_e$ method if only \OIIIwa\ and \OIIIw\ are used compared with those obtained using the self-consistent method (this work).
The four EMPG candidates are shown as circles; three move substantially above the EMPG regime after accounting for their high densities.
The shaded regions indicate the parameter space associated with genuine EMPGs ({blue}) and EMPG impostors ({red}).
The one-to-one relation is shown by the gray dashed line.
This figure indicates how traditional low-$n_e$ assumption in direct-$T_e$ method can misclassify high-$n_e$ galaxies as EMPGs.
}
\label{fig:ZZ}
\end{figure}

\begin{deluxetable*}{lcccc}
\tablecaption{\label{tab:prop}
R3 ratio, physical properties and physical conditions of four candidate EMPGs.}
\tablewidth{\columnwidth}
\tablehead{
\colhead{Properties} &
\colhead{SPURS-A2744-415} &
\colhead{SPURS-A2744-422} &
\colhead{SPURS-A2744-437} &
\colhead{SPURS-A2744-544} 
}
\startdata
\hline
R3(=\OIIIw/H$\beta$) & $3.0^{+0.1}_{-0.1}$ & $4.2^{+0.2}_{-0.2}$ & $4.3^{+0.3}_{-0.3}$ & $2.8^{+0.2}_{-0.2}$ \\
$E(B-V)$ & 0.00 & 0.00 & $0.04^{+0.06}_{-0.04}$ & 0.00 \\
$12+\log(\mathrm{O/H})_{R3}$ & $7.20^{+0.02}_{-0.02}$ & $7.41^{+0.05}_{-0.04}$ & $7.44^{+0.06}_{-0.05}$ & $7.16^{+0.03}_{-0.03}$ \\
\hline
\multicolumn{5}{c}{Traditional Direct $T_e$} \\
\hline
$T_e^{a}\,(n_e=10^3)/\rm{K}$ & $28000^{+2000}_{-4000}$ & $30000^{+0}_{-3000}$ & $24000^{+6000}_{-5000}$ & $25000^{+5000}_{-5000}$ \\
$\log\,n_e(\mathrm{CIII])/cm^{-3}}$ & $3.77^{+0.56}_{-0.13}$ & $4.14^{+0.43}_{-0.23}$ & $<3.76^{b}$ & \ldots \\
$12+\log(\mathrm{O/H})_{T_e}$ & $6.99^{+0.14}_{-0.06}$ & $7.06^{+0.11}_{-0.02}$ & $7.23^{+0.22}_{-0.19}$ & $7.07^{+0.23}_{-0.17}$ \\
$12+\log(\mathrm{O/H})_{T_e,n_e({\rm CIII]})}$ & $6.99^{+0.14}_{-0.07}$ & $7.06^{+0.13}_{-0.02}$ & $7.23^{+0.22}_{-0.19}$ & \ldots \\
\hline
\multicolumn{5}{c}{Self-Consistent $T_e$ and $n_e$} \\
\hline
$T_e$(self-consistent)/\rm{K} & $15000^{+1000}_{-1000}$ & $13000^{+2000}_{-2000}$ & $22000^{+4000}_{-3000}$ & $17000^{+2000}_{-2000}$ \\
$\log\,n_e(\mathrm{[OIII])/cm^{-3}}$ & $5.77^{+0.18}_{-0.22}$ & $5.96^{+0.22}_{-0.22}$ & $<5.14^{b}$ & $5.35^{+0.36}_{-1.20}$ \\
$12+\log(\mathrm{O/H})_{T_e,\,n_e({\rm [OIII]})}$ & $7.80^{+0.19}_{-0.18}$ & $8.14^{+0.30}_{-0.26}$ & $7.34^{+0.16}_{-0.13}$ & $7.49^{+0.24}_{-0.19}$ \\
\hline
\enddata
\tablenotetext{a}{$T_e$ is capped at $30,000\,{\rm K}$}
\tablenotetext{b}{1$\sigma$ upper limit}
%\textbf{Note.} 
%Different codes assume different IMFs (Table \ref{tab:sedfitting}). 
%For a fair comparison, we suggest readers to multiply 0.61 for stellar mass and 0.63 for SFR from \citet{Salpeter1955} to \citet{Chabrier2003} \citep{Madau2014}. 
%For a fair comparison, we \textcolor{red}{multiply} 0.94 for stellar mass from \citet{1993MNRAS.262..545K} to \citet{Chabrier2003} \citep{Madau2014}. 
\end{deluxetable*}
%\tablecomments{}
% https://docs.google.com/spreadsheets/d/1k5Xgf7vtZjnpvPa5rycbFL_WlGYhXsqBX5VMnrCTFos

\section{High Electron Density as Extremely Metal-Poor Galaxy Impostors?}
\label{sec:discussion}

\subsection{EMPG Impostors}
\label{sec:impostors}
% This paper has successfully derived the $n_e$ in the high-ionization gases to account for the stratification of ISM in four EMPG candidates.
We have provided the most robust direct-method metallicities of $z\gtrsim5$ EMPG candidates using the self-consistent method proposed by \citet{Berg2025_GISM} and \citet{Arellano2026}.
Among the four EMPG candidates, we find that three of them have $n_e\sim10^{5-6}\,{\rm cm^{-3}}$ in the high-ionization gas.
This high $n_e$ significantly affects the determination of the high-ionization zone $T_e$ and consequently the metallicity from the traditional direct method, whenever a low $n_e$ is instead adopted.
As shown in Figure \ref{fig:den}, a given \OIIIw/$\lambda4364$ ratio can be produced by high $T_e$ with low $n_e$, or low $T_e$ with high $n_e$. 
The blue curve in Figure \ref{fig:den} demonstrated that if a $T_e$ is derived from the traditional method with \OIIIw/$\lambda4364$, with low $n_e$ adopted, $T_e$ would in fact be overestimated, consistent with previous works \citep{Katz2023,Hayes2025,Martinez2025}.
In the high $n_e$ case, collisional de-excitation suppresses much of the \OIIIw\ emission, leading to an underestimated \OIIIw-based metallicity.
Hence, the increasing number of high-$n_e$ galaxies at high-redshift increases the chances of finding ``EMPG impostors'' where truly high-$n_e$ galaxies masquerade as EMPGs when using the traditional direct method.

Figure \ref{fig:ZZ} demonstrates the direct metallicities derived from the traditional $T_e$ and self-consistent methods.
Three of the candidate EMPGs are in fact impostors (one of them, 544, is also consistent with real EMPG), with high $n_e$ which leads to substantially higher (and more accurate) metallicities once density is properly accounted for.
Only one (437) has a low enough density to be a genuine EMPG.
Because these impostors' observed R3 ratios remain low despite their true, higher metallicities, strong-line diagnostics which rely on R3 as a metallicity proxy cannot distinguish them from real EMPGs and would misclassify them.
This suggests EMPG impostors could represent a substantial fraction of all EMPG candidates identified via strong-line methods ($50\%-75\%$ of the EMPG candidate samples in this analysis), although a more detailed completeness analysis is warranted (see \S\ref{sec:completeness}).
In other words, any EMPG candidate, whether identified purely via the R3 ratio or confirmed with the traditional direct-$T_e$ method with an assumption of low $n_e$, will still have an underestimated metallicity unless lines with very high critical densities are used.

{\citet{Arellano2026} applied the self-consistent method to six N-emitters, finding metallicity shifts of up to $\sim0.3\,$dex.
Although these shifts are smaller than those found for our EMPG impostors, the baseline density assumptions differ.
Their comparison uses previous literature metallicities, some of which were already derived assuming high densities (\(n_e \sim 10^5\ \mathrm{cm}^{-3}\)), by diagnostics such as \NIVdw.
In contrast, our comparison adopts a common low-$n_e$ baseline. The smaller published shifts therefore do not necessarily imply a weaker density-induced metallicity bias in these objects.
A consistent reanalysis using the same baseline assumptions is needed to establish how the magnitude of this bias varies across the metallicity range.
}
Although beyond the scope of this paper, we note that not only EMPG candidates but all direct-$T_e$ metallicity measurements at high redshift should be carefully re-examined (see \S\ref{sec:future}).

{As mentioned in \S\ref{sec:selection}, we do not explicitly exclude galaxies with potential AGN contributions from our sample.
Traditional BPT diagrams \citep[e.g.,][]{Baldwin1981,Kewley2001,Kauffmann2003} are not well suited for AGN classification at high redshift \citep[e.g.,][]{Kocevski2023,Ubler2023,Harikane2023,Maiolino2024,Scholtz2025}, and neither \NIIw\ nor \SIIdw\ is detected in our EMPG impostors.
We therefore adopt the AGN diagnostics proposed by \citet{Mazzolari2024} to assess potential AGN contributions. 
Two of the four galaxies show no clear AGN signatures in any of the three diagnostics.
The remaining two galaxies, SPURS-A2744-422 and SPURS-A2744-437, have lower limits that are consistent with both the AGN and SFG regions in one of the three diagnostics, due to the lack of \OII\ detections.
The other two diagnostics show no clear evidence of AGN activity in either galaxy.
Both galaxies also exhibit the high-ionization \HeIIw\ emission line, which may indicate AGN activity or another source of hard ionizing radiation.
Interestingly, SPURS-A2744-422 is classified as an EMPG impostor, whereas SPURS-A2744-437 is the most secure genuine EMPG in our sample. 
Therefore, we cannot establish whether AGN contributions play a significant role in producing EMPG impostors. 
Further investigation is needed to clarify their potential impact.
}

% We also demonstrate how a strong-line diagnostics fails and would severely underestimate the metallicity in Figure \ref{fig:R3}.

\subsection{Comparison to Strong-Line EMPG Diagnostics}
\label{sec:strong_line}
Few high-redshift spectra are deep enough to test EMPG candidates with the direct method.
Instead, the R3 (=\OIIIw/H$\beta$) strong-line ratio is the most widely used diagnostic to look for EMPG candidates \citep[e.g.,][]{Chemerynska2024,Hsiao2025,Maiolino2025,Morishita2025,Asada2026}.
We adopt the strong-line empirical relation from \citet{Isobe2026}, {because its calibration sample includes EMPGs with} $Z\sim1-2\%\,Z_{\odot}$:
\begin{equation}
    {\rm log(R3)} = -0.055x^{3}-0.4987x^{2}+0.9842x+0.3034,
\end{equation}
where $x=12+{\rm log(O/H)}-7$.
The SPURS galaxies studied here have R3 ratios of $2.8-4.3$, with the corresponding strong-line metallicities of 12+log(O/H) $\sim7.2-7.5$.
The metallicities estimated from R3 are listed in Table \ref{tab:prop}.

\begin{figure}
\centering
\includegraphics[width=\columnwidth]{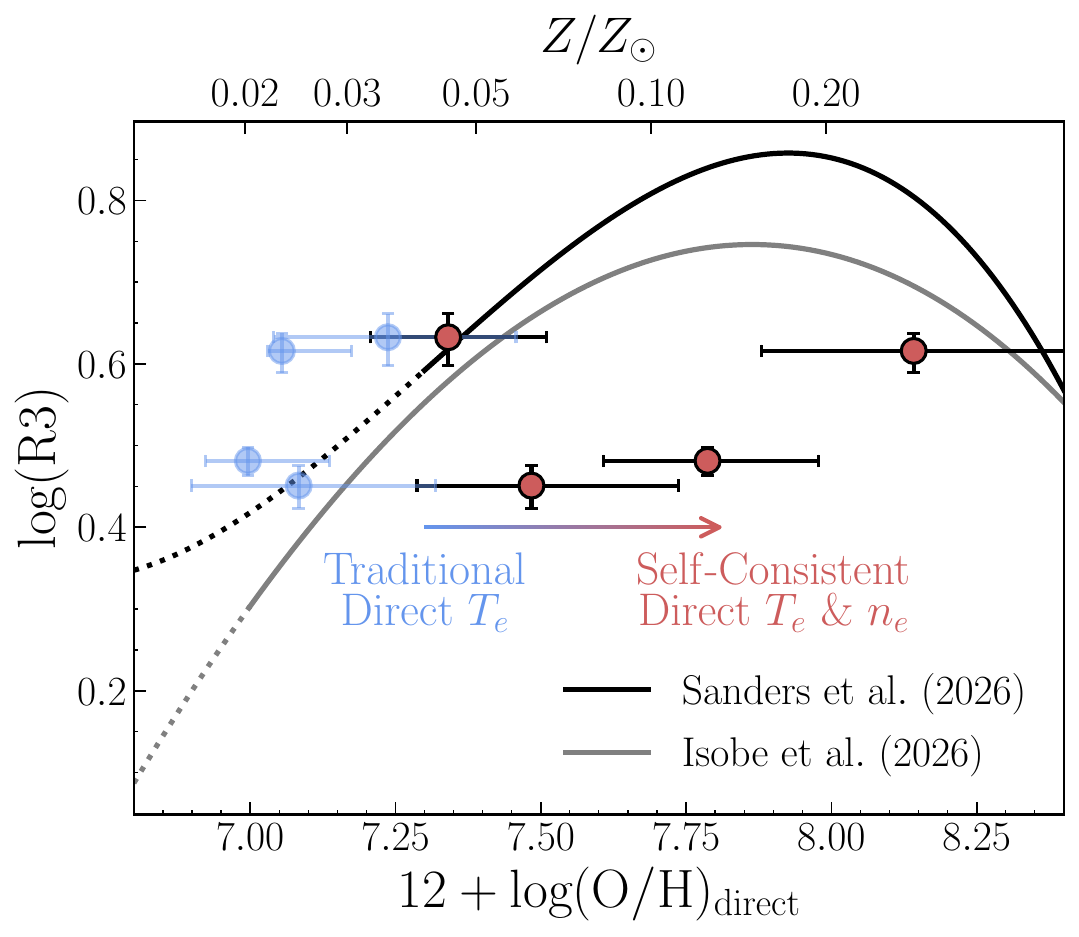}
\caption{Strong-line diagnostics of R3$=$\OIIIw/H$\beta$ vs metallicity.
The black and gray curves show the diagnostics calibrated by \citet{Sanders2026} and \citet{Isobe2026}, with the extrapolated regions marked as dotted lines.
The blue circles show traditional direct-method metallicities assuming $n_e=10^{3}\,{\rm cm^{-3}}$, while the red circles show metallicities obtained using the self-consistent method.
The arrow highlights how a low $n_e$ assumption can bias the metallicity measurement, and the importance of $n_e$ measurements at high-redshift and in high-ionization gas.
Both strong-line diagnostics and traditional direct-method measurements that assume low $n_e$ can misclassify high-density galaxies as EMPGs.}
\label{fig:R3}
\end{figure}

Figure \ref{fig:R3} shows the strong-line calibrations from \citet{Sanders2026} based on 139 high-redshift galaxies at $z\sim2-10$ with direct metallicity measurements, and \citet{Isobe2026} in gray curve, together with the R3 measurements and metallicities derived using both the traditional direct method (blue circles) and the new self-consistent direct method (red circles).
In the JWST era, several works have calibrated strong-line diagnostics at the high-redshift, low-metallicity end \citep[e.g.,][]{Sanders2024,Sanders2026,Hsiao2026,Isobe2026}.
% while no significant deviations from the local galaxies have been seen.
Again, if we use the traditional method with an assumption of low $n_e$, not surprisingly, the resulting metallicities are consistent with the strong-line empirical relations (blue circles in Figure \ref{fig:R3}).
However, once  $n_e$ and $T_e$ in the high-ionization zone are measured using the self-consistent method, the metallicities increase by up to $\sim1.1$ dex.
In addition to R3, many strong lines used in diagnostics have even lower critical densities than \OIIIw\ (e.g., \OIIdw), which should also be revisited if galaxies have high $n_e$ \citep{Martinez2025,Sanders2026,Hsiao2026,Isobe2026}.
{
We also explore other strong-line diagnostics calibrated by \citet{Isobe2026}, including R2 ($=$\OIIdw/H$\beta$), R23 ($=$(\OIIdw+\OIIIwc+\OIIIw)/H$\beta$), O32 ($=$\OIIIw/\OIIdw), Ne3 ($=$\NeIIIw/H$\gamma$), RO2Ne3 ($=$(\OIIdw+\allowbreak\NeIIIw)/H$\gamma$), and Ne3O2 ($=$\NeIIIw/\allowbreak\OIIdw).
We focus on SPURS-A2744-415 because it is one of the two plausible EMPG impostors and its \OIIdw\ detection allows us to apply all six diagnostics.
All yield $12+\log({\rm O/H})<7.25$, consistent with the conventional direct-method metallicity derived under a low-$n_e$ assumption, but none recovers the higher metallicity inferred from our self-consistent analysis.
One might expect diagnostics that do not involve \OIIIw\ to be less susceptible to the bias caused by collisional de-excitation of this line.
However, these diagnostics are also calibrated against direct-method metallicities derived under low-$n_e$ assumptions, so it is unsurprising that none recovers the self-consistent metallicity of this impostor.
}

\subsection{Implications for Chemical Evolution}
\label{sec:evo}
If metallicities are systematically underestimated in a significant fraction of high-redshift EMPGs or, more broadly, in high-redshift galaxies generally, then conclusions that rely on oxygen abundance measurements should be revisited accordingly. 
First of all, the mass-metallicity relation (MZR) has been established in the local universe, with a tight relation between stellar mass and metallicity \citep[e.g.,][]{Lequeux1979,Tremonti2004,Erb2006,Lee2006,Berg2012,Curti2020}, as well as the ``Fundamental Metallicity Relation'' (FMR) which is the inclusion of the SFR as a second dependence \citep{Ellison2008,Mannucci2010,Curti2020,Sanders2021}.
The MZR and FMR encode how galaxies regulate chemical enrichment through the balance among star formation, inflows, and outflows, and studying their evolution is key to understanding how galaxy chemical evolution and self-regulation proceed across cosmic time.
Further studies at high-redshift with JWST data have investigated the slope, normalization, and redshift evolution of these relations \citep[e.g.,][]{Heintz2023,Curti2024,Morishita2024,Chemerynska2024,Arellano2025,Sarkar2025,Chakraborty2025,Nishigaki2025,Pollock2026,Koller2026,Hsiao2026,Isobe2026}.
However, if a significant fraction of EMPG candidates are indeed impostors, the slope and normalization will be biased by these EMPG impostors.
Galaxies at high-redshift across all stellar masses should also be scrutinized.
{
% More broadly, density-dependent metallicity biases could contribute to the offsets from the local FMR reported at $z\gtrsim3$ \citep[e.g.,][]{Heintz2023,Nakajima2023,Curti2024,Hsiao2026,Lam2026,Stanton2026}.
Evolution of the FMR at $z\gtrsim3$ has been reported and may reflect differences in gas accretion, star formation, and feedback mechanisms \citep[e.g.,][]{Heintz2023,Nakajima2023,Curti2024,Hsiao2026,Lam2026,Stanton2026}.
Moreover, the applicability of the local FMR to low-mass galaxies ($M_*<10^{9}\,M_\odot$) remains uncertain even at low redshift \citep{Laseter2026}.
If high-$n_e$ nebulae are ubiquitous in high-redshift galaxies, metallicity biases from low-$n_e$ assumptions could also contribute to the apparent evolution of the FMR.
}
Although beyond the scope of this paper, {high $n_e$ should also be considered in SED fitting that includes nebular emission.
Density-dependent changes in nebular line strengths, including collisional de-excitation, can affect broadband fluxes and potentially bias inferred stellar masses if low-density models are assumed.
Accounting for these effects may therefore be important for accurately characterizing the MZR and FMR and their evolution.}

Not only is metallicity itself underestimated due to the inaccurate assumption of $n_e$, but other abundances relative to O are also biased.
Unusual chemical abundance patterns have also been discovered in high-redshift galaxies, including nitrogen enhancement \citep[e.g.,][]{Bunker2023,Cameron2023,Isobe2023b,Topping2024,Marques-Chaves2024,Morel2025,Naidu2026,Zhang2026,Berg2026,Cameron2026}, and, in a smaller number of cases, carbon enhancement \citep[e.g.,][]{DEugenio2024,Hsiao2025b,Nakajima2026,Scholtz2026,Pollock2026b}.
Some of the literature indeed considers the multi-zone ISM and takes high $n_e$ into account \citep[e.g.,][]{Berg2026}, but seldom do these studies measure $n_e$(\OIII) (see \S\ref{sec:ISM}).
\citet{Arellano2026} analyzed six nitrogen-enhanced high-z galaxies using the self-consistent method to measure $n_e$(\OIII), and demonstrated that N/O can be overestimated by up to $\sim0.8$ dex.
Comprehensive measurements of $n_e$ across multiple ionization zones are required to obtain accurate chemical abundances, and therefore determine the peculiar chemical patterns on a global scale.

\subsection{Density Stratification}
\label{sec:ISM}
\ion{H}{2} regions in the ISM do not seem to be homogeneous in some galaxies, including the Milky Way \citep[e.g.,][]{Franco2000,Simpson2004,Rubin2011} and some dwarf galaxies \citep[e.g.,][]{James2020}, while a chemical homogeneity is also seen in some dwarf metal-poor galaxies \citep[e.g.,][]{Kunth2000,Lee2006b}.
Therefore, it is crucial to test whether the ISM is stratified in early galaxies.
In young, compact star-forming regions, the density structure of the ionized gas may retain that of the dense natal molecular cloud before feedback-driven expansion establishes pressure equilibrium.
If density decreases outward from the youngest stellar populations, ionization stratification will cause higher-ionization species to preferentially trace denser gas closer to the ionizing sources, while lower-ionization species trace more diffuse gas at larger radii.
This scenario may be particularly relevant at high redshift, where electron densities and star-formation-rate surface densities are elevated and observations preferentially select young, high-equivalent-width starbursts \citep[e.g.,][]{Reddy2023,Topping2025}.
Thus, under this scenario, one might expect $n_e$(\OII: 13.6 eV)$<n_e$(\CIIId: 24.4 eV)$<n_e$(\OIII: 35.1 eV)$<n_e$(\NIV: 47.5 eV) \citep{Stanghellini1989,Gutierrez2010}.
As stated in \S\ref{sec:den_ciii}, in addition to the high-ionization $n_e$(\OIII), we also assess the $n_e$(\CIIId) in the intermediate ionization zone.
Figure \ref{fig:nn} exhibits the density stratification in the impostors, compared to literature measurements.
Most of the literature measurements do not have $n_e$ in high-ionization zones probed by \OIII.
Instead, we compare to $n_e$ measurements from other emission lines that trace high-ionization gas, such as \NIVdw.
Although N$^{3+}$ has a higher ionization potential ($47.5\,{\rm eV}$) than O$^{++}$ ($35.1\,{\rm eV}$), they are both considered representative tracers of high-ionization gases \citep[see, e.g., Figure 4 in][]{Berg2021}.

Despite \NIV\ having a higher ionization potential, the $n_e$(\OIII) values of two impostors are $\sim0.5-1.0$ dex higher than $n_e$(\NIV) in the literature samples. 
However, this simple uniform-shell structure has not yet been verified observationally, and the reverse trend (lower-ionization zones having higher $n_e$) has also been seen \citep[e.g.,][]{Maseda2017,Mingozzi2022}.
A simpler possibility is that EMPG impostors might have different ISM stratifications than the literature galaxies.
In other words, if $n_e$(\OIII) were measured directly in the literature galaxies, it might be smaller than the 
$n_e$(\NIV) actually reported for them, consistent with standard ionization stratification, since N$^{3+}$ traces a more interior zone than O$^{++}$.

The $n_e$(\OIII) values in two EMPG impostors are $\sim2$ dex higher than the $n_e$(\CIIId) in the intermediate ionization gases.
The contrast in $n_e$ between the high- and intermediate-ionization zones is $\sim1$ dex larger in our EMPG impostors than in the literature sample.
To avoid comparing apples to oranges, one should study $n_e$(\OIII) in other galaxies, and $n_e$(\NIV) in EMPG impostors for a fairer comparison, though none of the nitrogen lines are detected in this work.
At minimum, we demonstrate that the ISM can be strongly density-stratified and that this stratification can bias metallicity measurements.
We also present, to our knowledge, the first sample with measurements of both $n_e$(\OIII) and $n_e$(\CIIId) measured at such redshift.

\subsection{Does the \OIIIwb\ Detection Requirement Favor Impostors Over Real EMPGs?}
\label{sec:completeness}
\OIIIwb\ has a much higher excitation energy ($7.44\,$eV) than \OIIIw\ ($2.51\,$eV), so it is only seen in relatively metal poor galaxies with high temperatures.
Hence, one might expect it to be strongest in very metal poor galaxies (but eventually go away as Pop III is approached, similar to \OIIIwa\ and \OIIIw).
Therefore, in this section, we test whether, among galaxies already identified as EMPG candidates based on their low-R3 ratios, the additional requirement of an \OIIIwb\ detection preferentially selects EMPG impostors over real EMPGs.

Recall that our sample selection has two criteria: (1) a low R3 ratio, and (2) detection of all three lines, \OIIIw, \OIIIwa, and \OIIIwb.
In this section, we test whether our sample is biased toward high-density galaxies by one specific selection step: the requirement that \OIIIwb\ be detected.
Together these criteria admit two physically distinct populations: real EMPGs and EMPG impostors (high-density galaxies masquerading as metal-poor).
We consider the simplest scenario: fixing the observed \OIIIw\ and \OIIIwa\ fluxes, we vary the \OIIIwb\ flux and solve for the electron density implied at each value.

\begin{figure}
\centering
\includegraphics[width=\columnwidth]{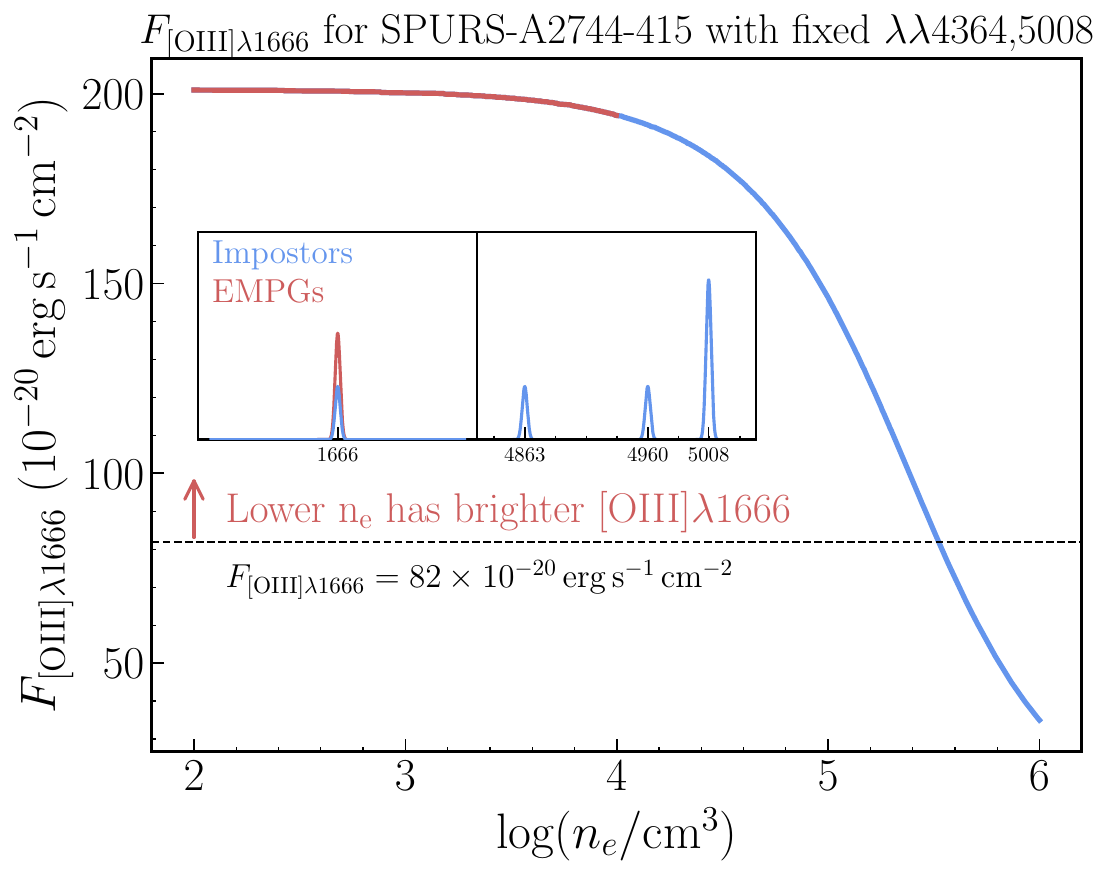}
\caption{{Predicted line fluxes of \OIIIwb\ as a function of $n_e$ with fixed \OIIIwa\ and \OIIIw\ for SPURS-A2744-415.
The inset panels show simulated spectra of a EMPG impostor (blue) and a real EMPG (red).
The black-dashed line indicate the measured \OIIIwb\ of SPURS-A2744-415.
If a galaxy is a true EMPG (low-$n_e$), \OIIIwb\ is brighter than a impostor (with fixed \OIIIwa\ and \OIIIw), and will be detected if it is a real EMPG.
Hence, the threshold of \OIIIwb\ does not preferentially select impostors.
}
}
\label{fig:bias}
\end{figure}

{Figure~\ref{fig:bias} shows how \OIIIwb\ varies with $n_e$ for SPURS-A2744-415, with the \OIIIw\ and \OIIIwa\ fluxes held fixed.}
We find that lower \OIIIwb\ fluxes correspond to higher inferred $n_e$.
Equivalently, a real low-density EMPG would produce a brighter \OIIIwb\ compared to an impostor if they have the same R3 and \OIIIw/$\lambda4364$, well above our detection threshold.
This means the \OIIIwb\ detection requirement does not preferentially exclude real EMPGs; a true low-density, low-metallicity galaxy would satisfy our selection criteria.
This test therefore shows that the bias toward high-density impostors in our low-R3 sample is not an artifact of the \OIIIwb\ detection cut.
Rather, it reflects the fact that, among galaxies with low R3, high-density impostors are physically common enough that requiring \OIIIwb\ does not filter them out.
Altogether, impostors, which have both higher oxygen abundance and higher electron density than genuine EMPGs, produce a brighter \OIIIwb\ line than a true EMPG with the same R3 ratio would (even though a real EMPG has a higher $T_e$).
As a result, \OIIIwb/\OIIIwa\ is a better $T_e$ diagnostic of EMPGs \citep[see also][]{Martinez2025}.

\subsection{Future Prospects}
\label{sec:future}
We have shown that three of our four EMPG candidates have high $n_e$ in the high-ionization zone, causing their metallicities to be underestimated by up to  $\sim1.1$ dex when a low $n_e$ is assumed, masquerading as genuine EMPGs even under the traditional direct-$T_e$ method.
This underscores the need to take  $n_e$ into account more seriously when measuring metallicities of EMPGs at high redshift.
In light of this finding, this analysis should also be performed for a larger sample, not limited to EMPGs alone.
A systematic analysis at high-z including massive/metal-rich galaxies is also needed before we can conclusively interpret any odd chemical abundances and chemical evolution mentioned in \S\ref{sec:evo}.
Therefore, to cover \OIIIwb, more JWST NIRSpec G140M/H coverage is required along with an archival analysis of all available high-redshift galaxies.

In addition, testing $n_e$ across different ionization zones and establishing empirical relations between them is another useful direction, serving as an additional calibration for strong-line diagnostics or for cases where high-ionization density measurements are unavailable.
For example, the EMPG impostors presented in this work have $n_e$ measured from both the \CIIIdw\ (intermediate-ionization) and from our O$^{++}$-based self-consistent diagnostic (high-ionization).
Low-ionization density measurements, such as $n_e$(\OIIdw), potentially obtained from the high-resolution G235H grating, would add another piece to the multi-zone ISM picture.
Relations among $n_e$ across ionization zones in high-z galaxies would provide the community with a scaling relation to apply when high-ionization density is not directly accessible.

Independent validation from other high-ionization density-sensitive lines would also strengthen this method.
For instance, \NIVdw, another density-sensitive doublet probing high-ionization gas ($>47.5\,{\rm eV}$), could serve as an additional or alternative tracer, while how tight the relation is between $n_e$(\NIV) and $n_e$(\OIII) should be also explored.
The Atacama Large Millimeter/submillimeter Array (ALMA) radio observations of \OIII$\lambda\,52\,{\rm \mu m}$ and \OIII$\lambda\,88\,{\rm \mu m}$ are also promising \citep{Harikane2025}, {although more sensitive to lower densities due to low critical densities of $\sim10^{3}\,{\rm cm^{-3}}$.}

\section{Conclusions}
\label{sec:conclusion}

In this paper, we identify four candidate EMPGs based initially on their low R3 ratios of $\rm{R3}<5$, and further supported by the ``direct-$T_e$'' metallicity with the traditional low-$n_e$ assumption with $Z\sim1-5\%\,Z_{\odot}$, consistent with the expectation from the strong-line R3 relation.
However, we adopt the self-consistent direct method to simultaneously constrain both $T_e$ and $n_e$ from \OIIIwb, \OIIIwa, and \OIIIw\ proposed by \citet{Berg2025_GISM} and \citet{Arellano2026}, and apply it to these EMPG candidates. 
We find that three of the four candidates have high electron densities 
{in the \OIII-emitting gas}, $n_e\sim10^{5}-10^{6}\,{\rm cm^{-3}}$ causing \OIIIw\ to be collisionally de-excited and biasing the inferred metallicities.

In these EMPG impostors, the metallicity is underestimated by up to $\sim1.1$ dex if the traditional direct method based on \OIIIw\ and \OIIIwa\ is used with a low-density assumption.
As a result, the strong-line diagnostics (e.g., the R3 ratio), would also severely underestimate the metallicities of these EMPG impostors and falsely categorize them into EMPGs.
These EMPG impostors could disguise themselves as genuine EMPGs in both strong-line diagnostics and even in traditional direct-method metallicity estimates when a low $n_e$ is assumed.
The inclusion of the rest-frame UV line \OIIIwb\ breaks this optical illusion, which arises from relying on rest-frame optical lines (\OIIIwa\ and \OIIIw) alone.
Hence, chemical abundance measurements in high-redshift galaxies should be revisited using accurately measured densities to take high-$n_e$ and the ISM stratification into account in future work.

\section{Acknowledgments}
%\begin{acknowledgments}
We appreciate the anonymous referee for their constructive comments.
TH thanks the University of Texas at Austin Cosmic Frontier Center and Department of Astronomy for supporting this work.
A.R.G. acknowledges support from the National Science Foundation through the NSF Graduate Research Fellowship Program under Grant No. DGE-2137420. Any opinions, findings, and conclusions or recommendations expressed in this material are those of the author(s) and do not necessarily reflect the views of the National Science Foundation.

This work is based in part on observations made with the NASA/ESA/CSA JWST.
The data were obtained from the Mikulski Archive for Space Telescopes at the Space Telescope Science Institute, which is operated by the Association of Universities for Research in Astronomy, Inc., under NASA contract NAS 5-03127 for JWST.
These observations are associated with program GO 9214 (PIs: C. Mason \& D. Stark), and we appreciate the SPURS team's efforts in designing the program and collecting the data.

%\end{acknowledgments}

%% To help institutions obtain information on the effectiveness of their 
%% telescopes the AAS Journals has created a group of keywords for telescope 
%% facilities.
%
%% Following the acknowledgments section, use the following syntax and the
%% \facility{} or \facilities{} macros to list the keywords of facilities used 
%% in the research for the paper.  Each keyword is check against the master 
%% list during copy editing.  Individual instruments can be provided in 
%% parentheses, after the keyword, but they are not verified.

\vspace{5mm}
%\facilities{JWST(NIRCam, NIRSpec, MIRI), HST(ACS, WFC3)}
%% Similar to \facility{}, there is the optional \software command to allow 
%% authors a place to specify which programs were used during the creation of 
%% the manuscript. Authors should list each code and include either a
%% citation or url to the code inside ()s when available.

%\software{STScI JWST pipeline;
%          \msaexp;
%          \grizli\ \citep{grizli};
%          \astropy\ \citep{astropy2022, astropy2018, astropy2013};
%          \piXedfit\ \citep{Abdurrouf2021,Abdurrouf2022};
%          \pyneb\ \citep{pyneb_Luridiana2015};
%          \HIIC\ \citep{Perez2014}
          %Prospector \citep{Leja2017,Johnson2021},
          %\jdaviz\ \citep{JDAviz}
%          }
%% Appendix material should be preceded with a single \appendix command.
%% There should be a \section command for each appendix. Mark appendix
%% subsections with the same markup you use in the main body of the paper.

%% Each Appendix (indicated with \section) will be lettered A, B, C, etc.
%% The equation counter will reset when it encounters the \appendix
%% command and will number appendix equations (A1), (A2), etc. The
%% Figure and Table counter will not reset.

%% For this sample we use BibTeX plus aasjournals.bst to generate the
%% the bibliography. The sample631.bib file was populated from ADS. To
%% get the citations to show in the compiled file do the following:
%%
%% pdflatex sample631.tex
%% bibtext sample631
%% pdflatex sample631.tex
%% pdflatex sample631.tex

\bibliography{papers}{}

@ARTICLE{horne86,
       author = {{Horne}, K.},
        title = "{An optimal extraction algorithm for CCD spectroscopy.}",
      journal = {\pasp},
         year = 1986,
        month = jun,
       volume = {98},
        pages = {609-617},
          doi = {10.1086/131801},
       adsurl = {https://ui.adsabs.harvard.edu/abs/1986PASP...98..609H}
}

@ARTICLE{Hsiao2025,
       author = {{Hsiao}, Tiger Yu-Yang and {Sun}, Fengwu and {Lin}, Xiaojing and {Coe}, Dan and {Egami}, Eiichi and {Eisenstein}, Daniel J. and {Fudamoto}, Yoshinobu and {Bunker}, Andrew J. and {Fan}, Xiaohui and {Harikane}, Yuichi and {Helton}, Jakob M. and {Kakiichi}, Koki and {Liu}, Yichen and {Liu}, Weizhe and {Maiolino}, Roberto and {Ouchi}, Masami and {Tee}, Wei Leong and {Wang}, Feige and {Wu}, Yunjing and {Xu}, Yi and {Yang}, Jinyi and {Zhu}, Yongda},
        title = "{SAPPHIRES: Extremely Metal-Poor Galaxy Candidates with $12+{\rm log(O/H)}<7.0$ at $z\sim5-7$ from Deep JWST/NIRCam Grism Observations}",
      journal = {arXiv e-prints},
         year = 2025,
        month = may,
          eid = {arXiv:2505.03873},
        pages = {arXiv:2505.03873},
          doi = {10.48550/arXiv.2505.03873},
archivePrefix = {arXiv},
       eprint = {2505.03873},
 primaryClass = {astro-ph.GA},
       adsurl = {https://ui.adsabs.harvard.edu/abs/2025arXiv250503873H}
}

@ARTICLE{Cullen2025,
       author = {{Cullen}, F. and {Carnall}, A.~C. and {Scholte}, D. and {McLeod}, D.~J. and {McLure}, R.~J. and {Arellano-C{\'o}rdova}, K.~Z. and {Stanton}, T.~M. and {Donnan}, C.~T. and {Dunlop}, J.~S. and {Shapley}, A.~E. and {Barrufet}, L. and {Begley}, R. and {Bondestam}, C. and {Cirasuolo}, M. and {Leung}, H. -H. and {Pollock}, C.~L. and {Stevenson}, S.},
        title = "{The JWST EXCELS survey: an extremely metal-poor galaxy at z = 8.271 hosting an unusual population of massive stars}",
      journal = {\mnras},
         year = 2025,
        month = jul,
       volume = {540},
       number = {3},
        pages = {2176-2194},
          doi = {10.1093/mnras/staf838},
archivePrefix = {arXiv},
       eprint = {2501.11099},
 primaryClass = {astro-ph.GA},
       adsurl = {https://ui.adsabs.harvard.edu/abs/2025MNRAS.540.2176C}
}

@ARTICLE{Willott2025,
       author = {{Willott}, Chris J. and {Asada}, Yoshihisa and {Iyer}, Kartheik G. and {Jude{\v{z}}}, Jon and {Rihtar{\v{s}}i{\v{c}}}, Gregor and {Martis}, Nicholas S. and {Sarrouh}, Ghassan T.~E. and {Desprez}, Guillaume and {Harshan}, Anishya and {Mowla}, Lamiya and {Noirot}, Ga{\"e}l and {Felicioni}, Giordano and {Brada{\v{c}}}, Maru{\v{s}}a and {Brammer}, Gabe and {Muzzin}, Adam and {Sawicki}, Marcin and {Antwi-Danso}, Jacqueline and {Markov}, Vladan and {Tripodi}, Roberta},
        title = "{In Search of the First Stars: An Ultra-compact and Very-low-metallicity Ly{\ensuremath{\alpha}} Emitter Deep within the Epoch of Reionization}",
      journal = {\apj},
         year = 2025,
        month = jul,
       volume = {988},
       number = {1},
          eid = {26},
        pages = {26},
          doi = {10.3847/1538-4357/addf49},
archivePrefix = {arXiv},
       eprint = {2502.07733},
 primaryClass = {astro-ph.GA},
       adsurl = {https://ui.adsabs.harvard.edu/abs/2025ApJ...988...26W}
}

@ARTICLE{Vanzella2023,
       author = {{Vanzella}, E. and {Loiacono}, F. and {Bergamini}, P. and {Me{\v{s}}tri{\'c}}, U. and {Castellano}, M. and {Rosati}, P. and {Meneghetti}, M. and {Grillo}, C. and {Calura}, F. and {Mignoli}, M. and {Brada{\v{c}}}, M. and {Adamo}, A. and {Rihtar{\v{s}}i{\v{c}}}, G. and {Dickinson}, M. and {Gronke}, M. and {Zanella}, A. and {Annibali}, F. and {Willott}, C. and {Messa}, M. and {Sani}, E. and {Acebron}, A. and {Bolamperti}, A. and {Comastri}, A. and {Gilli}, R. and {Caputi}, K.~I. and {Ricotti}, M. and {Gruppioni}, C. and {Ravindranath}, S. and {Mercurio}, A. and {Strait}, V. and {Martis}, N. and {Pascale}, R. and {Caminha}, G.~B. and {Annunziatella}, M. and {Nonino}, M.},
        title = "{An extremely metal-poor star complex in the reionization era: Approaching Population III stars with JWST}",
      journal = {\aap},
         year = 2023,
        month = oct,
       volume = {678},
          eid = {A173},
        pages = {A173},
          doi = {10.1051/0004-6361/202346981},
archivePrefix = {arXiv},
       eprint = {2305.14413},
 primaryClass = {astro-ph.GA},
       adsurl = {https://ui.adsabs.harvard.edu/abs/2023A&A...678A.173V}
}

@ARTICLE{Nakajima2025,
       author = {{Nakajima}, Kimihiko and {Ouchi}, Masami and {Harikane}, Yuichi and {Vanzella}, Eros and {Ono}, Yoshiaki and {Isobe}, Yuki and {Nishigaki}, Moka and {Tsujimoto}, Takuji and {Nakamura}, Fumitaka and {Xu}, Yi and {Umeda}, Hiroya and {Zhang}, Yechi},
        title = "{An Ultra-Faint, Chemically Primitive Galaxy Forming at the Epoch of Reionization}",
      journal = {arXiv e-prints},
         year = 2025,
        month = jun,
          eid = {arXiv:2506.11846},
        pages = {arXiv:2506.11846},
          doi = {10.48550/arXiv.2506.11846},
archivePrefix = {arXiv},
       eprint = {2506.11846},
 primaryClass = {astro-ph.GA},
       adsurl = {https://ui.adsabs.harvard.edu/abs/2025arXiv250611846N}
}

@ARTICLE{Fujimoto2025,
       author = {{Fujimoto}, Seiji and {Naidu}, Rohan P. and {Chisholm}, John and {Atek}, Hakim and {Endsley}, Ryan and {Kokorev}, Vasily and {Furtak}, Lukas J. and {Pan}, Richard and {Liu}, Boyuan and {Bromm}, Volker and {Venditti}, Alessandra and {Visbal}, Eli and {Sarmento}, Richard and {Weibel}, Andrea and {Oesch}, Pascal A. and {Brammer}, Gabriel and {Schaerer}, Daniel and {Adamo}, Angela and {Berg}, Danielle A. and {Bezanson}, Rachel and {Bouwens}, Rychard and {Chemerynska}, Iryna and {Claeyssens}, Ad{\'e}la{\"\i}de and {Dessauges-Zavadsky}, Miroslava and {Frebel}, Anna and {Korber}, Damien and {Labbe}, Ivo and {Marques-Chaves}, Rui and {Matthee}, Jorryt and {McQuinn}, Kristen B.~W. and {Mu{\~n}oz}, Julian B. and {Natarajan}, Priyamvada and {Saldana-Lopez}, Alberto and {Suess}, Katherine A. and {Volonteri}, Marta and {Zitrin}, Adi},
        title = "{GLIMPSE: An Ultrafaint ≃{}10$^{5}$ M$_{{\ensuremath{\odot}}}$ Pop III Galaxy Candidate and First Constraints on the Pop III UV Luminosity Function at z ≃ 6{\textendash}7}",
      journal = {\apj},
         year = 2025,
        month = aug,
       volume = {989},
       number = {1},
          eid = {46},
        pages = {46},
          doi = {10.3847/1538-4357/ade9a1},
archivePrefix = {arXiv},
       eprint = {2501.11678},
 primaryClass = {astro-ph.GA},
       adsurl = {https://ui.adsabs.harvard.edu/abs/2025ApJ...989...46F}
}

@ARTICLE{Maiolino2025,
       author = {{Maiolino}, Roberto and {Uebler}, Hannah and {D'Eugenio}, Francesco and {Scholtz}, Jan and {Juodzbalis}, Ignas and {Ji}, Xihan and {Perna}, Michele and {Bromm}, Volker and {Dayal}, Pratika and {Koudmani}, Sophie and {Liu}, Boyuan and {Schneider}, Raffaella and {Sijacki}, Debora and {Valiante}, Rosa and {Trinca}, Alessandro and {Zhang}, Saiyang and {Volonteri}, Marta and {Inayoshi}, Kohei and {Carniani}, Stefano and {Nakajima}, Kimihiko and {Isobe}, Yuki and {Witstok}, Joris and {Jones}, Gareth C. and {Tacchella}, Sandro and {Arribas}, Santiago and {Bunker}, Andrew and {Cataldi}, Elisa and {Charlot}, Stephane and {Cresci}, Giovanni and {Curti}, Mirko and {Fabian}, Andrew C. and {Katz}, Harley and {Kumari}, Nimisha and {Laporte}, Nicolas and {Mazzolari}, Giovanni and {Robertson}, Brant and {Sun}, Fengwu and {Rodriguez Del Pino}, Bruno and {Venturi}, Giacomo},
        title = "{A black hole in a near-pristine galaxy 700 million years after the Big Bang}",
      journal = {arXiv e-prints},
         year = 2025,
        month = may,
          eid = {arXiv:2505.22567},
        pages = {arXiv:2505.22567},
          doi = {10.48550/arXiv.2505.22567},
archivePrefix = {arXiv},
       eprint = {2505.22567},
 primaryClass = {astro-ph.GA},
       adsurl = {https://ui.adsabs.harvard.edu/abs/2025arXiv250522567M}
}

@ARTICLE{Morishita2025,
       author = {{Morishita}, Takahiro and {Liu}, Zhaoran and {Stiavelli}, Massimo and {Treu}, Tommaso and {Bergamini}, Pietro and {Zhang}, Yechi},
        title = "{Pristine Massive Star Formation Caught at the Break of Cosmic Dawn}",
      journal = {arXiv e-prints},
         year = 2025,
        month = jul,
          eid = {arXiv:2507.10521},
        pages = {arXiv:2507.10521},
          doi = {10.48550/arXiv.2507.10521},
archivePrefix = {arXiv},
       eprint = {2507.10521},
 primaryClass = {astro-ph.CO},
       adsurl = {https://ui.adsabs.harvard.edu/abs/2025arXiv250710521M}
}

@ARTICLE{Cai2025,
       author = {{Cai}, Sijia and {Li}, Mingyu and {Cai}, Zheng and {Wu}, Yunjing and {Yu}, Fujiang and {Dickinson}, Mark and {Sun}, Fengwu and {Fan}, Xiaohui and {Wang}, Ben and {Cullen}, Fergus and {Bian}, Fuyan and {Lin}, Xiaojing and {Zou}, Jiaqi},
        title = "{A Metal-Free Galaxy at $z = 3.19$? Evidence of Late Population III Star Formation at Cosmic Noon}",
      journal = {arXiv e-prints},
         year = 2025,
        month = jul,
          eid = {arXiv:2507.17820},
        pages = {arXiv:2507.17820},
          doi = {10.48550/arXiv.2507.17820},
archivePrefix = {arXiv},
       eprint = {2507.17820},
 primaryClass = {astro-ph.GA},
       adsurl = {https://ui.adsabs.harvard.edu/abs/2025arXiv250717820C}
}

@ARTICLE{Vanzella2025,
       author = {{Vanzella}, E. and {Messa}, M. and {Zanella}, A. and {Bolamperti}, A. and {Castellano}, M. and {Loiacono}, F. and {Bergamini}, P. and {Roberts-Borsani}, G. and {Adamo}, A. and {Fontana}, A. and {Treu}, T. and {Calura}, F. and {Grillo}, C. and {Lombardi}, M. and {Rosati}, P. and {Gilli}, R. and {Meneghetti}, M.},
        title = "{A Pristine Star-Forming Complex at z=4.19}",
      journal = {arXiv e-prints},
         year = 2025,
        month = sep,
          eid = {arXiv:2509.07073},
        pages = {arXiv:2509.07073},
          doi = {10.48550/arXiv.2509.07073},
archivePrefix = {arXiv},
       eprint = {2509.07073},
 primaryClass = {astro-ph.GA},
       adsurl = {https://ui.adsabs.harvard.edu/abs/2025arXiv250907073V}
}

@ARTICLE{Chemerynska2024,
       author = {{Chemerynska}, Iryna and {Atek}, Hakim and {Dayal}, Pratika and {Furtak}, Lukas J. and {Feldmann}, Robert and {Greene}, Jenny E. and {Maseda}, Michael V. and {Nanayakkara}, Themiya and {Oesch}, Pascal A. and {Fujimoto}, Seiji and {Labb{\'e}}, Ivo and {Bezanson}, Rachel and {Brammer}, Gabriel and {Cutler}, Sam E. and {Leja}, Joel and {Pan}, Richard and {Price}, Sedona H. and {Wang}, Bingjie and {Weaver}, John R. and {Whitaker}, Katherine E.},
        title = "{The Extreme Low-mass End of the Mass{\textendash}Metallicity Relation at z {\ensuremath{\sim}} 7}",
      journal = {\apjl},
         year = 2024,
        month = nov,
       volume = {976},
       number = {1},
          eid = {L15},
        pages = {L15},
          doi = {10.3847/2041-8213/ad8dc9},
archivePrefix = {arXiv},
       eprint = {2407.17110},
 primaryClass = {astro-ph.GA},
       adsurl = {https://ui.adsabs.harvard.edu/abs/2024ApJ...976L..15C}
}

@ARTICLE{Asada2026,
       author = {{Asada}, Yoshihisa and {Fujimoto}, Seiji and {Chisholm}, John and {Naidu}, Rohan P. and {Atek}, Hakim and {Brammer}, Gabriel and {Furtak}, Lukas J. and {Kokorev}, Vasily and {Pan}, Richard and {Basu}, Arghyadeep and {Bromm}, Volker and {Dessauges-Zavadsky}, Miroslava and {Hsiao}, Tiger Yu-Yang and {Jecmen}, Michelle and {Korber}, Damien and {Liu}, Boyuan and {McKinney}, Jed and {McQuinn}, Kristen B.~W. and {Schaerer}, Daniel},
        title = "{GLIMPSE-DDT spectroscopic properties of faint-end galaxies at $z\sim6$: Towards first metal enrichment, dust production, and ionizing photon production}",
      journal = {arXiv e-prints},
         year = 2026,
        month = jan,
          eid = {arXiv:2601.20045},
        pages = {arXiv:2601.20045},
          doi = {10.48550/arXiv.2601.20045},
archivePrefix = {arXiv},
       eprint = {2601.20045},
 primaryClass = {astro-ph.GA},
       adsurl = {https://ui.adsabs.harvard.edu/abs/2026arXiv260120045A}
}

@ARTICLE{Fujimoto2025b,
       author = {{Fujimoto}, Seiji and {Asada}, Yoshihisa and {Naidu}, Rohan P. and {Chisholm}, John and {Atek}, Hakim and {Brammer}, Gabriel and {Berg}, Danielle A. and {Schaerer}, Daniel and {Kokorev}, Vasily and {Furtak}, Lukas J. and {Richard}, Johan and {Venditti}, Alessandra and {Bromm}, Volker and {Adamo}, Angela and {Claeyssens}, Adelaide and {Dessauges-Zavadsky}, Miroslava and {Fei}, Qinyue and {Hsiao}, Tiger Yu-Yang and {Korber}, Damien and {Munoz}, Julian B. and {Pan}, Richard and {Saldana-Lopez}, Alberto},
        title = "{GLIMPSE-D: An Exotic Balmer-Jump Object at z=6.20? Revisiting Photometric Selection and the Cosmic Abundance of Pop III Galaxies}",
      journal = {arXiv e-prints},
         year = 2025,
        month = dec,
          eid = {arXiv:2512.11790},
        pages = {arXiv:2512.11790},
          doi = {10.48550/arXiv.2512.11790},
archivePrefix = {arXiv},
       eprint = {2512.11790},
 primaryClass = {astro-ph.GA},
       adsurl = {https://ui.adsabs.harvard.edu/abs/2025arXiv251211790F}
}

@ARTICLE{Hsiao2024,
       author = {{Hsiao}, Tiger Yu-Yang and {Abdurro'uf} and {Coe}, Dan and {Larson}, Rebecca L. and {Jung}, Intae and {Mingozzi}, Matilde and {Dayal}, Pratika and {Kumari}, Nimisha and {Kokorev}, Vasily and {Vikaeus}, Anton and {Brammer}, Gabriel and {Furtak}, Lukas J. and {Adamo}, Angela and {Andrade-Santos}, Felipe and {Antwi-Danso}, Jacqueline and {Brada{\v{c}}}, Maru{\v{s}}a and {Bradley}, Larry D. and {Broadhurst}, Tom and {Carnall}, Adam C. and {Conselice}, Christopher J. and {Diego}, Jose M. and {Donahue}, Megan and {Eldridge}, Jan J. and {Fujimoto}, Seiji and {Henry}, Alaina and {Hernandez}, Svea and {Hutchison}, Taylor A. and {James}, Bethan L. and {Norman}, Colin and {Park}, Hyunbae and {Pirzkal}, Norbert and {Postman}, Marc and {Ricotti}, Massimo and {Rigby}, Jane R. and {Vanzella}, Eros and {Welch}, Brian and {Wilkins}, Stephen M. and {Windhorst}, Rogier A. and {Xu}, Xinfeng and {Zackrisson}, Erik and {Zitrin}, Adi},
        title = "{JWST NIRSpec Spectroscopy of the Triply Lensed z = 10.17 Galaxy MACS0647─JD}",
      journal = {\apj},
         year = 2024,
        month = sep,
       volume = {973},
       number = {1},
          eid = {8},
        pages = {8},
          doi = {10.3847/1538-4357/ad5da8},
archivePrefix = {arXiv},
       eprint = {2305.03042},
 primaryClass = {astro-ph.GA},
       adsurl = {https://ui.adsabs.harvard.edu/abs/2024ApJ...973....8H}
}

@ARTICLE{Bunker2023,
       author = {{Bunker}, Andrew J. and {Saxena}, Aayush and {Cameron}, Alex J. and {Willott}, Chris J. and {Curtis-Lake}, Emma and {Jakobsen}, Peter and {Carniani}, Stefano and {Smit}, Renske and {Maiolino}, Roberto and {Witstok}, Joris and {Curti}, Mirko and {D'Eugenio}, Francesco and {Jones}, Gareth C. and {Ferruit}, Pierre and {Arribas}, Santiago and {Charlot}, Stephane and {Chevallard}, Jacopo and {Giardino}, Giovanna and {de Graaff}, Anna and {Looser}, Tobias J. and {L{\"u}tzgendorf}, Nora and {Maseda}, Michael V. and {Rawle}, Tim and {Rix}, Hans-Walter and {Del Pino}, Bruno Rodr{\'\i}guez and {Alberts}, Stacey and {Egami}, Eiichi and {Eisenstein}, Daniel J. and {Endsley}, Ryan and {Hainline}, Kevin and {Hausen}, Ryan and {Johnson}, Benjamin D. and {Rieke}, George and {Rieke}, Marcia and {Robertson}, Brant E. and {Shivaei}, Irene and {Stark}, Daniel P. and {Sun}, Fengwu and {Tacchella}, Sandro and {Tang}, Mengtao and {Williams}, Christina C. and {Willmer}, Christopher N.~A. and {Baker}, William M. and {Baum}, Stefi and {Bhatawdekar}, Rachana and {Bowler}, Rebecca and {Boyett}, Kristan and {Chen}, Zuyi and {Circosta}, Chiara and {Helton}, Jakob M. and {Ji}, Zhiyuan and {Kumari}, Nimisha and {Lyu}, Jianwei and {Nelson}, Erica and {Parlanti}, Eleonora and {Perna}, Michele and {Sandles}, Lester and {Scholtz}, Jan and {Suess}, Katherine A. and {Topping}, Michael W. and {{\"U}bler}, Hannah and {Wallace}, Imaan E.~B. and {Whitler}, Lily},
        title = "{JADES NIRSpec Spectroscopy of GN-z11: Lyman-{\ensuremath{\alpha}} emission and possible enhanced nitrogen abundance in a z = 10.60 luminous galaxy}",
      journal = {\aap},
         year = 2023,
        month = sep,
       volume = {677},
          eid = {A88},
        pages = {A88},
          doi = {10.1051/0004-6361/202346159},
archivePrefix = {arXiv},
       eprint = {2302.07256},
 primaryClass = {astro-ph.GA},
       adsurl = {https://ui.adsabs.harvard.edu/abs/2023A&A...677A..88B}
}

@ARTICLE{Naidu2026,
       author = {{Naidu}, Rohan P. and {Oesch}, Pascal A. and {Brammer}, Gabriel and {Weibel}, Andrea and {Li}, Yijia and {Matthee}, Jorryt and {Chisolm}, John and {Pollock}, Clara L. and {Heintz}, Kasper E. and {Johnson}, Benjamin D. and {Shen}, Xuejian and {Hviding}, Raphael E. and {Leja}, Joel and {Tacchella}, Sandro and {Ganguly}, Arpita and {Witten}, Callum and {Atek}, Hakim and {Belli}, Siro and {Bose}, Sownak and {Bouwens}, Rychard and {Dayal}, Pratika and {Decarli}, Roberto and {de Graaff}, Anna and {Fudamoto}, Yoshinobu and {Giovinazzo}, Emma and {Greene}, Jenny E. and {Illingworth}, Garth and {Inoue}, Akio K. and {Kane}, Sarah G. and {Labbe}, Ivo and {Leonova}, Ecaterina and {Marques-Chaves}, Rui and {Meyer}, Roman A. and {Nelson}, Erica J. and {Roberts-Borsani}, Guido and {Schaerer}, Daniel and {Simcoe}, Robert A. and {Stefanon}, Mauro and {Sugahara}, Yuma and {Toft}, Sune and {van der Wel}, Arjen and {van Dokkum}, Pieter and {Walter}, Fabian and {Watson}, Darrach and {Weaver}, John R. and {Whitaker}, Katherine E.},
        title = "{A Cosmic Miracle: A Remarkably Luminous Galaxy at zspec = 14.44 Confirmed with JWST}",
      journal = {The Open Journal of Astrophysics},
         year = 2026,
        month = jan,
       volume = {9},
        pages = {56033},
          doi = {10.33232/001c.156033},
archivePrefix = {arXiv},
       eprint = {2505.11263},
 primaryClass = {astro-ph.GA},
       adsurl = {https://ui.adsabs.harvard.edu/abs/2026OJAp....956033N}
}

@ARTICLE{CurtisLake2023,
       author = {{Curtis-Lake}, Emma and {Carniani}, Stefano and {Cameron}, Alex and {Charlot}, Stephane and {Jakobsen}, Peter and {Maiolino}, Roberto and {Bunker}, Andrew and {Witstok}, Joris and {Smit}, Renske and {Chevallard}, Jacopo and {Willott}, Chris and {Ferruit}, Pierre and {Arribas}, Santiago and {Bonaventura}, Nina and {Curti}, Mirko and {D'Eugenio}, Francesco and {Franx}, Marijn and {Giardino}, Giovanna and {Looser}, Tobias J. and {L{\"u}tzgendorf}, Nora and {Maseda}, Michael V. and {Rawle}, Tim and {Rix}, Hans-Walter and {Rodr{\'\i}guez del Pino}, Bruno and {{\"U}bler}, Hannah and {Sirianni}, Marco and {Dressler}, Alan and {Egami}, Eiichi and {Eisenstein}, Daniel J. and {Endsley}, Ryan and {Hainline}, Kevin and {Hausen}, Ryan and {Johnson}, Benjamin D. and {Rieke}, Marcia and {Robertson}, Brant and {Shivaei}, Irene and {Stark}, Daniel P. and {Tacchella}, Sandro and {Williams}, Christina C. and {Willmer}, Christopher N.~A. and {Bhatawdekar}, Rachana and {Bowler}, Rebecca and {Boyett}, Kristan and {Chen}, Zuyi and {de Graaff}, Anna and {Helton}, Jakob M. and {Hviding}, Raphael E. and {Jones}, Gareth C. and {Kumari}, Nimisha and {Lyu}, Jianwei and {Nelson}, Erica and {Perna}, Michele and {Sandles}, Lester and {Saxena}, Aayush and {Suess}, Katherine A. and {Sun}, Fengwu and {Topping}, Michael W. and {Wallace}, Imaan E.~B. and {Whitler}, Lily},
        title = "{Spectroscopic confirmation of four metal-poor galaxies at z = 10.3-13.2}",
      journal = {Nature Astronomy},
         year = 2023,
        month = may,
       volume = {7},
        pages = {622-632},
          doi = {10.1038/s41550-023-01918-w},
archivePrefix = {arXiv},
       eprint = {2212.04568},
 primaryClass = {astro-ph.GA},
       adsurl = {https://ui.adsabs.harvard.edu/abs/2023NatAs...7..622C}
}

@ARTICLE{Carniani2024,
       author = {{Carniani}, Stefano and {Hainline}, Kevin and {D'Eugenio}, Francesco and {Eisenstein}, Daniel J. and {Jakobsen}, Peter and {Witstok}, Joris and {Johnson}, Benjamin D. and {Chevallard}, Jacopo and {Maiolino}, Roberto and {Helton}, Jakob M. and {Willott}, Chris and {Robertson}, Brant and {Alberts}, Stacey and {Arribas}, Santiago and {Baker}, William M. and {Bhatawdekar}, Rachana and {Boyett}, Kristan and {Bunker}, Andrew J. and {Cameron}, Alex J. and {Cargile}, Phillip A. and {Charlot}, St{\'e}phane and {Curti}, Mirko and {Curtis-Lake}, Emma and {Egami}, Eiichi and {Giardino}, Giovanna and {Isaak}, Kate and {Ji}, Zhiyuan and {Jones}, Gareth C. and {Kumari}, Nimisha and {Maseda}, Michael V. and {Parlanti}, Eleonora and {P{\'e}rez-Gonz{\'a}lez}, Pablo G. and {Rawle}, Tim and {Rieke}, George and {Rieke}, Marcia and {Del Pino}, Bruno Rodr{\'\i}guez and {Saxena}, Aayush and {Scholtz}, Jan and {Smit}, Renske and {Sun}, Fengwu and {Tacchella}, Sandro and {{\"U}bler}, Hannah and {Venturi}, Giacomo and {Williams}, Christina C. and {Willmer}, Christopher N.~A.},
        title = "{Spectroscopic confirmation of two luminous galaxies at a redshift of 14}",
      journal = {\nat},
         year = 2024,
        month = sep,
       volume = {633},
       number = {8029},
        pages = {318-322},
          doi = {10.1038/s41586-024-07860-9},
archivePrefix = {arXiv},
       eprint = {2405.18485},
 primaryClass = {astro-ph.GA},
       adsurl = {https://ui.adsabs.harvard.edu/abs/2024Natur.633..318C}
}

@ARTICLE{Zavala2025,
       author = {{Zavala}, Jorge A. and {Castellano}, Marco and {Akins}, Hollis B. and {Bakx}, Tom J.~L.~C. and {Burgarella}, Denis and {Casey}, Caitlin M. and {Ch{\'a}vez Ortiz}, {\~A}. `scar A. and {Dickinson}, Mark and {Finkelstein}, Steven L. and {Mitsuhashi}, Ikki and {Nakajima}, Kimihiko and {P{\'e}rez-Gonz{\'a}lez}, Pablo G. and {Arrabal Haro}, Pablo and {Bergamini}, Pietro and {Buat}, Veronique and {Backhaus}, Bren and {Calabr{\`o}}, Antonello and {Cleri}, Nikko J. and {Fern{\'a}ndez-Arenas}, David and {Fontana}, Adriano and {Franco}, Maximilien and {Grillo}, Claudio and {Giavalisco}, Mauro and {Grogin}, Norman A. and {Hathi}, Nimish and {Hirschmann}, Michaela and {Ikeda}, Ryota and {Jung}, Intae and {Kartaltepe}, Jeyhan S. and {Koekemoer}, Anton M. and {Larson}, Rebeca L. and {McKinney}, Jed and {Papovich}, Casey and {Rosati}, Piero and {Saito}, Toshiki and {Santini}, Paola and {Terlevich}, Roberto and {Terlevich}, Elena and {Treu}, Tommaso and {Yung}, L.~Y. Aaron},
        title = "{A luminous and young galaxy at z = 12.33 revealed by a JWST/MIRI detection of H{\ensuremath{\alpha}} and [O III]}",
      journal = {Nature Astronomy},
         year = 2025,
        month = jan,
       volume = {9},
        pages = {155-164},
          doi = {10.1038/s41550-024-02397-3},
archivePrefix = {arXiv},
       eprint = {2403.10491},
 primaryClass = {astro-ph.GA},
       adsurl = {https://ui.adsabs.harvard.edu/abs/2025NatAs...9..155Z}
}

@ARTICLE{Wang2023,
       author = {{Wang}, Bingjie and {Fujimoto}, Seiji and {Labb{\'e}}, Ivo and {Furtak}, Lukas J. and {Miller}, Tim B. and {Setton}, David J. and {Zitrin}, Adi and {Atek}, Hakim and {Bezanson}, Rachel and {Brammer}, Gabriel and {Leja}, Joel and {Oesch}, Pascal A. and {Price}, Sedona H. and {Chemerynska}, Iryna and {Cutler}, Sam E. and {Dayal}, Pratika and {van Dokkum}, Pieter and {Goulding}, Andy D. and {Greene}, Jenny E. and {Fudamoto}, Y. and {Khullar}, Gourav and {Kokorev}, Vasily and {Marchesini}, Danilo and {Pan}, Richard and {Weaver}, John R. and {Whitaker}, Katherine E. and {Williams}, Christina C.},
        title = "{UNCOVER: Illuminating the Early Universe-JWST/NIRSpec Confirmation of z > 12 Galaxies}",
      journal = {\apjl},
         year = 2023,
        month = nov,
       volume = {957},
       number = {2},
          eid = {L34},
        pages = {L34},
          doi = {10.3847/2041-8213/acfe07},
archivePrefix = {arXiv},
       eprint = {2308.03745},
 primaryClass = {astro-ph.GA},
       adsurl = {https://ui.adsabs.harvard.edu/abs/2023ApJ...957L..34W}
}

@ARTICLE{Peimbert1967,
       author = {{Peimbert}, Manuel},
        title = "{Temperature Determinations of H II Regions}",
      journal = {\apj},
         year = 1967,
        month = dec,
       volume = {150},
        pages = {825},
          doi = {10.1086/149385},
       adsurl = {https://ui.adsabs.harvard.edu/abs/1967ApJ...150..825P}
}

@ARTICLE{Izotov2006,
       author = {{Izotov}, Y.~I. and {Stasi{\'n}ska}, G. and {Meynet}, G. and {Guseva}, N.~G. and {Thuan}, T.~X.},
        title = "{The chemical composition of metal-poor emission-line galaxies in the Data Release 3 of the Sloan Digital Sky Survey}",
      journal = {\aap},
         year = 2006,
        month = mar,
       volume = {448},
       number = {3},
        pages = {955-970},
          doi = {10.1051/0004-6361:20053763},
archivePrefix = {arXiv},
       eprint = {astro-ph/0511644},
 primaryClass = {astro-ph},
       adsurl = {https://ui.adsabs.harvard.edu/abs/2006A&A...448..955I}
}

@ARTICLE{Peimbert1969,
       author = {{Peimbert}, M. and {Costero}, R.},
        title = "{Chemical Abundances in Galactic HII Regions}",
      journal = {Boletin de los Observatorios Tonantzintla y Tacubaya},
         year = 1969,
        month = may,
       volume = {5},
        pages = {3-22},
       adsurl = {https://ui.adsabs.harvard.edu/abs/1969BOTT....5....3P}
}

@BOOK{Aller1984,
       author = {{Aller}, L.~H.},
        title = "{Physics of thermal gaseous nebulae}",
         year = 1984,
          doi = {10.1007/978-94-010-9639-3},
       adsurl = {https://ui.adsabs.harvard.edu/abs/1984ASSL..112.....A}
}

@BOOK{Osterbrock1989,
       author = {{Osterbrock}, Donald E.},
        title = "{Astrophysics of gaseous nebulae and active galactic nuclei}",
         year = 1989,
       adsurl = {https://ui.adsabs.harvard.edu/abs/1989agna.book.....O}
}

@ARTICLE{Martinez2025,
       author = {{Martinez}, Zorayda and {Berg}, Danielle A. and {James}, Bethan L. and {Arellano-C{\'o}rdova}, Karla Z. and {Stark}, Daniel P. and {Senchyna}, Peter and {Skillman}, Evan D. and {Rogers}, Noah S.~J. and {Chisholm}, John},
        title = "{Under Pressure: Decoding the Effect of High Densities on Derived Nebular Properties}",
      journal = {\apj},
         year = 2025,
        month = dec,
       volume = {995},
       number = {2},
          eid = {204},
        pages = {204},
          doi = {10.3847/1538-4357/ae17c6},
archivePrefix = {arXiv},
       eprint = {2510.21960},
 primaryClass = {astro-ph.GA},
       adsurl = {https://ui.adsabs.harvard.edu/abs/2025ApJ...995..204M}
}

@ARTICLE{Arellano2026,
       author = {{Arellano-C{\'o}rdova}, K.~Z. and {M{\'e}ndez-Delgado}, J.~E. and {Flury}, S.~R. and {Esteban}, C. and {Kreckel}, K. and {Garc{\'\i}a-Rojas}, J. and {Cullen}, F. and {Carigi}, L. and {Morisset}, C. and {Rosales-Ortega}, F.~F. and {Peimbert}, A. and {Stanton}, T.~M. and {Scholte}, D.},
        title = "{A self-consistent direct method for chemical abundances in high-z galaxies with JWST}",
      journal = {\mnras},
         year = 2026,
        month = apr,
       volume = {547},
       number = {4},
          eid = {stag380},
        pages = {stag380},
          doi = {10.1093/mnras/stag380},
archivePrefix = {arXiv},
       eprint = {2602.13007},
 primaryClass = {astro-ph.GA},
       adsurl = {https://ui.adsabs.harvard.edu/abs/2026MNRAS.547ag380A}
}

@ARTICLE{Isobe2023,
       author = {{Isobe}, Yuki and {Ouchi}, Masami and {Nakajima}, Kimihiko and {Harikane}, Yuichi and {Ono}, Yoshiaki and {Xu}, Yi and {Zhang}, Yechi and {Umeda}, Hiroya},
        title = "{Redshift Evolution of Electron Density in the Interstellar Medium at z   0-9 Uncovered with JWST/NIRSpec Spectra and Line-spread Function Determinations}",
      journal = {\apj},
         year = 2023,
        month = oct,
       volume = {956},
       number = {2},
          eid = {139},
        pages = {139},
          doi = {10.3847/1538-4357/acf376},
archivePrefix = {arXiv},
       eprint = {2301.06811},
 primaryClass = {astro-ph.GA},
       adsurl = {https://ui.adsabs.harvard.edu/abs/2023ApJ...956..139I}
}

@ARTICLE{Hayes2025,
       author = {{Hayes}, Matthew J. and {Saldana-Lopez}, Alberto and {Citro}, Annalisa and {James}, Bethan L. and {Mingozzi}, Matilde and {Scarlata}, Claudia and {Martinez}, Zorayda and {Berg}, Danielle A.},
        title = "{On the Average Ultraviolet Emission-line Spectra of High-redshift Galaxies: Hot and Cold, Carbon-poor, Nitrogen Modest, and Oozing Ionizing Photons}",
      journal = {\apj},
         year = 2025,
        month = mar,
       volume = {982},
       number = {1},
          eid = {14},
        pages = {14},
          doi = {10.3847/1538-4357/adaea1},
archivePrefix = {arXiv},
       eprint = {2411.09262},
 primaryClass = {astro-ph.GA},
       adsurl = {https://ui.adsabs.harvard.edu/abs/2025ApJ...982...14H}
}

@ARTICLE{Abdurrouf2024,
       author = {{Abdurro'uf} and {Larson}, Rebecca L. and {Coe}, Dan and {Hsiao}, Tiger Yu-Yang and {{\'A}lvarez-M{\'a}rquez}, Javier and {G{\'o}mez}, Alejandro Crespo and {Adamo}, Angela and {Bhatawdekar}, Rachana and {Bik}, Arjan and {Bradley}, Larry D. and {Conselice}, Christopher J. and {Dayal}, Pratika and {Diego}, Jose M. and {Fujimoto}, Seiji and {Furtak}, Lukas J. and {Hutchison}, Taylor A. and {Jung}, Intae and {Killi}, Meghana and {Kokorev}, Vasily and {Mingozzi}, Matilde and {Norman}, Colin and {Resseguier}, Tom and {Ricotti}, Massimo and {Rigby}, Jane R. and {Vanzella}, Eros and {Welch}, Brian and {Windhorst}, Rogier A. and {Xu}, Xinfeng and {Zitrin}, Adi},
        title = "{JWST NIRSpec High-resolution Spectroscopy of MACS0647─JD at z = 10.167: Resolved [O II] Doublet and Electron Density in an Early Galaxy}",
      journal = {\apj},
         year = 2024,
        month = sep,
       volume = {973},
       number = {1},
          eid = {47},
        pages = {47},
          doi = {10.3847/1538-4357/ad6001},
archivePrefix = {arXiv},
       eprint = {2404.16201},
 primaryClass = {astro-ph.GA},
       adsurl = {https://ui.adsabs.harvard.edu/abs/2024ApJ...973...47A}
}

@ARTICLE{Topping2025,
       author = {{Topping}, Michael W. and {Sanders}, Ryan L. and {Shapley}, Alice E. and {Pahl}, Anthony J. and {Reddy}, Naveen A. and {Stark}, Daniel P. and {Berg}, Danielle A. and {Clarke}, Leonardo and {Cullen}, Fergus and {Dunlop}, James S. and {Ellis}, Richard S. and {Schreiber}, N.~M. F{\"o}rster and {Illingworth}, Garth D. and {Jones}, Tucker and {Narayanan}, Desika and {Pettini}, Max and {Schaerer}, Daniel},
        title = "{The AURORA survey: the evolution of multiphase electron densities at high redshift}",
      journal = {\mnras},
         year = 2025,
        month = aug,
       volume = {541},
       number = {2},
        pages = {1707-1721},
          doi = {10.1093/mnras/staf903},
archivePrefix = {arXiv},
       eprint = {2502.08712},
 primaryClass = {astro-ph.GA},
       adsurl = {https://ui.adsabs.harvard.edu/abs/2025MNRAS.541.1707T}
}

@ARTICLE{Hsiao2026,
       author = {{Hsiao}, Tiger Yu-Yang and {Chisholm}, John and {Berg}, Danielle A. and {Finkelstein}, Steven L. and {Kokorev}, Vasily and {Atek}, Hakim and {Naidu}, Rohan P. and {Fujimoto}, Seiji and {Furtak}, Lukas J. and {Adamo}, Angela and {Aravindan}, Archana and {Asada}, Yoshihisa and {Basu}, Arghyadeep and {Blaizot}, Jeremy and {Choustikov}, Nicholas and {Dessauges-Zavadsky}, Miroslava and {Fei}, Qinyue and {Katz}, Harley and {Korber}, Damien and {McQuinn}, Kristen. B.~W. and {Mun}, Marcie and {Munoz}, Julian B. and {Natarajan}, Priyamvada and {Stephenson}, Mabel G. and {Schaerer}, Daniel},
        title = "{A Glimpse of the Low-Mass End of the Direct Mass-Metallicity Relation at $z\sim6-8$}",
      journal = {arXiv e-prints},
         year = 2026,
        month = may,
          eid = {arXiv:2605.06770},
        pages = {arXiv:2605.06770},
archivePrefix = {arXiv},
       eprint = {2605.06770},
 primaryClass = {astro-ph.GA},
       adsurl = {https://ui.adsabs.harvard.edu/abs/2026arXiv260506770H}
}

@ARTICLE{Berg2021,
       author = {{Berg}, Danielle A. and {Chisholm}, John and {Erb}, Dawn K. and {Skillman}, Evan D. and {Pogge}, Richard W. and {Olivier}, Grace M.},
        title = "{Characterizing Extreme Emission-line Galaxies. I. A Four-zone Ionization Model for Very High-ionization Emission}",
      journal = {\apj},
         year = 2021,
        month = dec,
       volume = {922},
       number = {2},
          eid = {170},
        pages = {170},
          doi = {10.3847/1538-4357/ac141b},
archivePrefix = {arXiv},
       eprint = {2105.12765},
 primaryClass = {astro-ph.GA},
       adsurl = {https://ui.adsabs.harvard.edu/abs/2021ApJ...922..170B}
}

@ARTICLE{Harikane2025,
       author = {{Harikane}, Yuichi and {Sanders}, Ryan L. and {Ellis}, Richard and {Jones}, Tucker and {Ouchi}, Masami and {Laporte}, Nicolas and {Roberts-Borsani}, Guido and {Katz}, Harley and {Nakajima}, Kimihiko and {Ono}, Yoshiaki and {Gupta}, Mansi},
        title = "{JWST and ALMA Joint Analysis with [O II] {\ensuremath{\lambda}}{\ensuremath{\lambda}}3726, 3729, [O III] {\ensuremath{\lambda}}4363, [O III] 88 {\ensuremath{\mu}}m, and [O III] 52 {\ensuremath{\mu}}m: Multizone Evolution of Electron Densities at z {\ensuremath{\sim}} 0─14 and its Impact on Metallicity Measurements}",
      journal = {\apj},
         year = 2025,
        month = nov,
       volume = {993},
       number = {2},
          eid = {204},
        pages = {204},
          doi = {10.3847/1538-4357/ae0e53},
archivePrefix = {arXiv},
       eprint = {2505.09186},
 primaryClass = {astro-ph.GA},
       adsurl = {https://ui.adsabs.harvard.edu/abs/2025ApJ...993..204H}
}

@ARTICLE{Mingozzi2022,
       author = {{Mingozzi}, Matilde and {James}, Bethan L. and {Arellano-C{\'o}rdova}, Karla Z. and {Berg}, Danielle A. and {Senchyna}, Peter and {Chisholm}, John and {Brinchmann}, Jarle and {Aloisi}, Alessandra and {Amor{\'\i}n}, Ricardo O. and {Charlot}, St{\'e}phane and {Feltre}, Anna and {Hayes}, Matthew and {Heckman}, Timothy and {Henry}, Alaina and {Hernandez}, Svea and {Kumari}, Nimisha and {Leitherer}, Claus and {Llerena}, Mario and {Martin}, Crystal L. and {Nanayakkara}, Themiya and {Ravindranath}, Swara and {Skillman}, Evan D. and {Sugahara}, Yuma and {Wofford}, Aida and {Xu}, Xinfeng},
        title = "{CLASSY IV. Exploring UV Diagnostics of the Interstellar Medium in Local High-z Analogs at the Dawn of the JWST Era}",
      journal = {\apj},
         year = 2022,
        month = nov,
       volume = {939},
       number = {2},
          eid = {110},
        pages = {110},
          doi = {10.3847/1538-4357/ac952c},
archivePrefix = {arXiv},
       eprint = {2209.09047},
 primaryClass = {astro-ph.GA},
       adsurl = {https://ui.adsabs.harvard.edu/abs/2022ApJ...939..110M}
}

@ARTICLE{Gardner2006,
       author = {{Gardner}, Jonathan P. and {Mather}, John C. and {Clampin}, Mark and {Doyon}, Rene and {Greenhouse}, Matthew A. and {Hammel}, Heidi B. and {Hutchings}, John B. and {Jakobsen}, Peter and {Lilly}, Simon J. and {Long}, Knox S. and {Lunine}, Jonathan I. and {McCaughrean}, Mark J. and {Mountain}, Matt and {Nella}, John and {Rieke}, George H. and {Rieke}, Marcia J. and {Rix}, Hans-Walter and {Smith}, Eric P. and {Sonneborn}, George and {Stiavelli}, Massimo and {Stockman}, H.~S. and {Windhorst}, Rogier A. and {Wright}, Gillian S.},
        title = "{The James Webb Space Telescope}",
      journal = {\ssr},
         year = 2006,
        month = apr,
       volume = {123},
       number = {4},
        pages = {485-606},
          doi = {10.1007/s11214-006-8315-7},
archivePrefix = {arXiv},
       eprint = {astro-ph/0606175},
 primaryClass = {astro-ph},
       adsurl = {https://ui.adsabs.harvard.edu/abs/2006SSRv..123..485G}
}

@ARTICLE{Gardner2023,
       author = {{Gardner}, Jonathan P. and {Mather}, John C. and {Abbott}, Randy and {Abell}, James S. and {Abernathy}, Mark and {Abney}, Faith E. and {Abraham}, John G. and {Abraham}, Roberto and {Abul-Huda}, Yasin M. and {Acton}, Scott and {Adams}, Cynthia K. and {Adams}, Evan and {Adler}, David S. and {Adriaensen}, Maarten and {Aguilar}, Jonathan Albert and {Ahmed}, Mansoor and {Ahmed}, Nasif S. and {Ahmed}, Tanjira and {Albat}, R{\"u}deger and {Albert}, Lo{\"\i}c and {Alberts}, Stacey and {Aldridge}, David and {Allen}, Mary Marsha and {Allen}, Shaune S. and {Altenburg}, Martin and {Altunc}, Serhat and {Alvarez}, Jose Lorenzo and {{\'A}lvarez-M{\'a}rquez}, Javier and {Alves de Oliveira}, Catarina and {Ambrose}, Leslie L. and {Anandakrishnan}, Satya M. and {Andersen}, Gregory C. and {Anderson}, Harry James and {Anderson}, Jay and {Anderson}, Kristen and {Anderson}, Sara M. and {Aprea}, Julio and {Archer}, Benita J. and {Arenberg}, Jonathan W. and {Argyriou}, Ioannis and {Arribas}, Santiago and {Artigau}, {\'E}tienne and {Arvai}, Amanda Rose and {Atcheson}, Paul and {Atkinson}, Charles B. and {Averbukh}, Jesse and {Aymergen}, Cagatay and {Bacinski}, John J. and {Baggett}, Wayne E. and {Bagnasco}, Giorgio and {Baker}, Lynn L. and {Balzano}, Vicki Ann and {Banks}, Kimberly A. and {Baran}, David A. and {Barker}, Elizabeth A. and {Barrett}, Larry K. and {Barringer}, Bruce O. and {Barto}, Allison and {Bast}, William and {Baudoz}, Pierre and {Baum}, Stefi and {Beatty}, Thomas G. and {Beaulieu}, Mathilde and {Bechtold}, Kathryn and {Beck}, Tracy and {Beddard}, Megan M. and {Beichman}, Charles and {Bellagama}, Larry and {Bely}, Pierre and {Berger}, Timothy W. and {Bergeron}, Louis E. and {Bernier}, Antoine-Darveau and {Bertch}, Maria D. and {Beskow}, Charlotte and {Betz}, Laura E. and {Biagetti}, Carl P. and {Birkmann}, Stephan and {Bjorklund}, Kurt F. and {Blackwood}, James D. and {Blazek}, Ronald Paul and {Blossfeld}, Stephen and {Bluth}, Marcel and {Boccaletti}, Anthony and {Boegner}, Jr., Martin E. and {Bohlin}, Ralph C. and {Boia}, John Joseph and {B{\"o}ker}, Torsten and {Bonaventura}, N. and {Bond}, Nicholas A. and {Bosley}, Kari Ann and {Boucarut}, Rene A. and {Bouchet}, Patrice and {Bouwman}, Jeroen and {Bower}, Gary and {Bowers}, Ariel S. and {Bowers}, Charles W. and {Boyce}, Leslye A. and {Boyer}, Christine T. and {Boyer}, Martha L. and {Boyer}, Michael and {Boyer}, Robert and {Bradley}, Larry D. and {Brady}, Gregory R. and {Brandl}, Bernhard R. and {Brannen}, Judith L. and {Breda}, David and {Bremmer}, Harold G. and {Brennan}, David and {Bresnahan}, Pamela A. and {Bright}, Stacey N. and {Broiles}, Brian J. and {Bromenschenkel}, Asa and {Brooks}, Brian H. and {Brooks}, Keira J. and {Brown}, Bob and {Brown}, Bruce and {Brown}, Thomas M. and {Bruce}, Barry W. and {Bryson}, Jonathan G. and {Bujanda}, Edwin D. and {Bullock}, Blake M. and {Bunker}, A.~J. and {Bureo}, Rafael and {Burt}, Irving J. and {Bush}, James Aaron and {Bushouse}, Howard A. and {Bussman}, Marie C. and {Cabaud}, Olivier and {Cale}, Steven and {Calhoon}, Charles D. and {Calvani}, Humberto and {Canipe}, Alicia M. and {Caputo}, Francis M. and {Cara}, Mihai and {Carey}, Larkin and {Case}, Michael Eli and {Cesari}, Thaddeus and {Cetorelli}, Lee D. and {Chance}, Don R. and {Chandler}, Lynn and {Chaney}, Dave and {Chapman}, George N. and {Charlot}, S. and {Chayer}, Pierre and {Cheezum}, Jeffrey I. and {Chen}, Bin and {Chen}, Christine H. and {Cherinka}, Brian and {Chichester}, Sarah C. and {Chilton}, Zachary S. and {Chittiraibalan}, Dharini and {Clampin}, Mark and {Clark}, Charles R. and {Clark}, Kerry W. and {Clark}, Stephanie M. and {Claybrooks}, Edward E. and {Cleveland}, Keith A. and {Cohen}, Andrew L. and {Cohen}, Lester M. and {Col{\'o}n}, Knicole D. and {Coleman}, Benee L. and {Colina}, Luis and {Comber}, Brian J. and {Comeau}, Thomas M. and {Comer}, Thomas and {Conde Reis}, Alain and {Connolly}, Dennis C. and {Conroy}, Kyle E. and {Contos}, Adam R. and {Contreras}, James and {Cook}, Neil J. and {Cooper}, James L. and {Cooper}, Rachel Aviva and {Correia}, Michael F. and {Correnti}, Matteo and {Cossou}, Christophe and {Costanza}, Brian F. and {Coulais}, Alain and {Cox}, Colin R. and {Coyle}, Ray T. and {Cracraft}, Misty M. and {Crew}, Keith A. and {Curtis}, Gary J. and {Cusveller}, Bianca and {Da Costa Maciel}, Cleyciane and {Dailey}, Christopher T. and {Daugeron}, Fr{\'e}d{\'e}ric and {Davidson}, Greg S. and {Davies}, James E. and {Davis}, Katherine Anne and {Davis}, Michael S. and {Day}, Ratna and {de Chambure}, Daniel and {de Jong}, Pauline and {De Marchi}, Guido and {Dean}, Bruce H. and {Decker}, John E. and {Delisa}, Amy S. and {Dell}, Lawrence C. and {Dellagatta}, Gail},
        title = "{The James Webb Space Telescope Mission}",
      journal = {\pasp},
         year = 2023,
        month = jun,
       volume = {135},
       number = {1048},
          eid = {068001},
        pages = {068001},
          doi = {10.1088/1538-3873/acd1b5},
archivePrefix = {arXiv},
       eprint = {2304.04869},
 primaryClass = {astro-ph.IM},
       adsurl = {https://ui.adsabs.harvard.edu/abs/2023PASP..135f8001G}
}

@ARTICLE{Rigby2023,
       author = {{Rigby}, Jane and {Perrin}, Marshall and {McElwain}, Michael and {Kimble}, Randy and {Friedman}, Scott and {Lallo}, Matt and {Doyon}, Ren{\'e} and {Feinberg}, Lee and {Ferruit}, Pierre and {Glasse}, Alistair and {Rieke}, Marcia and {Rieke}, George and {Wright}, Gillian and {Willott}, Chris and {Colon}, Knicole and {Milam}, Stefanie and {Neff}, Susan and {Stark}, Christopher and {Valenti}, Jeff and {Abell}, Jim and {Abney}, Faith and {Abul-Huda}, Yasin and {Acton}, D. Scott and {Adams}, Evan and {Adler}, David and {Aguilar}, Jonathan and {Ahmed}, Nasif and {Albert}, Lo{\"\i}c and {Alberts}, Stacey and {Aldridge}, David and {Allen}, Marsha and {Altenburg}, Martin and {{\'A}lvarez-M{\'a}rquez}, Javier and {Alves de Oliveira}, Catarina and {Andersen}, Greg and {Anderson}, Harry and {Anderson}, Sara and {Argyriou}, Ioannis and {Armstrong}, Amber and {Arribas}, Santiago and {Artigau}, Etienne and {Arvai}, Amanda and {Atkinson}, Charles and {Bacon}, Gregory and {Bair}, Thomas and {Banks}, Kimberly and {Barrientes}, Jaclyn and {Barringer}, Bruce and {Bartosik}, Peter and {Bast}, William and {Baudoz}, Pierre and {Beatty}, Thomas and {Bechtold}, Katie and {Beck}, Tracy and {Bergeron}, Eddie and {Bergkoetter}, Matthew and {Bhatawdekar}, Rachana and {Birkmann}, Stephan and {Blazek}, Ronald and {Blome}, Claire and {Boccaletti}, Anthony and {B{\"o}ker}, Torsten and {Boia}, John and {Bonaventura}, Nina and {Bond}, Nicholas and {Bosley}, Kari and {Boucarut}, Ray and {Bourque}, Matthew and {Bouwman}, Jeroen and {Bower}, Gary and {Bowers}, Charles and {Boyer}, Martha and {Bradley}, Larry and {Brady}, Greg and {Braun}, Hannah and {Breda}, David and {Bresnahan}, Pamela and {Bright}, Stacey and {Britt}, Christopher and {Bromenschenkel}, Asa and {Brooks}, Brian and {Brooks}, Keira and {Brown}, Bob and {Brown}, Matthew and {Brown}, Patricia and {Bunker}, Andy and {Burger}, Matthew and {Bushouse}, Howard and {Cale}, Steven and {Cameron}, Alex and {Cameron}, Peter and {Canipe}, Alicia and {Caplinger}, James and {Caputo}, Francis and {Cara}, Mihai and {Carey}, Larkin and {Carniani}, Stefano and {Carrasquilla}, Maria and {Carruthers}, Margaret and {Case}, Michael and {Catherine}, Riggs and {Chance}, Don and {Chapman}, George and {Charlot}, St{\'e}phane and {Charlow}, Brian and {Chayer}, Pierre and {Chen}, Bin and {Cherinka}, Brian and {Chichester}, Sarah and {Chilton}, Zack and {Chonis}, Taylor and {Clampin}, Mark and {Clark}, Charles and {Clark}, Kerry and {Coe}, Dan and {Coleman}, Benee and {Comber}, Brian and {Comeau}, Tom and {Connolly}, Dennis and {Cooper}, James and {Cooper}, Rachel and {Coppock}, Eric and {Correnti}, Matteo and {Cossou}, Christophe and {Coulais}, Alain and {Coyle}, Laura and {Cracraft}, Misty and {Curti}, Mirko and {Cuturic}, Steven and {Davis}, Katherine and {Davis}, Michael and {Dean}, Bruce and {DeLisa}, Amy and {deMeester}, Wim and {Dencheva}, Nadia and {Dencheva}, Nadezhda and {DePasquale}, Joseph and {Deschenes}, Jeremy and {Hunor Detre}, {\"O}rs and {Diaz}, Rosa and {Dicken}, Dan and {DiFelice}, Audrey and {Dillman}, Matthew and {Dixon}, William and {Doggett}, Jesse and {Donaldson}, Tom and {Douglas}, Rob and {DuPrie}, Kimberly and {Dupuis}, Jean and {Durning}, John and {Easmin}, Nilufar and {Eck}, Weston and {Edeani}, Chinwe and {Egami}, Eiichi and {Ehrenwinkler}, Ralf and {Eisenhamer}, Jonathan and {Eisenhower}, Michael and {Elie}, Michelle and {Elliott}, James and {Elliott}, Kyle and {Ellis}, Tracy and {Engesser}, Michael and {Espinoza}, Nestor and {Etienne}, Odessa and {Etxaluze}, Mireya and {Falini}, Patrick and {Feeney}, Matthew and {Ferry}, Malcolm and {Filippazzo}, Joseph and {Fincham}, Brian and {Fix}, Mees and {Flagey}, Nicolas and {Florian}, Michael and {Flynn}, Jim and {Fontanella}, Erin and {Ford}, Terrance and {Forshay}, Peter and {Fox}, Ori and {Franz}, David and {Fu}, Henry and {Fullerton}, Alexander and {Galkin}, Sergey and {Galyer}, Anthony and {Garc{\'\i}a Mar{\'\i}n}, Macarena and {Gardner}, Jonathan P. and {Gardner}, Lisa and {Garland}, Dennis and {Garrett}, Bruce and {Gasman}, Danny and {Gaspar}, Andras and {Gaudreau}, Daniel and {Gauthier}, Peter and {Geers}, Vincent and {Geithner}, Paul and {Gennaro}, Mario and {Giardino}, Giovanna and {Girard}, Julien and {Giuliano}, Mark and {Glassmire}, Kirk and {Glauser}, Adrian},
        title = "{The Science Performance of JWST as Characterized in Commissioning}",
      journal = {\pasp},
         year = 2023,
        month = apr,
       volume = {135},
       number = {1046},
          eid = {048001},
        pages = {048001},
          doi = {10.1088/1538-3873/acb293},
archivePrefix = {arXiv},
       eprint = {2207.05632},
 primaryClass = {astro-ph.IM},
       adsurl = {https://ui.adsabs.harvard.edu/abs/2023PASP..135d8001R}
}

@ARTICLE{Tang2026,
       author = {{Tang}, Mengtao and {Stark}, Daniel P. and {Mason}, Charlotte A. and {Chen}, Zuyi and {Katz}, Harley and {Gronke}, Max and {Furtak}, Lukas J. and {Chang}, Seok-Jun and {Matthee}, Jorryt and {Whitler}, Lily and {Zitrin}, Adi and {Endsley}, Ryan and {Gelli}, Viola and {Roychowdhury}, Tamojeet and {Senchyna}, Peter and {Topping}, Michael W. and {Zhang}, Meng},
        title = "{SPURS: Evidence for Clumpy Neutral Envelopes and Ionized IGM Surrounding Little Red Dots in Abell 2744 from Ultra-Deep Rest-UV Spectroscopy}",
      journal = {arXiv e-prints},
         year = 2026,
        month = apr,
          eid = {arXiv:2604.03563},
        pages = {arXiv:2604.03563},
          doi = {10.48550/arXiv.2604.03563},
archivePrefix = {arXiv},
       eprint = {2604.03563},
 primaryClass = {astro-ph.GA},
       adsurl = {https://ui.adsabs.harvard.edu/abs/2026arXiv260403563T}
}

@ARTICLE{Foreman-Mackey2013,
       author = {{Foreman-Mackey}, Daniel and {Hogg}, David W. and {Lang}, Dustin and {Goodman}, Jonathan},
        title = "{emcee: The MCMC Hammer}",
      journal = {\pasp},
         year = 2013,
        month = mar,
       volume = {125},
       number = {925},
        pages = {306},
          doi = {10.1086/670067},
archivePrefix = {arXiv},
       eprint = {1202.3665},
 primaryClass = {astro-ph.IM},
       adsurl = {https://ui.adsabs.harvard.edu/abs/2013PASP..125..306F}
}

@ARTICLE{Luridiana2015,
       author = {{Luridiana}, V. and {Morisset}, C. and {Shaw}, R.~A.},
        title = "{PyNeb: a new tool for analyzing emission lines. I. Code description and validation of results}",
      journal = {\aap},
         year = 2015,
        month = jan,
       volume = {573},
          eid = {A42},
        pages = {A42},
          doi = {10.1051/0004-6361/201323152},
archivePrefix = {arXiv},
       eprint = {1410.6662},
 primaryClass = {astro-ph.IM},
       adsurl = {https://ui.adsabs.harvard.edu/abs/2015A&A...573A..42L}
}

@ARTICLE{Calzetti2000,
       author = {{Calzetti}, Daniela and {Armus}, Lee and {Bohlin}, Ralph C. and {Kinney}, Anne L. and {Koornneef}, Jan and {Storchi-Bergmann}, Thaisa},
        title = "{The Dust Content and Opacity of Actively Star-forming Galaxies}",
      journal = {\apj},
         year = 2000,
        month = apr,
       volume = {533},
       number = {2},
        pages = {682-695},
          doi = {10.1086/308692},
archivePrefix = {arXiv},
       eprint = {astro-ph/9911459},
 primaryClass = {astro-ph},
       adsurl = {https://ui.adsabs.harvard.edu/abs/2000ApJ...533..682C}
}

@ARTICLE{Storey1995,
       author = {{Storey}, P.~J. and {Hummer}, D.~G.},
        title = "{Recombination line intensities for hydrogenic ions-IV. Total recombination coefficients and machine-readable tables for Z=1 to 8}",
      journal = {\mnras},
         year = 1995,
        month = jan,
       volume = {272},
       number = {1},
        pages = {41-48},
          doi = {10.1093/mnras/272.1.41},
       adsurl = {https://ui.adsabs.harvard.edu/abs/1995MNRAS.272...41S}
}

@ARTICLE{Aggarwal1999,
       author = {{Aggarwal}, Kanti M. and {Keenan}, Francis P.},
        title = "{Excitation Rate Coefficients for Fine-Structure Transitions in O III}",
      journal = {\apjs},
         year = 1999,
        month = jul,
       volume = {123},
       number = {1},
        pages = {311-349},
          doi = {10.1086/313232},
       adsurl = {https://ui.adsabs.harvard.edu/abs/1999ApJS..123..311A}
}

@ARTICLE{Isobe2026,
       author = {{Isobe}, Yuki and {Curti}, Mirko and {Maiolino}, Roberto and {Duan}, Qiao and {McClymont}, William and {Pusk{\'a}s}, D{\'a}vid and {D'Eugenio}, Francesco and {Rinaldi}, Pierluigi and {Trussler}, James A.~A. and {Scholtz}, Jan and {Looser}, Tobias J. and {Nelson}, Erica and {Ji}, Xihan and {Langeroodi}, Danial and {Tacchella}, Sandro and {Jones}, Gareth C. and {Juod{\v{z}}balis}, Ignas and {Pascalau}, Robert G. and {Hsiao}, Tiger Yu-Yang and {{\"U}bler}, Hannah and {Baker}, William M. and {Bunker}, Andrew J. and {Carniani}, Stefano and {Charlot}, St{\'e}phane and {Curtis-Lake}, Emma and {Geris}, Sophia and {Koller}, Maria and {Lyu}, Jianwei and {Robertson}, Brant and {Williams}, Christina C. and {Wu}, Zihao},
        title = "{JADES: the mass-metallicity relation at $z=1-10$. New calibrations, extremely metal-poor galaxies, and chemical diversity}",
      journal = {arXiv e-prints},
         year = 2026,
        month = jun,
          eid = {arXiv:2606.11345},
        pages = {arXiv:2606.11345},
          doi = {10.48550/arXiv.2606.11345},
archivePrefix = {arXiv},
       eprint = {2606.11345},
 primaryClass = {astro-ph.GA},
       adsurl = {https://ui.adsabs.harvard.edu/abs/2026arXiv260611345I}
}

@ARTICLE{Sanders2026,
       author = {{Sanders}, Ryan L. and {Shapley}, Alice E. and {Topping}, Michael W. and {Reddy}, Naveen A. and {Berg}, Danielle A. and {Khostovan}, Ali Ahmad and {Bouwens}, Rychard J. and {Brammer}, Gabriel and {Carnall}, Adam C. and {Cullen}, Fergus and {Dav{\'e}}, Romeel and {Dunlop}, James S. and {Ellis}, Richard S. and {F{\"o}rster Schreiber}, N.~M. and {Furlanetto}, Steven R. and {Glazebrook}, Karl and {Illingworth}, Garth D. and {Jones}, Tucker and {Kriek}, Mariska and {McLeod}, Derek J. and {McLure}, Ross J. and {Narayanan}, Desika and {Oesch}, Pascal A. and {Pahl}, Anthony J. and {Pettini}, Max and {Schaerer}, Daniel and {Stark}, Daniel P. and {Steidel}, Charles C. and {Tang}, Mengtao and {Clarke}, Leonardo and {Donnan}, Callum T. and {Kehoe}, Emily},
        title = "{The AURORA Survey: High-redshift Empirical Metallicity Calibrations from Electron Temperature Measurements at z = 2─10}",
      journal = {\apj},
         year = 2026,
        month = jun,
       volume = {1003},
       number = {2},
          eid = {228},
        pages = {228},
          doi = {10.3847/1538-4357/ae66e2},
archivePrefix = {arXiv},
       eprint = {2508.10099},
 primaryClass = {astro-ph.GA},
       adsurl = {https://ui.adsabs.harvard.edu/abs/2026ApJ..1003..228S}
}

@ARTICLE{Sanders2024,
       author = {{Sanders}, Ryan L. and {Shapley}, Alice E. and {Topping}, Michael W. and {Reddy}, Naveen A. and {Brammer}, Gabriel B.},
        title = "{Direct T $_{e}$-based Metallicities of z = 2─9 Galaxies with JWST/NIRSpec: Empirical Metallicity Calibrations Applicable from Reionization to Cosmic Noon}",
      journal = {\apj},
         year = 2024,
        month = feb,
       volume = {962},
       number = {1},
          eid = {24},
        pages = {24},
          doi = {10.3847/1538-4357/ad15fc},
archivePrefix = {arXiv},
       eprint = {2303.08149},
 primaryClass = {astro-ph.GA},
       adsurl = {https://ui.adsabs.harvard.edu/abs/2024ApJ...962...24S}
}

@ARTICLE{Tremonti2004,
       author = {{Tremonti}, Christy A. and {Heckman}, Timothy M. and {Kauffmann}, Guinevere and {Brinchmann}, Jarle and {Charlot}, St{\'e}phane and {White}, Simon D.~M. and {Seibert}, Mark and {Peng}, Eric W. and {Schlegel}, David J. and {Uomoto}, Alan and {Fukugita}, Masataka and {Brinkmann}, Jon},
        title = "{The Origin of the Mass-Metallicity Relation: Insights from 53,000 Star-forming Galaxies in the Sloan Digital Sky Survey}",
      journal = {\apj},
         year = 2004,
        month = oct,
       volume = {613},
       number = {2},
        pages = {898-913},
          doi = {10.1086/423264},
archivePrefix = {arXiv},
       eprint = {astro-ph/0405537},
 primaryClass = {astro-ph},
       adsurl = {https://ui.adsabs.harvard.edu/abs/2004ApJ...613..898T}
}

@ARTICLE{Lee2006,
       author = {{Lee}, Henry and {Skillman}, Evan D. and {Cannon}, John M. and {Jackson}, Dale C. and {Gehrz}, Robert D. and {Polomski}, Elisha F. and {Woodward}, Charles E.},
        title = "{On Extending the Mass-Metallicity Relation of Galaxies by 2.5 Decades in Stellar Mass}",
      journal = {\apj},
         year = 2006,
        month = aug,
       volume = {647},
       number = {2},
        pages = {970-983},
          doi = {10.1086/505573},
archivePrefix = {arXiv},
       eprint = {astro-ph/0605036},
 primaryClass = {astro-ph},
       adsurl = {https://ui.adsabs.harvard.edu/abs/2006ApJ...647..970L}
}

@ARTICLE{Erb2006,
       author = {{Erb}, Dawn K. and {Shapley}, Alice E. and {Pettini}, Max and {Steidel}, Charles C. and {Reddy}, Naveen A. and {Adelberger}, Kurt L.},
        title = "{The Mass-Metallicity Relation at z>\raisebox{-0.5ex}\textasciitilde2}",
      journal = {\apj},
         year = 2006,
        month = jun,
       volume = {644},
       number = {2},
        pages = {813-828},
          doi = {10.1086/503623},
archivePrefix = {arXiv},
       eprint = {astro-ph/0602473},
 primaryClass = {astro-ph},
       adsurl = {https://ui.adsabs.harvard.edu/abs/2006ApJ...644..813E}
}

@ARTICLE{Berg2012,
       author = {{Berg}, Danielle A. and {Skillman}, Evan D. and {Marble}, Andrew R. and {van Zee}, Liese and {Engelbracht}, Charles W. and {Lee}, Janice C. and {Kennicutt}, Jr., Robert C. and {Calzetti}, Daniela and {Dale}, Daniel A. and {Johnson}, Benjamin D.},
        title = "{Direct Oxygen Abundances for Low-luminosity LVL Galaxies}",
      journal = {\apj},
         year = 2012,
        month = aug,
       volume = {754},
       number = {2},
          eid = {98},
        pages = {98},
          doi = {10.1088/0004-637X/754/2/98},
archivePrefix = {arXiv},
       eprint = {1205.6782},
 primaryClass = {astro-ph.CO},
       adsurl = {https://ui.adsabs.harvard.edu/abs/2012ApJ...754...98B}
}

@ARTICLE{Curti2020,
       author = {{Curti}, Mirko and {Mannucci}, Filippo and {Cresci}, Giovanni and {Maiolino}, Roberto},
        title = "{The mass-metallicity and the fundamental metallicity relation revisited on a fully T$_{e}$-based abundance scale for galaxies}",
      journal = {\mnras},
         year = 2020,
        month = jan,
       volume = {491},
       number = {1},
        pages = {944-964},
          doi = {10.1093/mnras/stz2910},
archivePrefix = {arXiv},
       eprint = {1910.00597},
 primaryClass = {astro-ph.GA},
       adsurl = {https://ui.adsabs.harvard.edu/abs/2020MNRAS.491..944C}
}

@ARTICLE{Lequeux1979,
       author = {{Lequeux}, J. and {Peimbert}, M. and {Rayo}, J.~F. and {Serrano}, A. and {Torres-Peimbert}, S.},
        title = "{Chemical Composition and Evolution of Irregular and Blue Compact Galaxies}",
      journal = {\aap},
         year = 1979,
        month = dec,
       volume = {80},
        pages = {155},
       adsurl = {https://ui.adsabs.harvard.edu/abs/1979A&A....80..155L}
}

@ARTICLE{Morishita2024,
       author = {{Morishita}, Takahiro and {Stiavelli}, Massimo and {Grillo}, Claudio and {Rosati}, Piero and {Schuldt}, Stefan and {Trenti}, Michele and {Bergamini}, Pietro and {Boyett}, Kit and {Chary}, Ranga-Ram and {Leethochawalit}, Nicha and {Roberts-Borsani}, Guido and {Treu}, Tommaso and {Vanzella}, Eros},
        title = "{Diverse Oxygen Abundance in Early Galaxies Unveiled by Auroral Line Analysis with JWST}",
      journal = {\apj},
         year = 2024,
        month = aug,
       volume = {971},
       number = {1},
          eid = {43},
        pages = {43},
          doi = {10.3847/1538-4357/ad5290},
archivePrefix = {arXiv},
       eprint = {2402.14084},
 primaryClass = {astro-ph.GA},
       adsurl = {https://ui.adsabs.harvard.edu/abs/2024ApJ...971...43M}
}

@ARTICLE{Curti2024,
       author = {{Curti}, Mirko and {Maiolino}, Roberto and {Curtis-Lake}, Emma and {Chevallard}, Jacopo and {Carniani}, Stefano and {D'Eugenio}, Francesco and {Looser}, Tobias J. and {Scholtz}, Jan and {Charlot}, Stephane and {Cameron}, Alex and {{\"U}bler}, Hannah and {Witstok}, Joris and {Boyett}, Kristian and {Laseter}, Isaac and {Sandles}, Lester and {Arribas}, Santiago and {Bunker}, Andrew and {Giardino}, Giovanna and {Maseda}, Michael V. and {Rawle}, Tim and {Rodr{\'\i}guez Del Pino}, Bruno and {Smit}, Renske and {Willott}, Chris J. and {Eisenstein}, Daniel J. and {Hausen}, Ryan and {Johnson}, Benjamin and {Rieke}, Marcia and {Robertson}, Brant and {Tacchella}, Sandro and {Williams}, Christina C. and {Willmer}, Christopher and {Baker}, William M. and {Bhatawdekar}, Rachana and {Egami}, Eiichi and {Helton}, Jakob M. and {Ji}, Zhiyuan and {Kumari}, Nimisha and {Perna}, Michele and {Shivaei}, Irene and {Sun}, Fengwu},
        title = "{JADES: Insights into the low-mass end of the mass-metallicity-SFR relation at 3 < z < 10 from deep JWST/NIRSpec spectroscopy}",
      journal = {\aap},
         year = 2024,
        month = apr,
       volume = {684},
          eid = {A75},
        pages = {A75},
          doi = {10.1051/0004-6361/202346698},
archivePrefix = {arXiv},
       eprint = {2304.08516},
 primaryClass = {astro-ph.GA},
       adsurl = {https://ui.adsabs.harvard.edu/abs/2024A&A...684A..75C}
}

@ARTICLE{Heintz2023,
       author = {{Heintz}, Kasper E. and {Brammer}, Gabriel B. and {Gim{\'e}nez-Arteaga}, Clara and {Strait}, Victoria B. and {Lagos}, Claudia del P. and {Vijayan}, Aswin P. and {Matthee}, Jorryt and {Watson}, Darach and {Mason}, Charlotte A. and {Hutter}, Anne and {Toft}, Sune and {Fynbo}, Johan P.~U. and {Oesch}, Pascal A.},
        title = "{Dilution of chemical enrichment in galaxies 600 Myr after the Big Bang}",
      journal = {Nature Astronomy},
         year = 2023,
        month = dec,
       volume = {7},
        pages = {1517-1524},
          doi = {10.1038/s41550-023-02078-7},
archivePrefix = {arXiv},
       eprint = {2212.02890},
 primaryClass = {astro-ph.GA},
       adsurl = {https://ui.adsabs.harvard.edu/abs/2023NatAs...7.1517H}
}

@ARTICLE{Chakraborty2025,
       author = {{Chakraborty}, Priyanka and {Sarkar}, Arnab and {Smith}, Randall and {Ferland}, Gary J. and {McDonald}, Michael and {Forman}, William and {Vogelsberger}, Mark and {Torrey}, Paul and {Garcia}, Alex M. and {Bautz}, Mark and {Foster}, Adam and {Miller}, Eric and {Grant}, Catherine},
        title = "{Unveiling the Cosmic Chemistry. II. ``Direct'' T$_{e}$-based Metallicity of Galaxies at 3 < z < 10 with JWST/NIRSpec}",
      journal = {\apj},
         year = 2025,
        month = may,
       volume = {985},
       number = {1},
          eid = {24},
        pages = {24},
          doi = {10.3847/1538-4357/adc7b5},
archivePrefix = {arXiv},
       eprint = {2412.15435},
 primaryClass = {astro-ph.GA},
       adsurl = {https://ui.adsabs.harvard.edu/abs/2025ApJ...985...24C}
}

@ARTICLE{Sarkar2025,
       author = {{Sarkar}, Arnab and {Chakraborty}, Priyanka and {Vogelsberger}, Mark and {McDonald}, Michael and {Torrey}, Paul and {Garcia}, Alex M. and {Khullar}, Gourav and {Ferland}, Gary J. and {Forman}, William and {Wolk}, Scott and {Schneider}, Benjamin and {Bautz}, Mark and {Miller}, Eric and {Grant}, Catherine and {ZuHone}, John},
        title = "{Unveiling the Cosmic Chemistry: Revisiting the Mass{\textendash}Metallicity Relation with JWST/NIRSpec at 4 < z < 10}",
      journal = {\apj},
         year = 2025,
        month = jan,
       volume = {978},
       number = {2},
          eid = {136},
        pages = {136},
          doi = {10.3847/1538-4357/ad8f32},
archivePrefix = {arXiv},
       eprint = {2408.07974},
 primaryClass = {astro-ph.GA},
       adsurl = {https://ui.adsabs.harvard.edu/abs/2025ApJ...978..136S}
}

@ARTICLE{Arellano2025,
       author = {{Arellano-C{\'o}rdova}, K.~Z. and {Cullen}, F. and {Carnall}, A.~C. and {Scholte}, D. and {Stanton}, T.~M. and {Kobayashi}, C. and {Martinez}, Z. and {Berg}, D.~A. and {Barrufet}, L. and {Begley}, R. and {Donnan}, C.~T. and {Dunlop}, J.~S. and {Hamadouche}, M.~L. and {McLeod}, D.~J. and {McLure}, R.~J. and {Rowlands}, K. and {Shapley}, A.~E.},
        title = "{The JWST EXCELS survey: direct estimates of C, N, and O abundances in two relatively metal-rich galaxies at z ≃ 5}",
      journal = {\mnras},
         year = 2025,
        month = jul,
       volume = {540},
       number = {4},
        pages = {2991-3007},
          doi = {10.1093/mnras/staf855},
archivePrefix = {arXiv},
       eprint = {2412.10557},
 primaryClass = {astro-ph.GA},
       adsurl = {https://ui.adsabs.harvard.edu/abs/2025MNRAS.540.2991A}
}

@ARTICLE{Pollock2026,
       author = {{Pollock}, Clara L. and {Gottumukkala}, Rashmi and {Heintz}, Kasper E. and {Brammer}, Gabriel B. and {Roberts-Borsani}, Guido and {Oesch}, Pascal A. and {Witstok}, Joris and {Arellano-C{\'o}rdova}, Karla Z. and {Cullen}, Fergus and {Scholte}, Dirk and {Terp}, Chamilla and {Rowland}, Lucie and {Sneppen}, Albert and {Ito}, Kei and {Valentino}, Francesco and {Matthee}, Jorryt and {Watson}, Darach and {Toft}, Sune},
        title = "{Novel z {\ensuremath{\sim}} 10 auroral line measurements extend the gradual offset of the fundamental metallicity relation deep into the first gigayear of cosmic time}",
      journal = {\aap},
         year = 2026,
        month = apr,
       volume = {708},
          eid = {A203},
        pages = {A203},
          doi = {10.1051/0004-6361/202556032},
archivePrefix = {arXiv},
       eprint = {2506.15779},
 primaryClass = {astro-ph.GA},
       adsurl = {https://ui.adsabs.harvard.edu/abs/2026A&A...708A.203P}
}

@ARTICLE{Koller2026,
       author = {{Koller}, Maria and {Maiolino}, Roberto and {{\"U}bler}, Hannah and {Duan}, Qiao and {Scholtz}, Jan and {Arribas}, Santiago and {Baker}, William M. and {Carniani}, Stefano and {Charlot}, Stephane and {Curti}, Mirko and {Graziani}, Luca and {Jones}, Gareth and {McClymont}, William and {Perna}, Michele and {Rodr{\'\i}guez Del Pino}, Bruno and {Tacchella}, Sandro and {Venditti}, Alessandra and {Venturi}, Giacomo and {Witstok}, Joris},
        title = "{Metal Mayhem at $\rm z \sim 7-10$: Diversity and Evolution of Gas-Phase Metallicity Gradients}",
      journal = {arXiv e-prints},
         year = 2026,
        month = apr,
          eid = {arXiv:2604.07076},
        pages = {arXiv:2604.07076},
          doi = {10.48550/arXiv.2604.07076},
archivePrefix = {arXiv},
       eprint = {2604.07076},
 primaryClass = {astro-ph.GA},
       adsurl = {https://ui.adsabs.harvard.edu/abs/2026arXiv260407076K}
}

@ARTICLE{Nishigaki2025,
       author = {{Nishigaki}, Moka and {Nakajima}, Kimihiko and {Ouchi}, Masami and {Behroozi}, Peter and {Nakane}, Minami and {Takeda}, Yui and {Umeda}, Hiroya and {Yajima}, Hidenobu and {Yanagisawa}, Hiroto},
        title = "{DREAMS.II. Galaxy Demographics from Direct Te-Based Metallicities at z\raisebox{-0.5ex}\textasciitilde2-10: Tracing the Evolution of the Mass-Metallicity and Fundamental Relations}",
      journal = {arXiv e-prints},
         year = 2025,
        month = dec,
          eid = {arXiv:2512.12983},
        pages = {arXiv:2512.12983},
          doi = {10.48550/arXiv.2512.12983},
archivePrefix = {arXiv},
       eprint = {2512.12983},
 primaryClass = {astro-ph.GA},
       adsurl = {https://ui.adsabs.harvard.edu/abs/2025arXiv251212983N}
}

@ARTICLE{Ellison2008,
       author = {{Ellison}, Sara L. and {Patton}, David R. and {Simard}, Luc and {McConnachie}, Alan W.},
        title = "{Galaxy Pairs in the Sloan Digital Sky Survey. I. Star Formation, Active Galactic Nucleus Fraction, and the Mass-Metallicity Relation}",
      journal = {\aj},
         year = 2008,
        month = may,
       volume = {135},
       number = {5},
        pages = {1877-1899},
          doi = {10.1088/0004-6256/135/5/1877},
archivePrefix = {arXiv},
       eprint = {0803.0161},
 primaryClass = {astro-ph},
       adsurl = {https://ui.adsabs.harvard.edu/abs/2008AJ....135.1877E}
}

@ARTICLE{Mannucci2010,
       author = {{Mannucci}, F. and {Cresci}, G. and {Maiolino}, R. and {Marconi}, A. and {Gnerucci}, A.},
        title = "{A fundamental relation between mass, star formation rate and metallicity in local and high-redshift galaxies}",
      journal = {\mnras},
         year = 2010,
        month = nov,
       volume = {408},
       number = {4},
        pages = {2115-2127},
          doi = {10.1111/j.1365-2966.2010.17291.x},
archivePrefix = {arXiv},
       eprint = {1005.0006},
 primaryClass = {astro-ph.CO},
       adsurl = {https://ui.adsabs.harvard.edu/abs/2010MNRAS.408.2115M}
}

@ARTICLE{Sanders2021,
       author = {{Sanders}, Ryan L. and {Shapley}, Alice E. and {Jones}, Tucker and {Reddy}, Naveen A. and {Kriek}, Mariska and {Siana}, Brian and {Coil}, Alison L. and {Mobasher}, Bahram and {Shivaei}, Irene and {Dav{\'e}}, Romeel and {Azadi}, Mojegan and {Price}, Sedona H. and {Leung}, Gene and {Freeman}, William R. and {Fetherolf}, Tara and {de Groot}, Laura and {Zick}, Tom and {Barro}, Guillermo},
        title = "{The MOSDEF Survey: The Evolution of the Mass-Metallicity Relation from z = 0 to z 3.3}",
      journal = {\apj},
         year = 2021,
        month = jun,
       volume = {914},
       number = {1},
          eid = {19},
        pages = {19},
          doi = {10.3847/1538-4357/abf4c1},
archivePrefix = {arXiv},
       eprint = {2009.07292},
 primaryClass = {astro-ph.GA},
       adsurl = {https://ui.adsabs.harvard.edu/abs/2021ApJ...914...19S}
}

@ARTICLE{Nakajima2023,
       author = {{Nakajima}, Kimihiko and {Ouchi}, Masami and {Isobe}, Yuki and {Harikane}, Yuichi and {Zhang}, Yechi and {Ono}, Yoshiaki and {Umeda}, Hiroya and {Oguri}, Masamune},
        title = "{JWST Census for the Mass-Metallicity Star Formation Relations at z = 4-10 with Self-consistent Flux Calibration and Proper Metallicity Calibrators}",
      journal = {\apjs},
         year = 2023,
        month = dec,
       volume = {269},
       number = {2},
          eid = {33},
        pages = {33},
          doi = {10.3847/1538-4365/acd556},
archivePrefix = {arXiv},
       eprint = {2301.12825},
 primaryClass = {astro-ph.GA},
       adsurl = {https://ui.adsabs.harvard.edu/abs/2023ApJS..269...33N}
}

@ARTICLE{Hsiao2025b,
       author = {{Hsiao}, Tiger Yu-Yang and {Topping}, Michael W. and {Coe}, Dan and {Chisholm}, John and {Berg}, Danielle A. and {Abdurro'uf} and {{\'A}lvarez-M{\'a}rquez}, Javier and {Maiolino}, Roberto and {Dayal}, Pratika and {Furtak}, Lukas J.},
        title = "{First Direct Carbon Abundance Measured at z > 10 in the Lensed Galaxy MACS0647-JD}",
      journal = {\apj},
         year = 2025,
        month = nov,
       volume = {993},
       number = {1},
          eid = {70},
        pages = {70},
          doi = {10.3847/1538-4357/ae07d7},
archivePrefix = {arXiv},
       eprint = {2409.04625},
 primaryClass = {astro-ph.GA},
       adsurl = {https://ui.adsabs.harvard.edu/abs/2025ApJ...993...70H}
}

@ARTICLE{DEugenio2024,
       author = {{D'Eugenio}, Francesco and {Maiolino}, Roberto and {Carniani}, Stefano and {Chevallard}, Jacopo and {Curtis-Lake}, Emma and {Witstok}, Joris and {Charlot}, Stephane and {Baker}, William M. and {Arribas}, Santiago and {Boyett}, Kristan and {Bunker}, Andrew J. and {Curti}, Mirko and {Eisenstein}, Daniel J. and {Hainline}, Kevin and {Ji}, Zhiyuan and {Johnson}, Benjamin D. and {Kumari}, Nimisha and {Looser}, Tobias J. and {Nakajima}, Kimihiko and {Nelson}, Erica and {Rieke}, Marcia and {Robertson}, Brant and {Scholtz}, Jan and {Smit}, Renske and {Sun}, Fengwu and {Venturi}, Giacomo and {Tacchella}, Sandro and {{\"U}bler}, Hannah and {Willmer}, Christopher N.~A. and {Willott}, Chris},
        title = "{JADES: Carbon enrichment 350 Myr after the Big Bang}",
      journal = {\aap},
         year = 2024,
        month = sep,
       volume = {689},
          eid = {A152},
        pages = {A152},
          doi = {10.1051/0004-6361/202348636},
archivePrefix = {arXiv},
       eprint = {2311.09908},
 primaryClass = {astro-ph.GA},
       adsurl = {https://ui.adsabs.harvard.edu/abs/2024A&A...689A.152D}
}

@ARTICLE{Nakajima2026,
       author = {{Nakajima}, Kimihiko and {Ouchi}, Masami and {Harikane}, Yuichi and {Vanzella}, Eros and {Ono}, Yoshiaki and {Isobe}, Yuki and {Nishigaki}, Moka and {Tsujimoto}, Takuji and {Nakamura}, Fumitaka and {Xu}, Yi and {Umeda}, Hiroya and {Zhang}, Yechi},
        title = "{An ultra-faint, chemically primitive galaxy forming in the reionization era}",
      journal = {\nat},
         year = 2026,
        month = may,
       volume = {653},
       number = {8114},
        pages = {363-367},
          doi = {10.1038/s41586-026-10374-1},
archivePrefix = {arXiv},
       eprint = {2506.11846},
 primaryClass = {astro-ph.GA},
       adsurl = {https://ui.adsabs.harvard.edu/abs/2026Natur.653..363N}
}

@ARTICLE{Scholtz2026,
       author = {{Scholtz}, J. and {Silcock}, M.~S. and {Curtis-Lake}, E. and {Maiolino}, R. and {Carniani}, S. and {D'Eugenio}, F. and {Ji}, X. and {Jakobsen}, P. and {Hainline}, K. and {Arribas}, S. and {Baker}, W.~M. and {Bhatawdekar}, R. and {Bunker}, A.~J. and {Charlot}, S. and {Chevallard}, J. and {Curti}, M. and {Eisenstein}, Daniel J. and {Isobe}, Y. and {Jones}, G.~C. and {Parlanti}, E. and {P{\'e}rez-Gonz{\'a}lez}, P.~G. and {Rinaldi}, P. and {Robertson}, B. and {Tacchella}, S. and {{\"U}bler}, H. and {Williams}, C.~C. and {Willott}, C. and {Witstok}, J.},
        title = "{JADES: carbon-enhanced, nitrogen-normal compact galaxy at z = 11.2}",
      journal = {\mnras},
         year = 2026,
        month = jan,
       volume = {545},
       number = {3},
          eid = {staf2107},
        pages = {staf2107},
          doi = {10.1093/mnras/staf2107},
archivePrefix = {arXiv},
       eprint = {2507.17809},
 primaryClass = {astro-ph.GA},
       adsurl = {https://ui.adsabs.harvard.edu/abs/2026MNRAS.545f2107S}
}

@ARTICLE{Pollock2026b,
       author = {{Pollock}, Clara L. and {Heintz}, Kasper E. and {Rusta}, Elka and {Watson}, Darach and {Salvadori}, Stefania and {Witten}, Callum and {Witstok}, Joris and {Koutsouridou}, Ioanna and {Gelli}, Viola and {Oesch}, Pascal A. and {Brammer}, Gabriel B. and {Saccardi}, Andrea and {Ito}, Kei and {Gottumukkala}, Rashmi and {Bergholt}, Dagmar and {Brooksby}, Kasper R. and {Terp}, Chamilla and {Valentino}, Francesco},
        title = "{Chemical signatures from the first stars embedded in metal-poor gas in galaxies at cosmic dawn}",
      journal = {arXiv e-prints},
         year = 2026,
        month = jun,
          eid = {arXiv:2606.13078},
        pages = {arXiv:2606.13078},
          doi = {10.48550/arXiv.2606.13078},
archivePrefix = {arXiv},
       eprint = {2606.13078},
 primaryClass = {astro-ph.GA},
       adsurl = {https://ui.adsabs.harvard.edu/abs/2026arXiv260613078P}
}

@ARTICLE{Cameron2023,
       author = {{Cameron}, Alex J. and {Katz}, Harley and {Rey}, Martin P. and {Saxena}, Aayush},
        title = "{Nitrogen enhancements 440 Myr after the big bang: supersolar N/O, a tidal disruption event, or a dense stellar cluster in GN-z11?}",
      journal = {\mnras},
         year = 2023,
        month = aug,
       volume = {523},
       number = {3},
        pages = {3516-3525},
          doi = {10.1093/mnras/stad1579},
archivePrefix = {arXiv},
       eprint = {2302.10142},
 primaryClass = {astro-ph.GA},
       adsurl = {https://ui.adsabs.harvard.edu/abs/2023MNRAS.523.3516C}
}

@ARTICLE{Isobe2023b,
       author = {{Isobe}, Yuki and {Ouchi}, Masami and {Tominaga}, Nozomu and {Watanabe}, Kuria and {Nakajima}, Kimihiko and {Umeda}, Hiroya and {Yajima}, Hidenobu and {Harikane}, Yuichi and {Fukushima}, Hajime and {Xu}, Yi and {Ono}, Yoshiaki and {Zhang}, Yechi},
        title = "{JWST Identification of Extremely Low C/N Galaxies with [N/O] {\ensuremath{\gtrsim}} 0.5 at z 6-10 Evidencing the Early CNO-cycle Enrichment and a Connection with Globular Cluster Formation}",
      journal = {\apj},
         year = 2023,
        month = dec,
       volume = {959},
       number = {2},
          eid = {100},
        pages = {100},
          doi = {10.3847/1538-4357/ad09be},
archivePrefix = {arXiv},
       eprint = {2307.00710},
 primaryClass = {astro-ph.GA},
       adsurl = {https://ui.adsabs.harvard.edu/abs/2023ApJ...959..100I}
}

@ARTICLE{Topping2024,
       author = {{Topping}, Michael W. and {Stark}, Daniel P. and {Senchyna}, Peter and {Plat}, Adele and {Zitrin}, Adi and {Endsley}, Ryan and {Charlot}, St{\'e}phane and {Furtak}, Lukas J. and {Maseda}, Michael V. and {Smit}, Renske and {Mainali}, Ramesh and {Chevallard}, Jacopo and {Molyneux}, Stephen and {Rigby}, Jane R.},
        title = "{Metal-poor star formation at z > 6 with JWST: new insight into hard radiation fields and nitrogen enrichment on 20 pc scales}",
      journal = {\mnras},
         year = 2024,
        month = apr,
       volume = {529},
       number = {4},
        pages = {3301-3322},
          doi = {10.1093/mnras/stae682},
archivePrefix = {arXiv},
       eprint = {2401.08764},
 primaryClass = {astro-ph.GA},
       adsurl = {https://ui.adsabs.harvard.edu/abs/2024MNRAS.529.3301T}
}

@ARTICLE{Zhang2026,
       author = {{Zhang}, Yechi and {Morishita}, Takahiro and {Stiavelli}, Massimo},
        title = "{Potential Nitrogen Enrichment via Direct-collapse Wolf─Rayet Stars in a z = 4.7 Star-forming Galaxy}",
      journal = {\apj},
         year = 2026,
        month = feb,
       volume = {998},
       number = {1},
          eid = {141},
        pages = {141},
          doi = {10.3847/1538-4357/ae3825},
archivePrefix = {arXiv},
       eprint = {2502.04817},
 primaryClass = {astro-ph.GA},
       adsurl = {https://ui.adsabs.harvard.edu/abs/2026ApJ...998..141Z}
}

@ARTICLE{Marques-Chaves2024,
       author = {{Marques-Chaves}, R. and {Schaerer}, D. and {Kuruvanthodi}, A. and {Korber}, D. and {Prantzos}, N. and {Charbonnel}, C. and {Weibel}, A. and {Izotov}, Y.~I. and {Messa}, M. and {Brammer}, G. and {Dessauges-Zavadsky}, M. and {Oesch}, P.},
        title = "{Extreme N-emitters at high redshift: Possible signatures of supermassive stars and globular cluster or black hole formation in action}",
      journal = {\aap},
         year = 2024,
        month = jan,
       volume = {681},
          eid = {A30},
        pages = {A30},
          doi = {10.1051/0004-6361/202347411},
archivePrefix = {arXiv},
       eprint = {2307.04234},
 primaryClass = {astro-ph.GA},
       adsurl = {https://ui.adsabs.harvard.edu/abs/2024A&A...681A..30M}
}

@ARTICLE{Berg2026,
       author = {{Berg}, Danielle A. and {Naidu}, Rohan P. and {Chisholm}, John and {Atek}, Hakim and {Fujimoto}, Seiji and {Kokorev}, Vasily and {Furtak}, Lukas J. and {Kobayashi}, Chiaki and {Schaerer}, Daniel and {Adamo}, Angela and {Fei}, Qinyue and {Korber}, Damien and {Matthee}, Jorryt and {Marques-Chaves}, Rui and {Martinez}, Zorayda and {McQuinn}, Kristen. B.~W. and {Mu{\~n}oz}, Julian B. and {Oesch}, Pascal A. and {Saldana-Lopez}, Alberto and {Stark}, Daniel P. and {Stephenson}, Mabel G. and {Hsiao}, Tiger Yu-Yang},
        title = "{A Fleeting GLIMPSE of N/O Enrichment at Cosmic Dawn: Evidence for Wolf Rayet N Stars in a z = 6.1 Galaxy}",
      journal = {\apj},
         year = 2026,
        month = jun,
       volume = {1003},
       number = {2},
          eid = {112},
        pages = {112},
          doi = {10.3847/1538-4357/ae5e4c},
archivePrefix = {arXiv},
       eprint = {2511.13591},
 primaryClass = {astro-ph.GA},
       adsurl = {https://ui.adsabs.harvard.edu/abs/2026ApJ..1003..112B}
}

@ARTICLE{Morel2025,
       author = {{Morel}, I. and {Schaerer}, D. and {Marques-Chaves}, R. and {Prantzos}, N. and {Charbonnel}, C. and {Brammer}, G. and {Xiao}, M. and {Dessauges-Zavadsky}, M.},
        title = "{Discovery of new N-emitters over a wide redshift range}",
      journal = {arXiv e-prints},
         year = 2025,
        month = nov,
          eid = {arXiv:2511.20484},
        pages = {arXiv:2511.20484},
          doi = {10.48550/arXiv.2511.20484},
archivePrefix = {arXiv},
       eprint = {2511.20484},
 primaryClass = {astro-ph.GA},
       adsurl = {https://ui.adsabs.harvard.edu/abs/2025arXiv251120484M}
}

@ARTICLE{Cameron2026,
       author = {{Cameron}, Alex J. and {Carreira}, Courtney and {Simmonds}, Charlotte and {Bunker}, Andrew J. and {Saxena}, Aayush and {Carniani}, Stefano and {Charlot}, St{\'e}phane and {Chevallard}, Jacopo and {Curtis-Lake}, Emma and {Hainline}, Kevin and {Hausen}, Ryan and {Ji}, Xihan and {Ji}, Zhiyuan and {Johnson}, Benjamin D. and {Rinaldi}, Pierluigi and {Robertson}, Brant and {Scholtz}, Jan and {Silcock}, Maddie S. and {Tacchella}, Sandro and {Trussler}, James A.~A. and {{\"U}bler}, Hannah and {Williams}, Christina C. and {Willmer}, Christopher N.~A. and {Willott}, Chris and {Witstok}, Joris},
        title = "{JADES: Evolution of nitrogen abundances in star-forming galaxies from z \raisebox{-0.5ex}\textasciitilde 1.5-7}",
      journal = {arXiv e-prints},
         year = 2026,
        month = jan,
          eid = {arXiv:2601.15964},
        pages = {arXiv:2601.15964},
          doi = {10.48550/arXiv.2601.15964},
archivePrefix = {arXiv},
       eprint = {2601.15964},
 primaryClass = {astro-ph.GA},
       adsurl = {https://ui.adsabs.harvard.edu/abs/2026arXiv260115964C}
}

@ARTICLE{Stanghellini1989,
       author = {{Stanghellini}, Letizia and {Kaler}, James B.},
        title = "{Electron Densities in Planetary Nebulae}",
      journal = {\apj},
         year = 1989,
        month = aug,
       volume = {343},
        pages = {811},
          doi = {10.1086/167751},
       adsurl = {https://ui.adsabs.harvard.edu/abs/1989ApJ...343..811S}
}

@ARTICLE{Gutierrez2010,
       author = {{Guti{\'e}rrez}, Leonel and {Beckman}, John E.},
        title = "{The Galaxy-wide Distributions of Mean Electron Density in the H II Regions of M51 and NGC 4449}",
      journal = {\apjl},
         year = 2010,
        month = feb,
       volume = {710},
       number = {1},
        pages = {L44-L48},
          doi = {10.1088/2041-8205/710/1/L44},
archivePrefix = {arXiv},
       eprint = {1001.0276},
 primaryClass = {astro-ph.CO},
       adsurl = {https://ui.adsabs.harvard.edu/abs/2010ApJ...710L..44G}
}

@ARTICLE{Maseda2017,
       author = {{Maseda}, Michael V. and {Brinchmann}, Jarle and {Franx}, Marijn and {Bacon}, Roland and {Bouwens}, Rychard J. and {Schmidt}, Kasper B. and {Boogaard}, Leindert A. and {Contini}, Thierry and {Feltre}, Anna and {Inami}, Hanae and {Kollatschny}, Wolfram and {Marino}, Raffaella A. and {Richard}, Johan and {Verhamme}, Anne and {Wisotzki}, Lutz},
        title = "{The MUSE Hubble Ultra Deep Field Survey. IV. Global properties of C III] emitters}",
      journal = {\aap},
         year = 2017,
        month = dec,
       volume = {608},
          eid = {A4},
        pages = {A4},
          doi = {10.1051/0004-6361/201730985},
archivePrefix = {arXiv},
       eprint = {1710.06432},
 primaryClass = {astro-ph.GA},
       adsurl = {https://ui.adsabs.harvard.edu/abs/2017A&A...608A...4M}
}

@ARTICLE{Simpson2004,
       author = {{Simpson}, Janet P. and {Rubin}, Robert H. and {Colgan}, Sean W.~J. and {Erickson}, Edwin F. and {Haas}, Michael R.},
        title = "{On the Measurement of Elemental Abundance Ratios in Inner Galaxy H II Regions}",
      journal = {\apj},
         year = 2004,
        month = aug,
       volume = {611},
       number = {1},
        pages = {338-352},
          doi = {10.1086/422028},
archivePrefix = {arXiv},
       eprint = {astro-ph/0404404},
 primaryClass = {astro-ph},
       adsurl = {https://ui.adsabs.harvard.edu/abs/2004ApJ...611..338S}
}

@ARTICLE{Rubin2011,
       author = {{Rubin}, Robert H. and {Simpson}, Janet P. and {O'Dell}, C.~R. and {McNabb}, Ian A. and {Colgan}, Sean W.~J. and {Zhuge}, Scott Y. and {Ferland}, Gary J. and {Hidalgo}, Sergio A.},
        title = "{Spitzer reveals what is behind Orion's Bar}",
      journal = {\mnras},
         year = 2011,
        month = jan,
       volume = {410},
       number = {2},
        pages = {1320-1348},
          doi = {10.1111/j.1365-2966.2010.17522.x},
archivePrefix = {arXiv},
       eprint = {1008.2736},
 primaryClass = {astro-ph.GA},
       adsurl = {https://ui.adsabs.harvard.edu/abs/2011MNRAS.410.1320R}
}

@ARTICLE{Franco2000,
       author = {{Franco}, Jos{\'e} and {Kurtz}, Stan and {Hofner}, Peter and {Testi}, Leonardo and {Garc{\'\i}a-Segura}, Guillermo and {Martos}, Marco},
        title = "{The Density Structure of Highly Compact H II Regions}",
      journal = {\apjl},
         year = 2000,
        month = oct,
       volume = {542},
       number = {2},
        pages = {L143-L146},
          doi = {10.1086/312938},
       adsurl = {https://ui.adsabs.harvard.edu/abs/2000ApJ...542L.143F}
}

@ARTICLE{Asplund2021,
       author = {{Asplund}, M. and {Amarsi}, A.~M. and {Grevesse}, N.},
        title = "{The chemical make-up of the Sun: A 2020 vision}",
      journal = {\aap},
         year = 2021,
        month = sep,
       volume = {653},
          eid = {A141},
        pages = {A141},
          doi = {10.1051/0004-6361/202140445},
archivePrefix = {arXiv},
       eprint = {2105.01661},
 primaryClass = {astro-ph.SR},
       adsurl = {https://ui.adsabs.harvard.edu/abs/2021A&A...653A.141A}
}

@ARTICLE{Campbell1986,
       author = {{Campbell}, Alison and {Terlevich}, Roberto and {Melnick}, Jorge},
        title = "{The stellar populations and evolution of H II galaxies - I. High signal-to-noise optical spectroscopy.}",
      journal = {\mnras},
         year = 1986,
        month = dec,
       volume = {223},
        pages = {811-825},
          doi = {10.1093/mnras/223.4.811},
       adsurl = {https://ui.adsabs.harvard.edu/abs/1986MNRAS.223..811C}
}

@ARTICLE{Chen2026,
       author = {{Chen}, Zuyi and {Stark}, Daniel P. and {Mason}, Charlotte A. and {Plat}, Adele and {Gelli}, Viola and {Senchyna}, Peter and {Keerthi Vasan G.}, C. and {Endsley}, Ryan and {Tang}, Mengtao and {Topping}, Michael W. and {Whitler}, Lily},
        title = "{SPURS: Bursty Star Formation in an Extremely Luminous Weak Emission Line Galaxy at $z=9.3$}",
      journal = {arXiv e-prints},
         year = 2026,
        month = apr,
          eid = {arXiv:2604.21516},
        pages = {arXiv:2604.21516},
          doi = {10.48550/arXiv.2604.21516},
archivePrefix = {arXiv},
       eprint = {2604.21516},
 primaryClass = {astro-ph.GA},
       adsurl = {https://ui.adsabs.harvard.edu/abs/2026arXiv260421516C}
}

@BOOK{Osterbrock2006,
       author = {{Osterbrock}, Donald E. and {Ferland}, Gary J.},
        title = "{Astrophysics of gaseous nebulae and active galactic nuclei}",
         year = 2006,
       adsurl = {https://ui.adsabs.harvard.edu/abs/2006agna.book.....O}
}
\bibliographystyle{aasjournal}

%% This command is needed to show the entire author+affiliation list when
%% the collaboration and author truncation commands are used.  It has to
%% go at the end of the manuscript.
%\allauthors

%% Include this line if you are using the \added, \replaced, \deleted
%% commands to see a summary list of all changes at the end of the article.
%\listofchanges

\end{document}